\documentclass[]{aa}

\usepackage{graphicx}
\usepackage{natbib}
\usepackage{subfigure}

\def\gtrsim{\mathrel{\hbox{\rlap{\hbox{\lower4pt\hbox{$\sim$}}}\hbox{$>$}}}}
\def\lesssim{\mathrel{\hbox{\rlap{\hbox{\lower4pt\hbox{$\sim$}}}\hbox{$<$}}}}

\bibpunct{(}{)}{;}{a}{}{,} 

\begin{document}

\title{The ESO-Spitzer Imaging extragalactic Survey (ESIS) II:\\
VIMOS I,z wide field imaging of ELAIS-S1 and selection of 
distant massive galaxies
\thanks{Based on observations collected at the European Southern
Observatory, Chile, ESO No. 168.A-0322(A).} \thanks{ESIS web page:
http://www.astro.unipd.it/esis}}

\author{Stefano Berta\inst{1}\fnmsep\inst{2}
        \and
        Stefano Rubele\inst{2}\fnmsep\inst{3}
	\and
	Alberto Franceschini\inst{2}
	\and
	Enrico Held\inst{3}
	\and
	Luca Rizzi\inst{4}
	\and
	Giulia Rodighiero\inst{2}
	\and
	Andrea Cimatti\inst{5}
	\and
	Jo\~ao E. Dias\inst{6}\fnmsep\inst{7}
	\and
	Chiara Feruglio\inst{8}
	\and
	Fabio La Franca\inst{9}
	\and
	Carol~J. Lonsdale\inst{10}\fnmsep\inst{11}
	\and
	Roberto Maiolino\inst{12}
	\and
	Israel Matute\inst{6}
	\and
	Micheal Rowan-Robinson\inst{13}
	\and
	Nicola Sacchi\inst{9}
	\and
	Giovanni Zamorani\inst{14}
}

\offprints{Stefano Berta, \email{berta@mpe.mpg.de, ste\_atreb@yahoo.it}}

\institute{Max-Planck-Institut f\"{u}r Extraterrestrische Physik (MPE),
Postfach 1312, 85741 Garching, Germany.
\and
Dipartimento di Astronomia, Universit\`a di Padova, Vicolo dell'Osservatorio 3, 
35122 Padova, Italy.
\and 
INAF -- Osservatorio Astronomico di Padova, Vicolo dell'Osservatorio 5,
35122 Padova, Italy.
\and
Joint Astronomy Centre, University Park, 660 N. Aohoku Place, Hilo, HI 96720, USA.
\and
Dipartimento di Astronomia, Universit\`a di Bologna, Via Ranzani 1,
40127 Bologna, Italy.
\and
INAF -- Osservatorio Astronomico di Arcetri, Largo E. Fermi 5, 
50125 Firenze, Italy. 
\and
Centro de Astronomia \& Astrofisica da Universidade de Lisboa, Tapada da Ajuda, 1349-018, Lisboa, Portugal. 
\and
CEA / Service d'Astrophysique, Bat 709 CE-Saclay, Orme des Merisiers, F 91 191 Gif-sur-Yvette, France.
\and
Dipartimento di Fisica, Universit\`a\ degli Studi `Roma Tre', Via della Vasca Navale 84, I-00146 Roma, Italy. 
\and
Department of Astronomy, University of Virginia, Charlottesville, VA 22904, USA.
\and
Infrared Processing \& Analysis Center, California Institute of Technology
100-22, Pasadena, CA 91125, USA. 
\and
INAF -- Osservatorio Astronomico di Roma, Via Frascati 33, I-00044 
Monteporzio Catone, Italy. 
\and
Astrophysics Group, Blackett Laboratory, Imperial College of Science 
Technology and Medicine, Prince Consort Road, London SW7 2AZ.
\and
INAF-Osservatorio Astronomico di Bologna, via Ranzani 1, I-40127 Bologna, Italy.
}

\date{Received April 25th, 2008; accepted June 21st, 2008.}

\titlerunning{ESIS-VIMOS}
\authorrunning{Berta S., et al. }

 
  \abstract
   {The \begin{em}ESO-Spitzer Imaging extragalactic Survey\end{em} (ESIS) is the optical 
   follow up of the \begin{em}Spitzer Wide-area Infra-Red Extragalactic\end{em} survey (SWIRE)
   in the ELAIS-S1 region of the sky.}
   {In the era of observational cosmology, the main efforts are focused on 
   the study of galaxy evolution and its environmental 
   dependence.  Wide area, multiwavelength, extragalactic surveys 
   are needed in order to probe sufficiently large volumes,
   minimize cosmic variance and find significant numbers of rare objects.
   }
   {We present VIMOS $I$ and $z$ band imaging belonging to the ESIS survey. 
   A total of $\sim4$ deg$^2$ were targeted in $I$ and $\sim1$ deg$^2$ in $z$.
   Accurate data processing includes removal of fringing, 
   and mosaicking of the complex observing pattern. Completeness levels and 
   photometric uncertainties are estimated through simulations.
   The multi-wavelength data available in the area are exploited to identify 
   high-redshift galaxies, using the IR-peak technique.}
   {More than 300000 galaxies have been detected in the $I$ band and $\sim50000$ in the 
   $z$ band. Object coordinates are defined within an uncertainty of $\sim0.2$ arcsec r.m.s., 
   with respect to GSC 2.2. We reach a 90\% average completeness at 
   23.1 and 22.5 mag (Vega) in the $I$ and $z$ bands, respectively.
   On the basis of IRAC colors, we identified galaxies having the 1.6$\mu$m stellar 
   peak shifted to $z=1-3$. The new $I,\ z$ band data provide reliable constraints 
   to avoid low-redshift interlopers and reinforce this selection. 
   Roughly 1000 galaxies between $z=2-3$ were identified over the ESIS $\sim4$ deg$^2$, 
   at the SWIRE 5.8$\mu$m depth (25.8 $\mu$Jy at 3$\sigma$). 
   These are the best galaxy candidates to dominate the massive tail ($M>10^{11}$ M$_\odot$)
   of the $z>2$ mass function.}
   {}


   \maketitle



\section{Introduction}\label{sect:intro}

In the recent years, the observational effort in cosmology has focused on 
the understanding of galaxy formation and tracing the processes 
that transformed the smooth and homogeneous Universe detected right after the
Big Bang into the clumpy, clustered structures we see at $z=0$.

With this respect, the critical parameters to be probed turned out to be 
baryonic mass and the epoch of its major assembly 
\citep[e.g.][]{pozzetti2007,rudnick2006,rudnick2003,
dickinson2003b,brinchmann2000}.
In this respect, particularly controversial is the origin of the recently discovered population
of massive ($M>10^{11}$ M$_\odot$) galaxies at redshift $z>1.5-2.0$
\citep[e.g.][]{berta2007b,fontana2006,fontana2004,franceschini2006,
bundy2005,daddi2004,cimatti2002c}.

This puzzle has been approached with two opposite ``philosophies'':
very-deep, pencil-beam surveys, or wide-area, shallower surveys. 

The first approach has mostly relied on space-based and 8m telescopes observations,
that provided the deepest and clearest views of the distant Universe so far.
Some examples, are the Hubble Deep Fields
\citep{williams1996,williams2000}, the GOODS survey 
\citep{dickinson2003a} and the Hubble Ultra Deep Field \citep{beckwith2003},
pushing the detection limit to the 30th magnitude.

Due to their small areas, these surveys are affected by non negligible cosmic variance, 
which can be a significant source of uncertainty in the study of galaxy
statistical properties, even at high redshift \citep[see for
example][]{somerville2004}.

Moreover the search for rare objects, such as massive high-redshift galaxies, 
requires large volumes to be sampled, in order to detect them in representative 
numbers. For example, \citet{berta2007b} find a number density of $\sim 3\times
10^{-5}$ $[$h$_{70}^3$ Mpc$^{-3}]$ for $M>1.6\times10^{11}$ M$_\odot$ galaxies 
at $z=2-3$, which translates in only a few hundreds sources per square degree.

In overcoming these effects, wide field surveys, such as COSMOS \citep[2
deg$^2$][]{scoville2007a}, VVDS \citep[16 deg$^2$][]{lefevre2004}, CFHTLS
\citep[4 to 410 deg$^2$ at different depths, e.g.][]{nuijten2005} play 
a crucial role. 

In this framework, the {\em Spitzer Wide-area Infra-Red Extragalactic} survey
(SWIRE, 49 deg$^2$) is quite unique.
The availability of Spitzer IRAC (3.6, 4.5, 5.8, 8.0 $\mu$m) 
data samples the restframe near-IR emission of galaxies up to 
redshift $z=3$ at least, directly probing the stellar mass assembly of galaxies in the 
distant Universe. Despite SWIRE being a rather shallow survey (reaching 5$\sigma$ depths of 3.7 and 43 $\mu$Jy
at 3.6 and 5.8 $\mu$m, respectively), it detects $2\times10^{11}$ M$_\odot$
galaxies up to $z=3$ \citep{berta2007b}.
MIPS (mainly 24$\mu$m) data secure the identification of star forming systems 
in the same redshift range.

An extensive multi-wavelength coverage is required. 
When dealing with large datasets over such a wide area, a spectroscopic
characterization of the survey is beyond the multiplexing 
capabilities of current instrumentation, and photometric investigation 
is clearly the only viable approach to be carried out in a reasonable amount of
time.

The {\em ESO-Spitzer wide-area Imaging Survey} (ESIS) 
is an ESO Large Programme (P.I. Alberto Franceschini), targeting 
the SWIRE ELAIS-S1 field.
This region includes the absolute minimum of the Galactic 100 $\mu$m
emission in the Southern sky (0.37 $[$MJy/sr$]$). 

In paper-I \citep{berta2006} we presented the ESIS WFI/2.2m
observations in the $B$, $V$, $R$ bands, illustrating also the potential of
panchromatic studies of galaxies. 
Here VIMOS $I$ and $z$ band observations are presented, completing the optical
ESIS survey. 

The paper is organized as follows. Section \ref{sect:observations} presents the
survey and the observing strategy. Section \ref{sect:datared} and Appendix
\ref{sect:app_fringing} describe the main steps in the data reduction, including
astrometric, photometric calibration and mosaicking. The quality of data
products is tested in Sect. \ref{sect:quality}, while the contents of the
released catalogs are described in Sect. \ref{sect:catalogs}. 
In Sect. \ref{sect:selection} we show how VIMOS $I$ and $z$ band data help in
the selection and in constraining the spectral energy distributions (SEDs) of
high-redshift galaxies. Finally, Sect. \ref{sect:summary} summarizes our
findings.


\section{Observations}\label{sect:observations}

The ESIS VIMOS project covers a total area of $\sim$4 deg$^2$ in the $I$ band and
$\sim$1 deg$^2$ in the $z$ band. 
A total of 116 hours have been allocated, during periods P71-P73 (from April 2003 to
September 2004); 
observations were carried over also in period P74 (from October 2004 to March 2005).

\begin{figure}[!ht]
\centering
\includegraphics[width=0.48\textwidth]{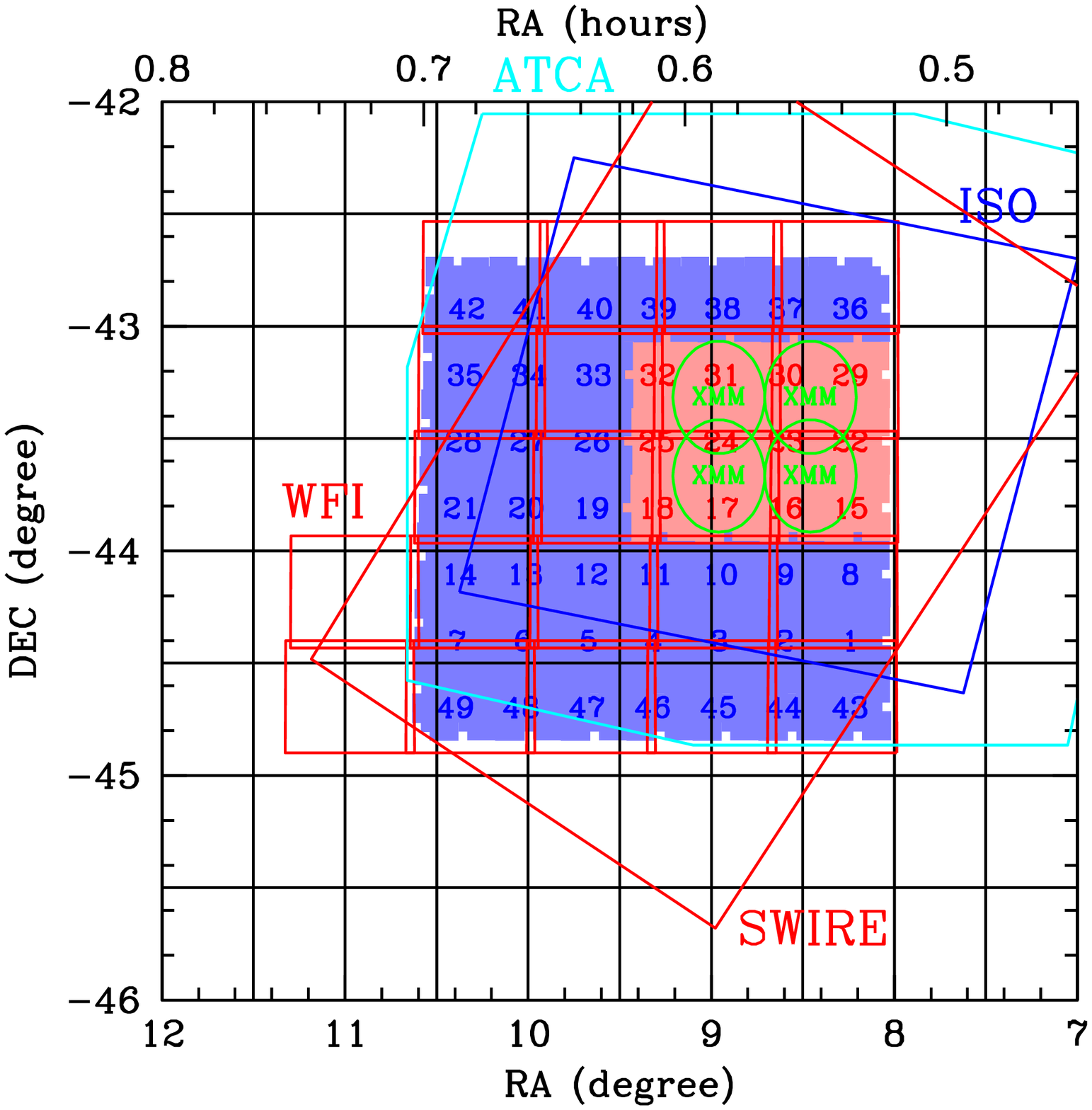}
\caption{The ESIS VIMOS survey in ELAIS-S1 and the multiwavelength coverage
of the area. The shaded area represents the 49 ESIS VIMOS pointings in the $I$
(blue/dark) and $I+z$ (red/light) bands. Over plotted are: the ESIS WFI survey
\citep[$\sim30^{\prime}\times30^{\prime}$ squares][]{berta2006} and the SWIRE 
\citep{lonsdale2003,lonsdale2004}, XMM \citep{puccetti2006},
ELAIS ISO \citep[e.g.][]{rowanrobinson2004}, ATCA 1.4 GHz
\citep{middelberg2007,gruppioni1999} areas.}
\label{fig:vimos_fields}
\end{figure}

Figure \ref{fig:vimos_fields} shows the position of the surveyed area, with respect to other 
available observations. VIMOS fields are plotted as shaded areas 
(in blue/dark $I$ band only and in red/light $I+z$ pointings). Multiwavelength imaging includes Spitzer/SWIRE
\citep{lonsdale2003,lonsdale2004}, optical ESIS WFI \citep{berta2006}, 
near-IR $JK$ \citep{dias2007}, GALEX \citep{martin2005,burgarella2005}, XMM \citep{puccetti2006},
Chandra (Feruglio et al., sub.), ELAIS ISO \citep[e.g.][]{rowanrobinson2004},
ATCA 1.4 GHz \citep{middelberg2007,gruppioni1999}.  
See \citet{berta2006} for a summary and description of all available data.
The VIMOS $z$ band observations are centered on the sub-area targeted by X-ray
and $JK$ observations.

The VIMOS \citep{lefevre2003} array is made of 4 CCDs, each one covering $\sim 6.5^\prime \times
8^\prime$ in the sky, separated by large gaps ($\sim 120-140$ arcsec wide, 
depending of which CCDs are considered). The pixel scale is 0.205 arcsec/pixel.

\begin{figure*}[!ht]
\centering
\includegraphics[width=0.40\textwidth]{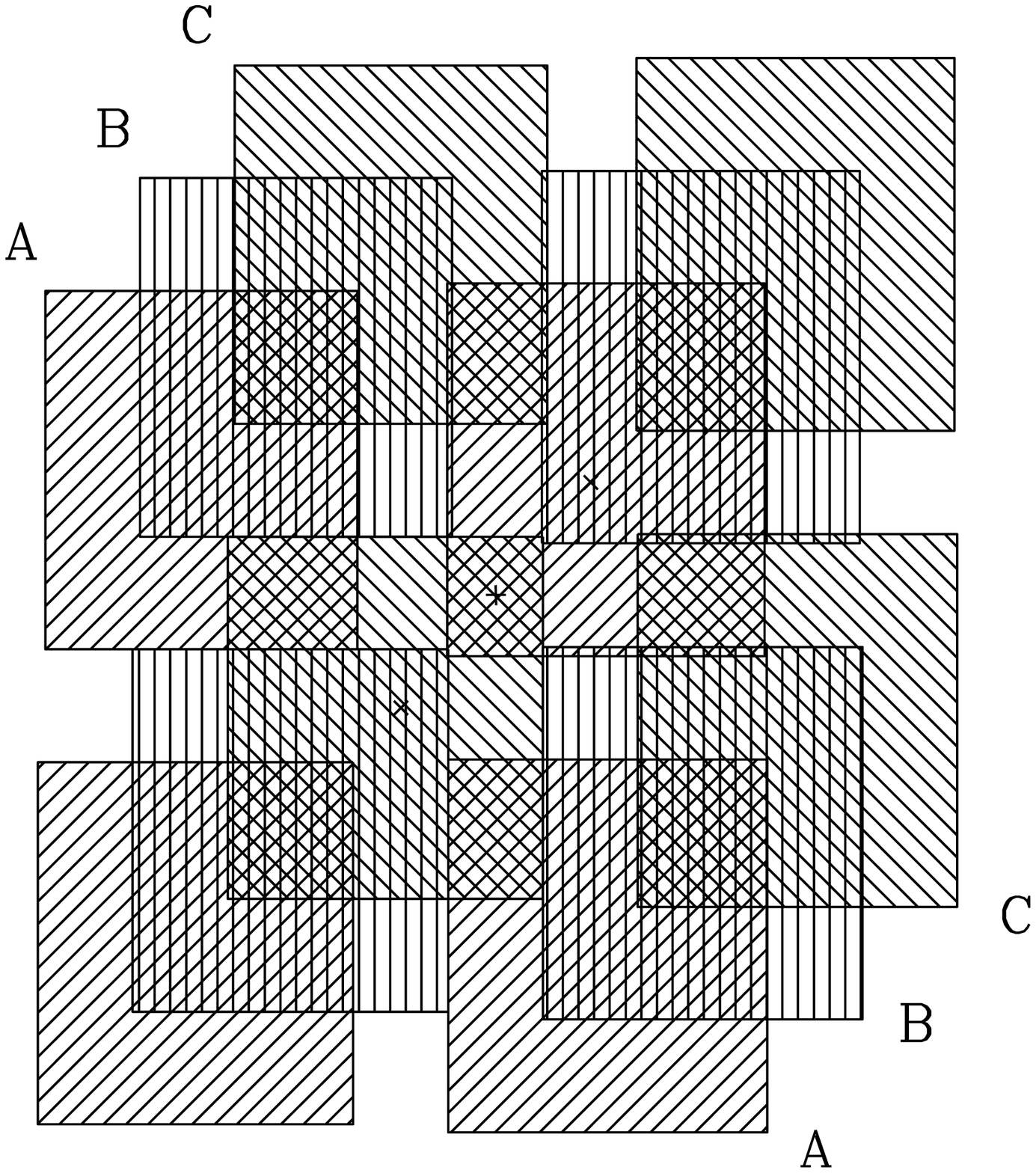}
\includegraphics[width=0.40\textwidth]{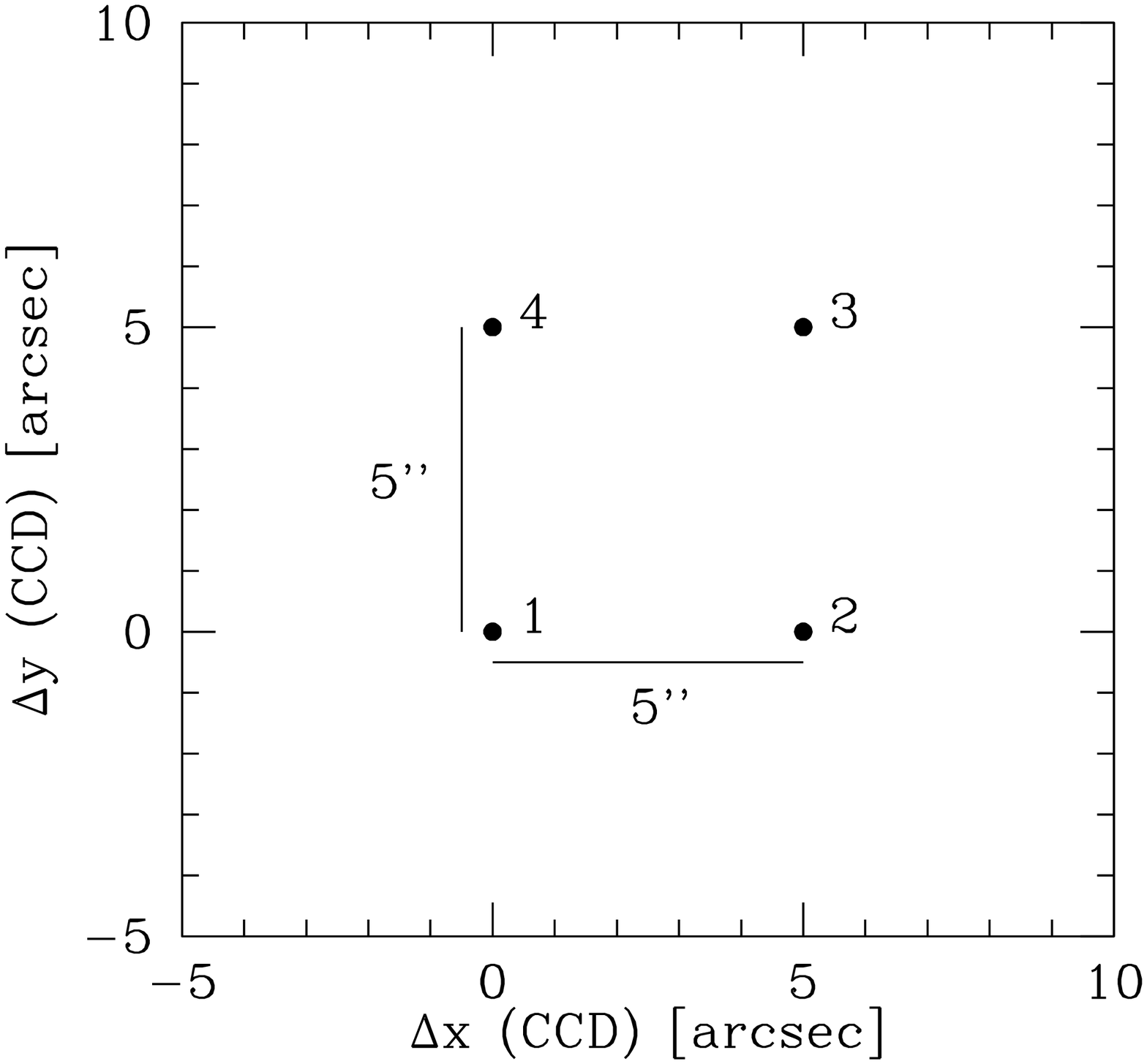}
\caption{Scheme of the ESIS VIMOS observing strategy. {\em Left panel}: each
pointing is made of three different OBs, shifted in the sky in order to fill
gaps between CCDs. {\em Right panel}: each OB consist of four dithered exposures,
for optimal bad pixels masking.}
\label{fig:1OB}
\end{figure*}

In order to cover a contiguous area and minimize
overheads, the observing strategy was set as follows:  
\begin{enumerate}
\item the total area is divided into 37 $I$ and 12 $I+z$ pointings (see Fig. \ref{fig:vimos_fields}).
\item each pointing is effectively made of 3 different Observing Blocks (OBs), 
named ``A, B, C''. Between pointings A and B one-gap-wide x,y offsets exist; the
same holds between B and C pointings (see Fig. \ref{fig:1OB}, left panel). 
\item each one of these 3 OBs is made of 4 dithered exposures, in order to get
rid of bad pixels and gap signatures (see Fig. \ref{fig:1OB}, right panel).  
\item care has been taken in order to observe at least one of the 3 OBs per pointing during 
a photometric night. In this way, it is possible to compute photometric offsets and 
correct the whole field to photometric conditions.
\item the total number of individual science
exposures is 588 and 144 in the $I$ and $z$ band respectively.
\item the total exposure time per OB is 600s in the $I$ band and 1200s in the $z$
band, causing saturation at $I\sim16$ $[$Vega$]$.
\item as a result of offsetting (3 OBs per pointing), the final coverage is
not homogeneous, but there are regions with a 600s coverage, others with
1200s (the majority) and finally some with 1800s coverage. These values refer to
the $I$ band; as far as the $z$ band is concerned, the effective exposure times are
1200s, 2400s, 3600s. 
\end{enumerate}

\begin{figure}[!ht]
\centering
\rotatebox{-90}{
\includegraphics[height=0.48\textwidth]{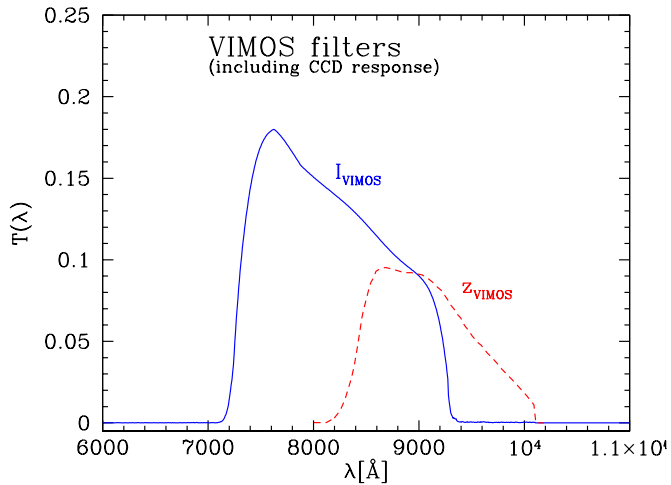}
}
\caption{The VIMOS $I$ and $z$ passbands average transmissivity.}
\label{fig:vimos_filters}
\end{figure}

Figure \ref{fig:vimos_filters} shows the average VIMOS $I$ and $z$ passbands,
defined as filters convolved with CCD and optics response. 
Actually the VIMOS transmission depends slightly on which of the 
four quadrants is considered. Here here we adopt an average transmission.
Tables
\ref{tab:log_I} and \ref{tab:log_z} (available in electronic version)
list the observation logs of the ESIS VIMOS survey.

\section{Data reduction}\label{sect:datared}

The ESIS VIMOS data were reduced within the IRAF\footnote{The package IRAF is
distributed by the National Optical  
Astronomy Observatory which is operated by the Association of
Universities for Research in Astronomy, Inc., under cooperative agreement
with the National Science Foundation.} environment, 
exploiting the capabilities of the WFPDRED package, developed in Padova 
and adapted to the VIMOS case. 

\begin{figure*}[!ht]
\centering
\subfigure[Image before de-fringing]{\includegraphics[width=0.28\textwidth]{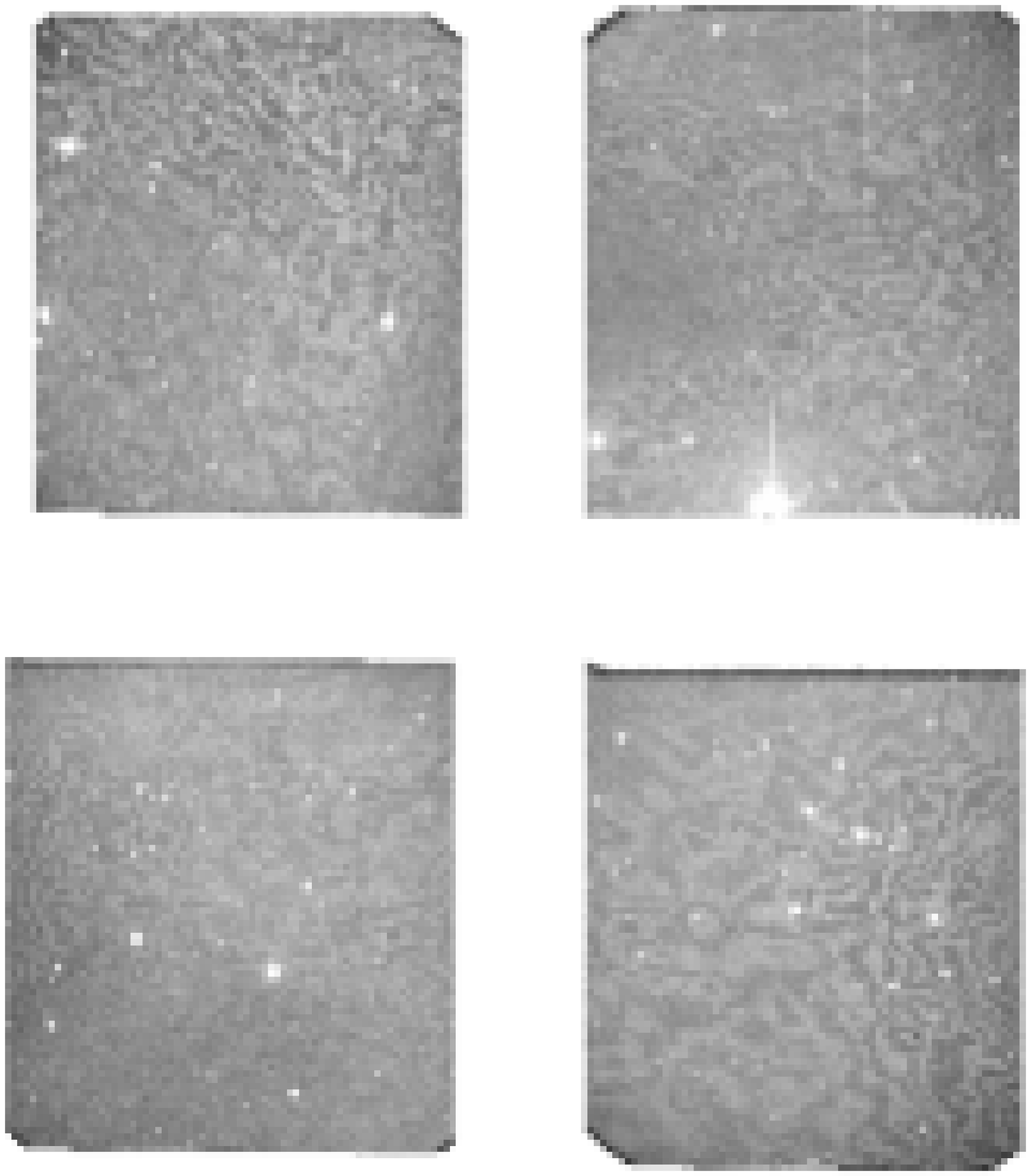}}
\hspace{0.04\textwidth}
\subfigure[Fringe pattern]{\includegraphics[width=0.28\textwidth]{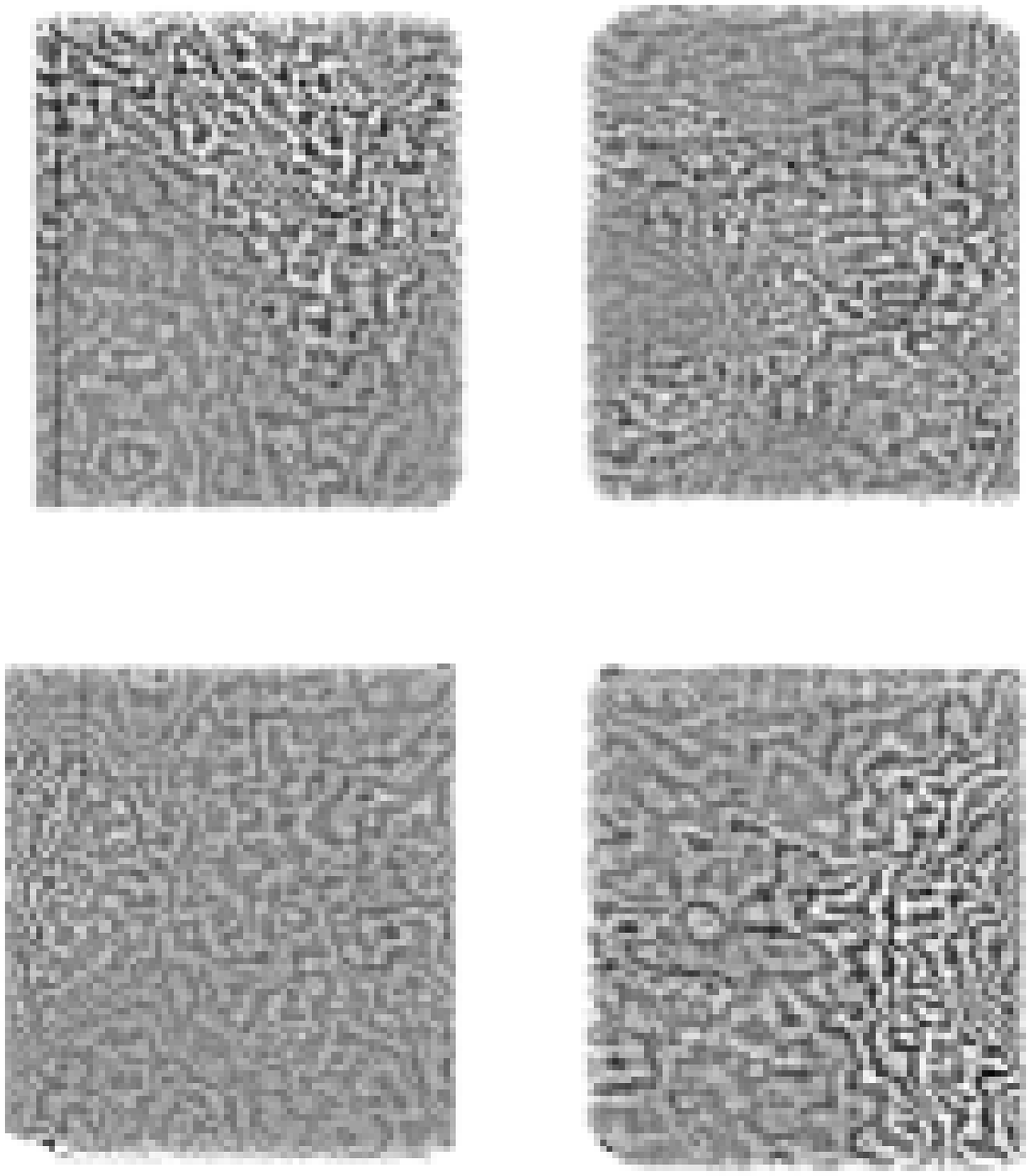}}
\hspace{0.04\textwidth}
\subfigure[Image after de-fringing]{\includegraphics[width=0.28\textwidth]{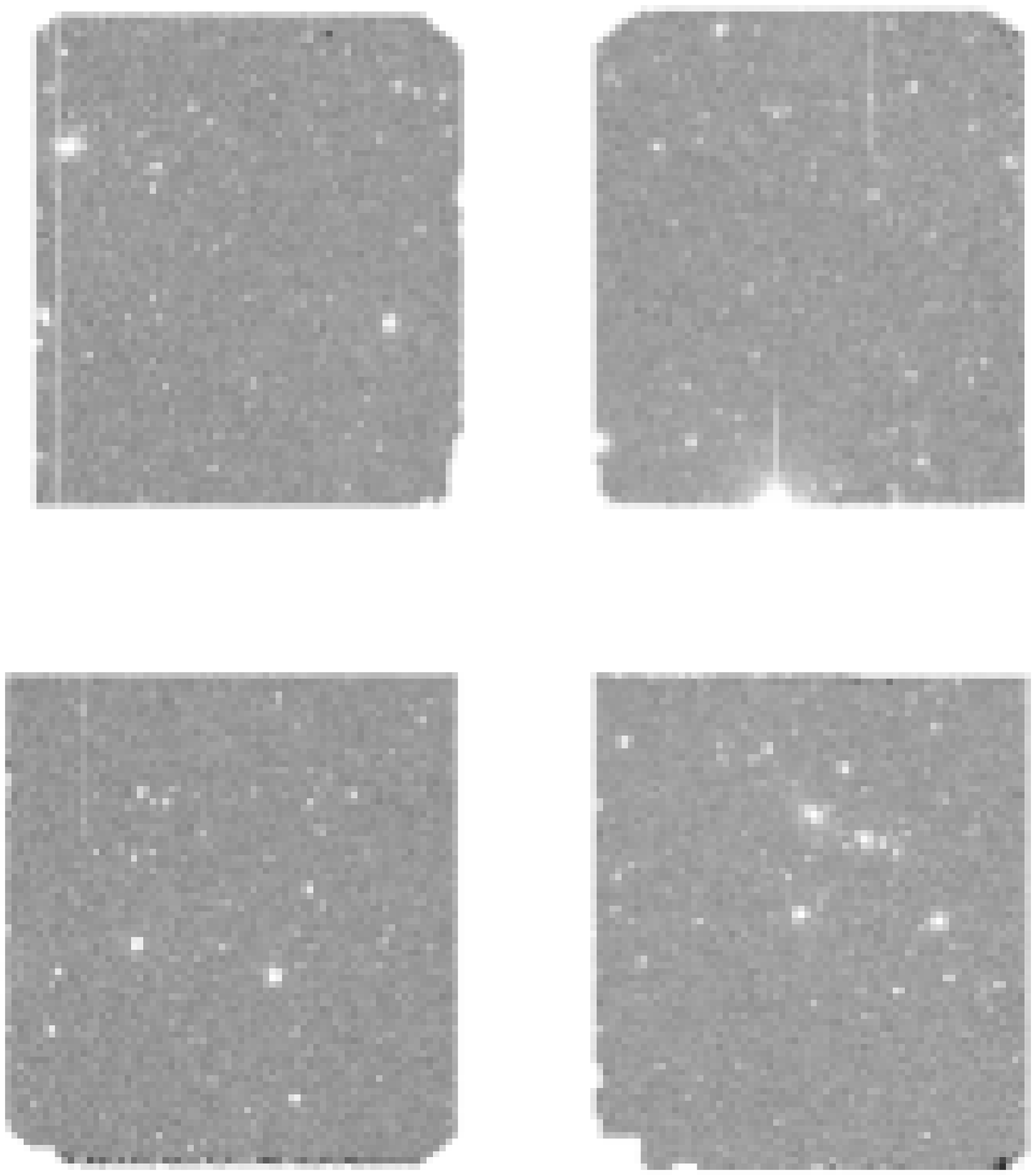}}
\caption{Example of de-fringing process in the $I$ band.}
\label{fig:defringing}
\end{figure*}

De-biasing and flat-fielding were performed in the standard manner; sky flat
field frames taken during each night were used.
VIMOS $I$ and $z$ band images are affected by strong fringing, that needs to be
corrected before photometric calibration and mosaicking. Appendix
\ref{sect:app_fringing} describes the main causes of fringing and the procedure
adopted to correct ESIS frames. Figure \ref{fig:defringing} 
shows one VIMOS $I$ band frame before (left panel) and after (right panel)
subtracting the fringe pattern (central panel).

\subsection{Astrometric calibration}\label{sect:astrom_map}

The ESIS survey targets a large area in the sky ($\sim$4 deg$^2$ in 
the $I$ band), therefore it is necessary to accurately map the 
astrometric distortion of VIMOS, in order to correctly build wide-field 
mosaics.

The survey includes observations of the \citet{stone1999} astrometric
fields D and E, obtained both with the WFI and VIMOS instruments, 
and aimed at properly mapping pixel into celestial coordinates.

We have chosen a TNX mapping \citep[see][]{berta2006}, which combines a linear
projection of the sky sphere onto a tangent plane and a polynomial function for
distortions. In order to obtain the best pixels-sky mapping, we have exploited
the VIMOS astrometric observations and extended the resulting catalog with the
WFI observations.

The best results were obtained using 5th order distortion
polynomials. Before mosaicking, the astrometric solution was applied to 
individual science images, and each frame was re-centered on the GSC 2.2
(STScI \& OaTO, \citeyear{gsc2}) catalog.
Section \ref{sect:astrom_accuracy} reports on the accuracy of ESIS VIMOS
astrometry.

\subsection{Photometric zeropoints}\label{sect:phot_zp}

As part of the ESIS VIMOS observations, 
\citet{landolt1992} Ru149 or TPHE spectrophotometric
standard fields were targeted during photometric nights, 
in order to perform a dedicated photometric calibration 
of ESIS science frames. These observation have been 
defined with four different exposures per observing block, 
including the main standard stars on each individual CCD of the
VIMOS array.	

Unfortunately, these fields are not rich in stars and the VIMOS chips 
are relatively small, therefore only few objects fall on the CCD, 
and the estimate of the zeropoints is not optimal.
Nevertheless, after the reduction of standard fields and correction of single CCD frames 
for differences in {\em gain}, the $I$ band zeropoints were computed. 
The resulting zeropoint $zp_{I}=26.96 \pm 0.07$ (in units of electrons) is consistent with 
the ESO\footnote{http://www.eso.org/observing/dfo/quality/ VIMOS/qc/zeropoints.html} 
value (26.99 on average), computed on richer fields.

As far as the $z$ band is concerned, 
despite the fact that ESIS included observations of \citet{landolt1992} fields,
no reference standard catalog is available in these fields, 
therefore we relied on a different procedure to determine the zeropoint.

The ELAIS-S1 field benefits from an extensive multiwavelength 
follow up from the X-rays to the radio \citep[see][]{berta2006}.
We have therefore selected stars with available {\em BVRJK}+IRAC photometry
in the ESIS field and performed SED fitting to estimate the expected 
$z$ band magnitudes and determine the VIMOS zeropoint:
\begin{equation}\label{eq:z_zeropoint}
ZP = m_{expected} - m_{measured}\textrm{.}
\end{equation}
Source extraction has been performed with SExtractor \citep{bertin1996}
on individual science frames 
observed during photometric nights and reduced as described above.
Each image was transformed into e$^-$/s, by using the proper {\em
gain} for the given CCD. 

\begin{figure*}[!ht]
\centering
\includegraphics[width=0.35\textwidth]{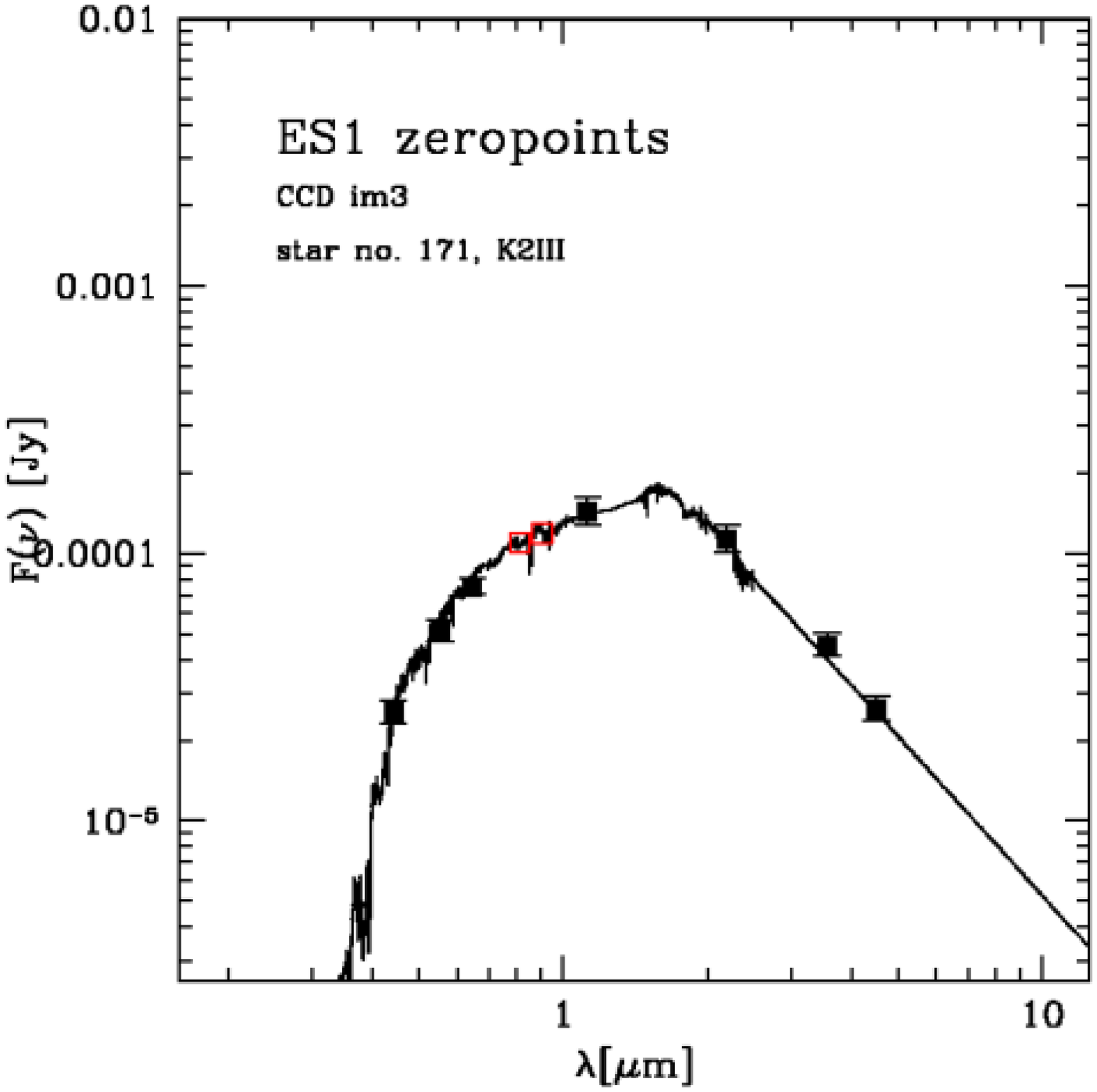}
\includegraphics[width=0.35\textwidth]{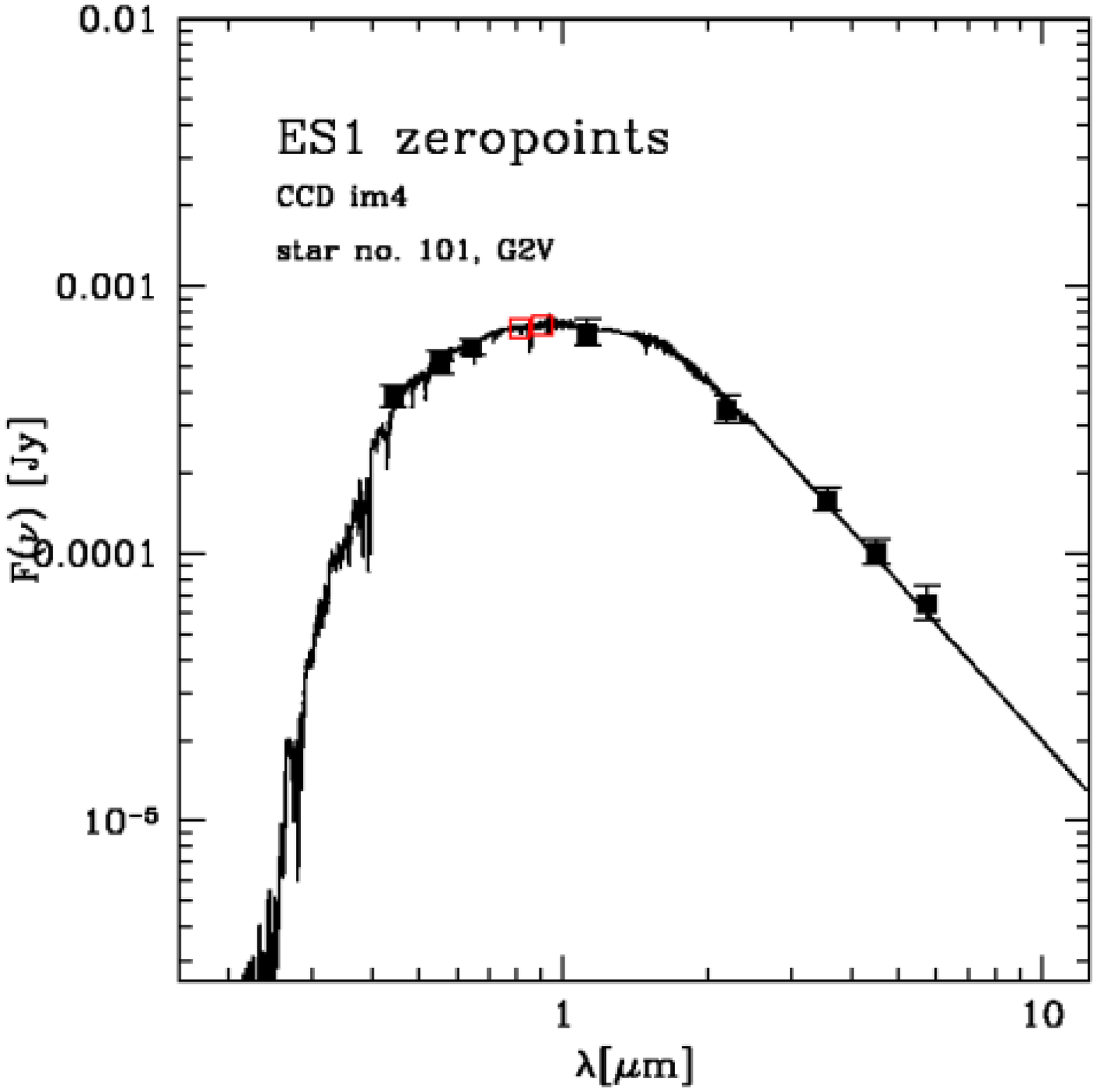}
\includegraphics[width=0.35\textwidth]{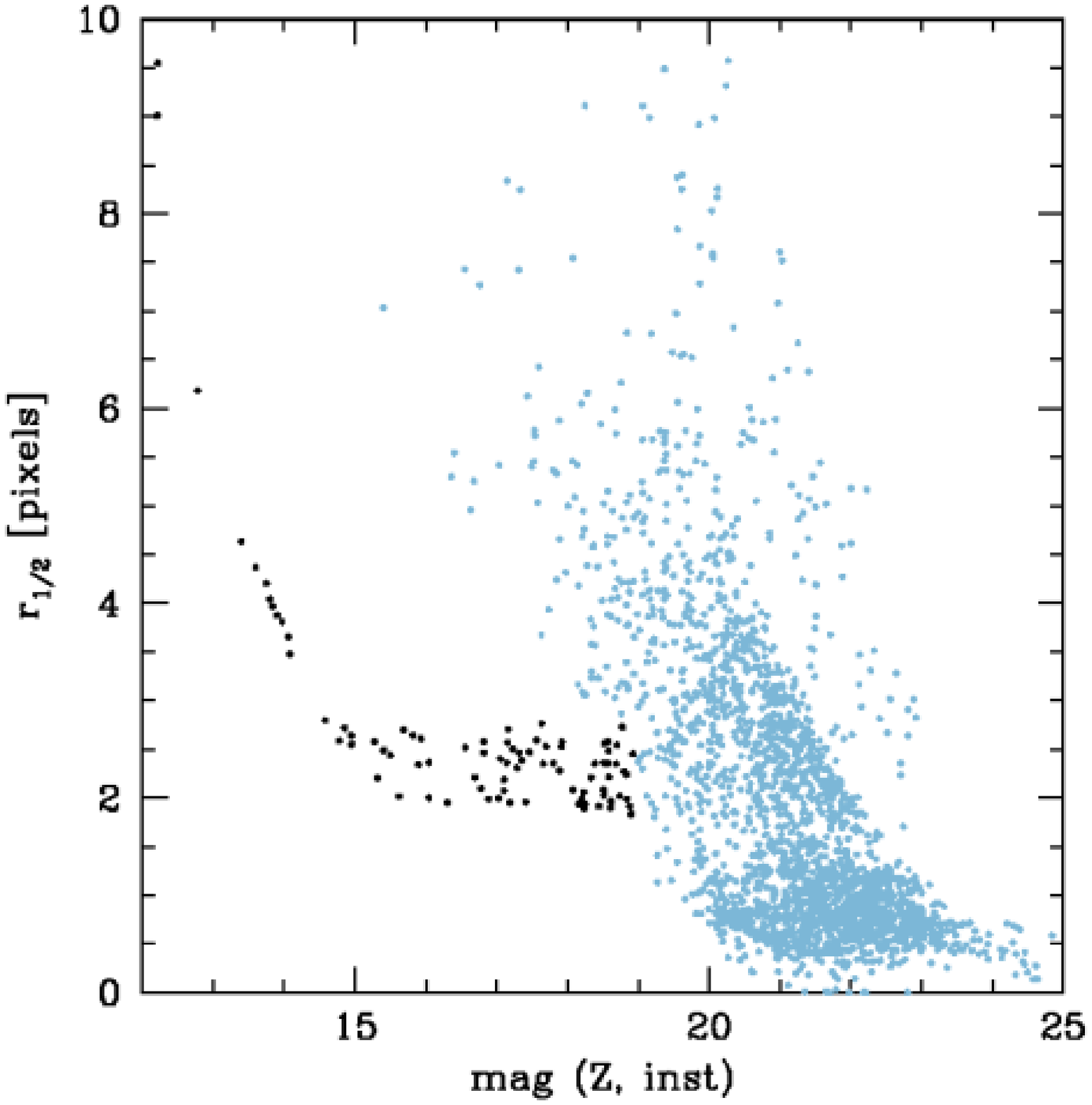}
\includegraphics[width=0.35\textwidth]{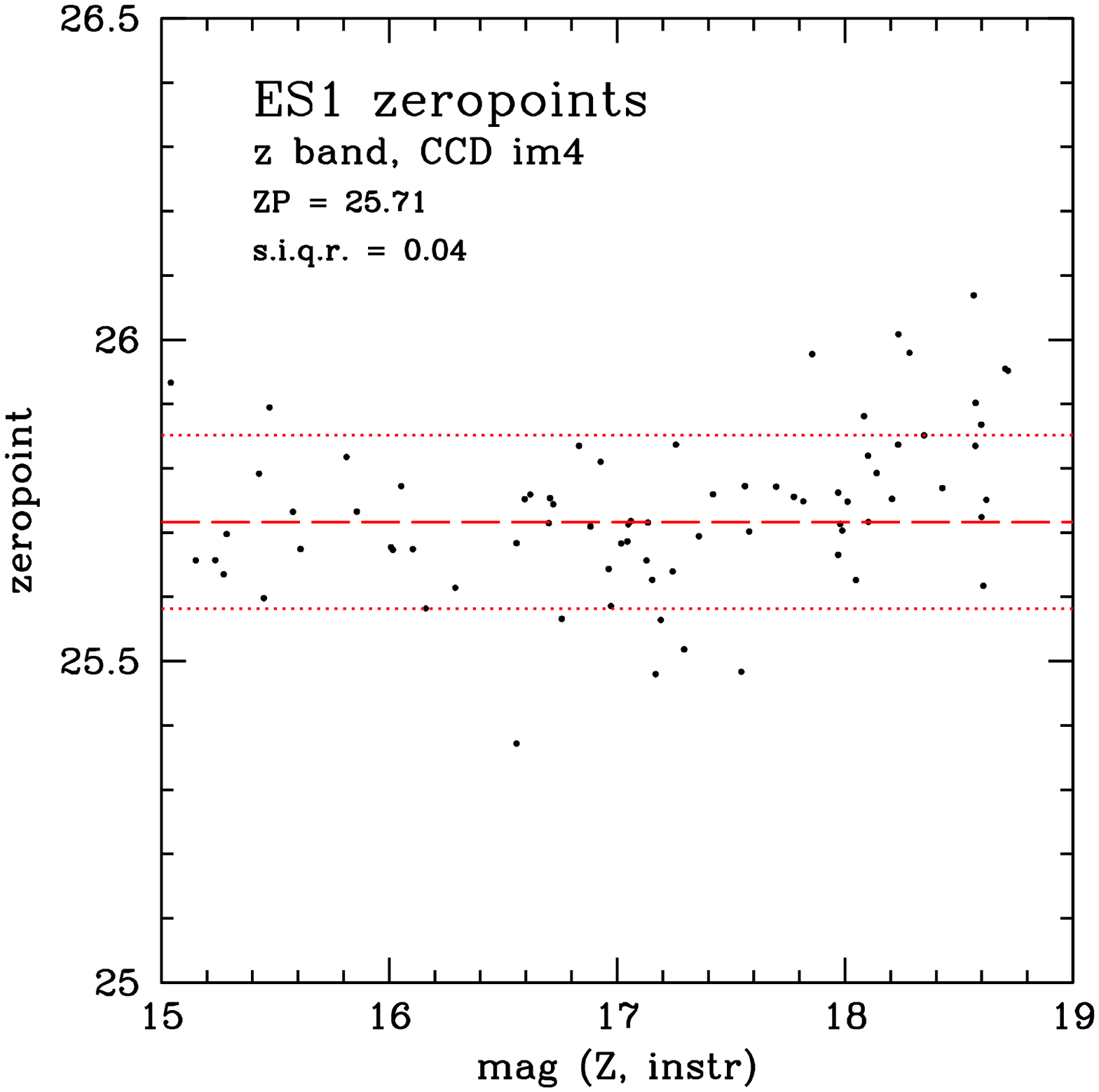}
\caption{Determination of VIMOS zeropoints. {\em Top panels}: two examples of SED fitting of stars in ESIS
photometric pointings. The open squares represent the $I$ and $z$ band expected
fluxes derived from the best fit. {\em Bottom left panel}: selection of point-like objects on the $z$ band photometric images
used to derive zeropoints. The dark dots represent the ``stellar'' locus. {\em
Bottom right panel}: $z$ band zeropoint for CCD 4. The dashed line represents
the median, the dotted lines are the $\pm 3\ s.i.q.r.$ levels used for clipping
the data.}
\label{fig:zeropoints_Z}
\end{figure*}

Point-like objects were selected on the basis of the dependence of their 
half-flux radius on instrumental magnitude (see Fig. \ref{fig:zeropoints_Z},
bottom left panel, and Sect. \ref{sect:catalogs}). 
This procedure has the advantage 
of well identifying the magnitude range over which point-like sources are 
well defined.

The $z$ band extracted catalog was matched to the other wavelengths source lists
with a simple closest-neighbor algorithm, using a 1 arcsec matching radius
\citep[][]{berta2006}. The final catalog contains $\sim$50-100 objects per CCD.
Since we are using a sub-area of the entire ESIS VIMOS survey (i.e. the region in
the sky actually covered by JK data), these photometric images were taken
during two epochs only, namely December 2003 and Aug.-Sept. 2004.

The adopted stellar library is based on \citet{pickles1998} stellar
spectra\footnote{http://www.ifa.hawaii.edu/users/pickles/AJP/hilib.html}, 
extended to $\lambda>2.5\mu$m with a Rayleigh-Jeans law.

The observed spectral energy distributions (SEDs) of stellar objects 
were fitted taking into account optical and near-IR photometry up to the IRAC
4.5$\mu$m band. 
Best fit is sought by means of $\chi^2$ minimization, among all templates in the
stellar library and varying their normalization.
Two examples of SED fits are shown in Fig. \ref{fig:zeropoints_Z}.
The 5.8$\mu$m flux is also shown (when available), even if it was not included
in the estimation of $\chi^2$.
The 1,2,3 $\sigma$ uncertainty on the expected $z$ band magnitude, as derived from the best
fit, is computed as $\chi_{best}^2+1,4,9$. 

The zeropoint for a given CCD is given by the median of all zeropoints (one
per star, see Fig. \ref{fig:zeropoints_Z}). We used a 3 s.i.q.r.\footnote{semi inter-quartile range} clipping
(corresponding to a 2 $\sigma$ clipping for a normal distribution).

As zeropoint differences between different chips are 
mainly due to gain differences, the resulting zeropoints for the 4
VIMOS quadrants are consistent with each other. The $z$ band zeropoint 
obtained in this way is (in units of electrons):
\begin{equation}\label{eq:final_zp}
ZP (Z\ \textrm{band}) = 25.74 \pm 0.11
\end{equation}
where the uncertainty is given by $4\times s.i.q.r.$.

Similarly, we have taken advantage of the selected stellar catalogs to 
newly compute the zeropoint in the $I$ band, directly from ESIS 
data, obtaining a zeropoint of $26.93\pm0.13$ mag (in units of electrons),
consistent with our previous and ESO's estimates.
No significant difference in zeropoints between the two epochs is detected.

\begin{figure*}[!ht]
\centering
\includegraphics[width=0.8\textwidth]{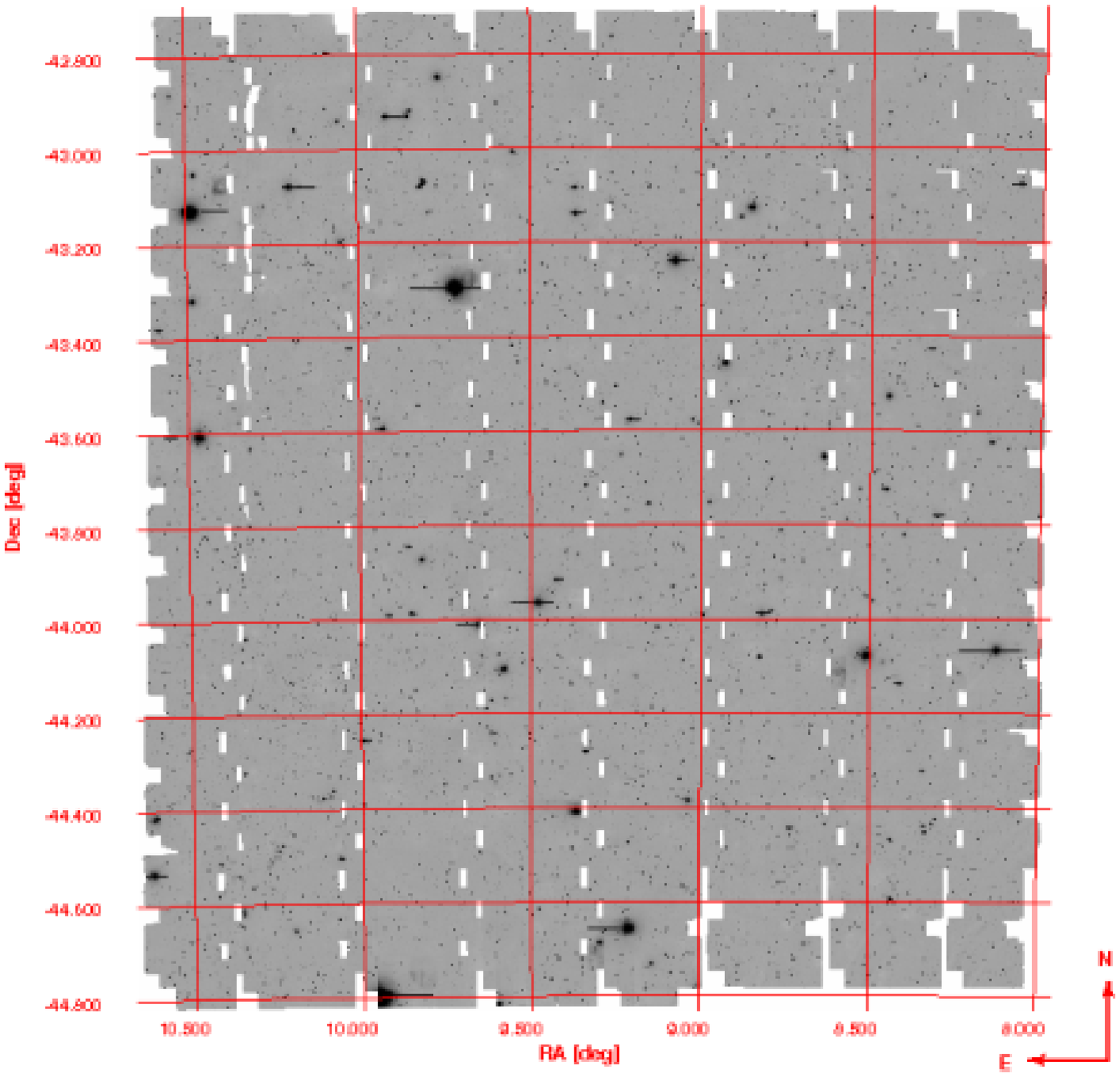}
\caption{The ESIS VIMOS final $I$ band mosaic.}
\label{fig:whole_I}
\end{figure*}

\begin{figure*}[!ht]
\centering
\includegraphics[width=0.8\textwidth]{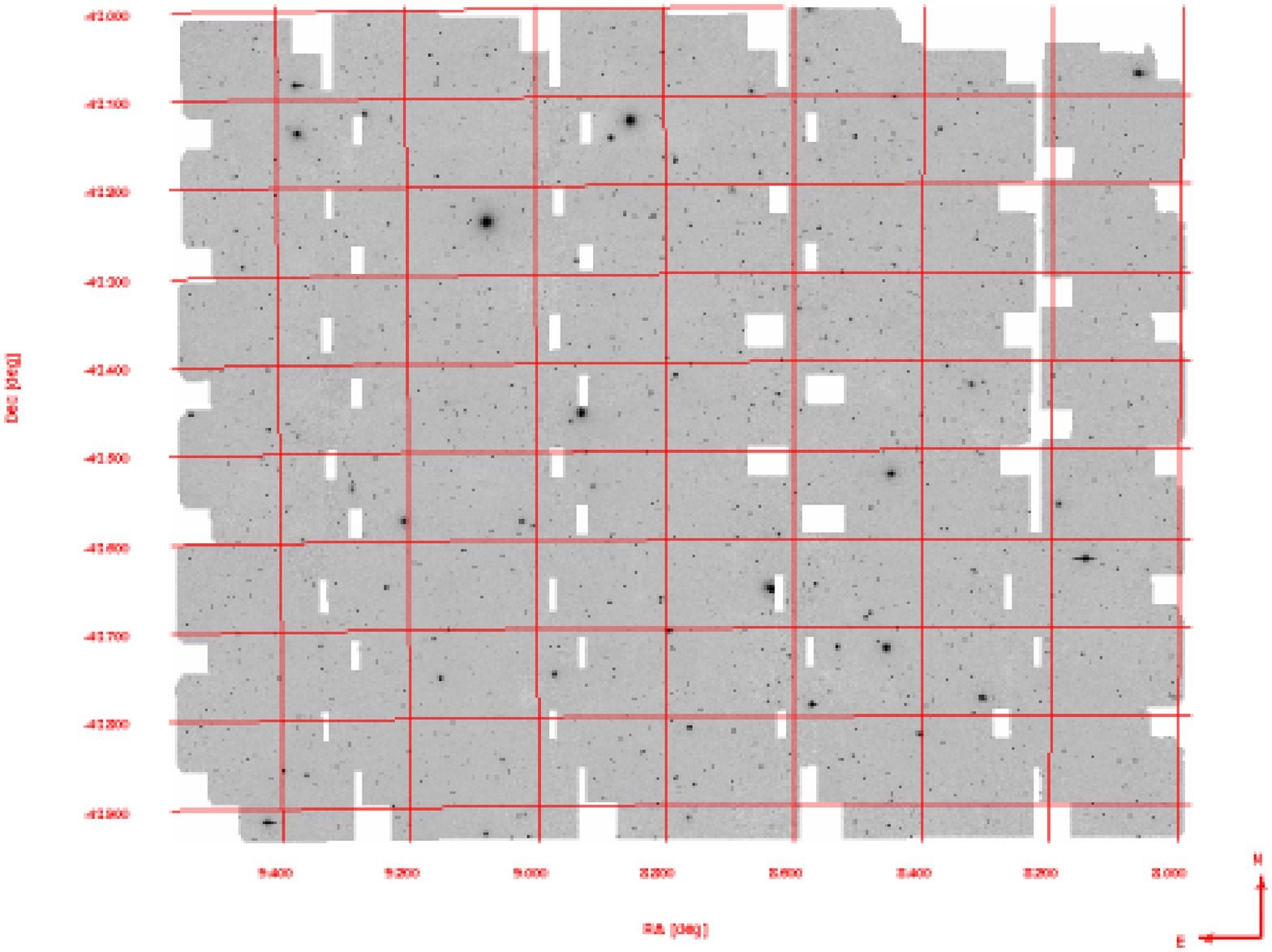}
\caption{The ESIS VIMOS final $z$ band mosaic.}
\label{fig:whole_z}
\end{figure*}

\begin{figure*}[!ht]
\centering
\includegraphics[width=0.55\textwidth]{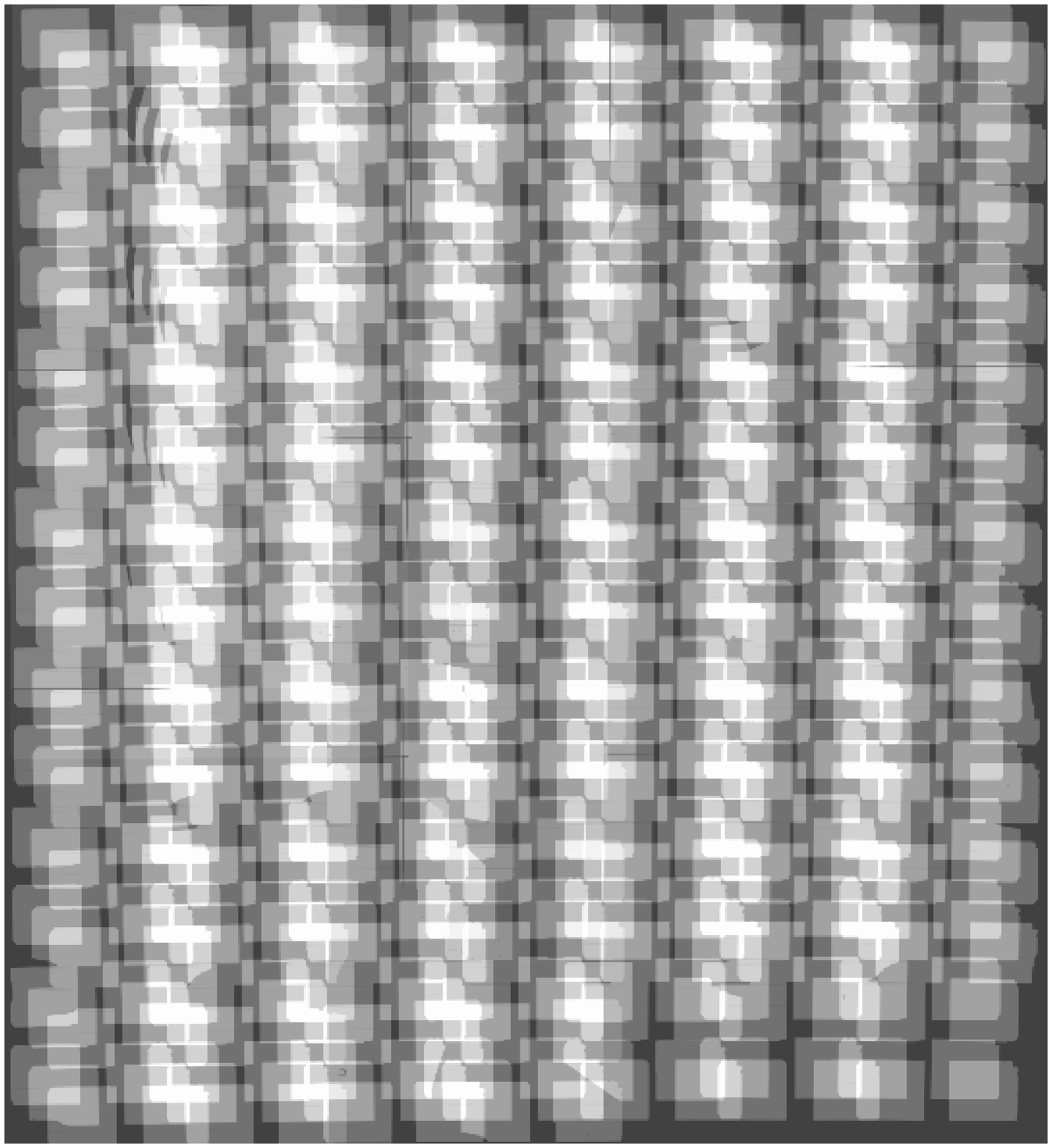}
\includegraphics[width=0.05\textwidth]{white.eps}
\includegraphics[width=0.30\textwidth]{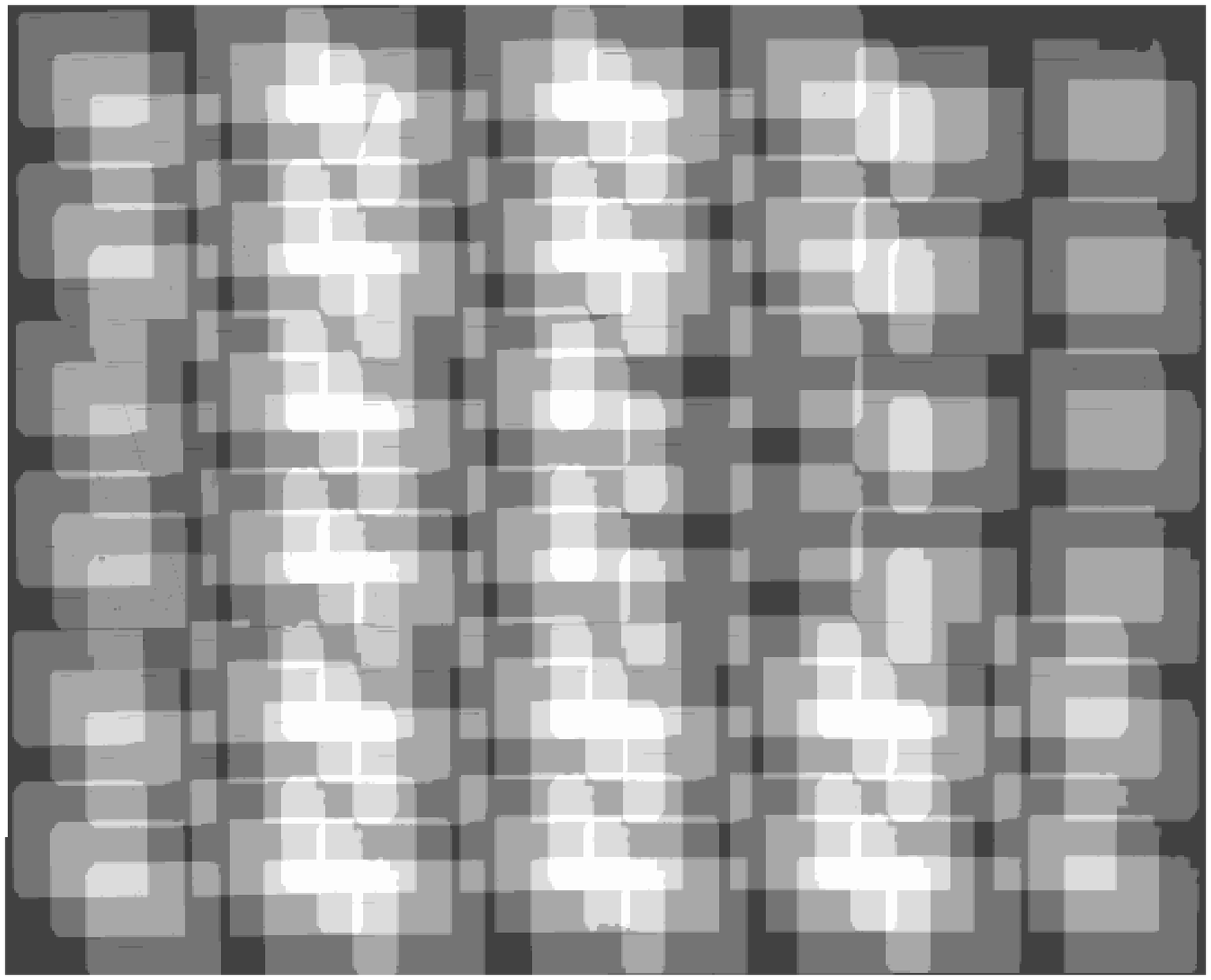}
\caption{The ESIS VIMOS final $I$ (left) and $z$ band (right) exposure maps. Brighter 
regions correspond to deeper coverage. North is up, east is left; the two images 
have roughly the same scale.}
\label{fig:whole_weights}
\end{figure*}

\subsection{Photometric calibration}\label{sect:phot_cal}

As already pointed out, the observations of ESIS VIMOS fields are spread across
several nights, and are characterized by different sky conditions.

The requirement that at least one OB per pointing was observed in photometric
conditions facilitates the task of absolute calibration of the whole dataset;
we call these photometric images ``reference frames''.
We have identified point-like objects on each science image, in common with 
any reference frame and then computed the zeropoint shift due to
non-photometric conditions: $\Delta mag=mag_{any}-mag_{ref}$.

Hence, each individual frame has been corrected to photometric conditions, 
normalized to unitary airmass and exposure time, and converted to electrons:
\begin{equation}\label{eq:deltamag}
I_{phot}=\frac{I_{any}\times 10^{0.4\left( AM\cdot K_\lambda +\Delta
mag\right)}}{t_{exp}}\times \textrm{\em CONAD} \textrm{,}
\end{equation}  
where AM is the airmass of the given frame, $K_\lambda$ is the atmospheric
extinction coefficient for Paranal ($K_I=0.027$, $K_z=0.02$), and {\em CONAD} is
the gain in e$^-$/ADU.

No color corrections are applied, and final catalogs are given in 
calibrated VIMOS magnitudes (i.e. not transformed to any standard photometric
system). The effective wavelengths for the $I$ and $z$ passbands are 8140 and
9050 \AA\ respectively. The flux zeropoints and AB
shifts\footnote{$mag_{AB}=-2.5\times \log \left(S_\nu[\textrm{erg s}^{-1}
\textrm{cm}^{-2} \textrm{Hz}^{-1}]\right)-48.58$} for these magnitudes
are: 
\begin{eqnarray}\label{eq:Jy_transform}
S(I,\ Jy)&=&2385\times 10^{0.4\cdot mag(I,\ Vega)}\\ 
S(z,\ Jy)&=&2207\times 10^{0.4\cdot mag(z,\ Vega)}
\end{eqnarray}
\begin{eqnarray}\label{eq:AB_transform}
mag(I,\ AB)&=&mag(I,\ Vega)+0.4763\\
mag(z,\ AB)&=&mag(z,\ Vega)+0.5604
\end{eqnarray}
as derived convolving the spectrum of Vega with VIMOS passbands (see Fig.
\ref{fig:vimos_filters}).

\subsection{Mosaicking}\label{sect:mosaic}

The astrometrically- and photometrically-calibrated frames
were finally combined together into the final ESIS mosaics.

In the $I$ band due to the large surveyed area and due to 
image size limitations, nine different 
mosaics were produced, in a $3\times3$ pattern covering the whole
ESIS-VIMOS field. 
For the $z$ band it was possible to 
assemble one single final mosaic, $\sim1$ deg$^2$ wide.

\begin{figure*}[!ht]
\centering
\includegraphics[height=0.3\textwidth]{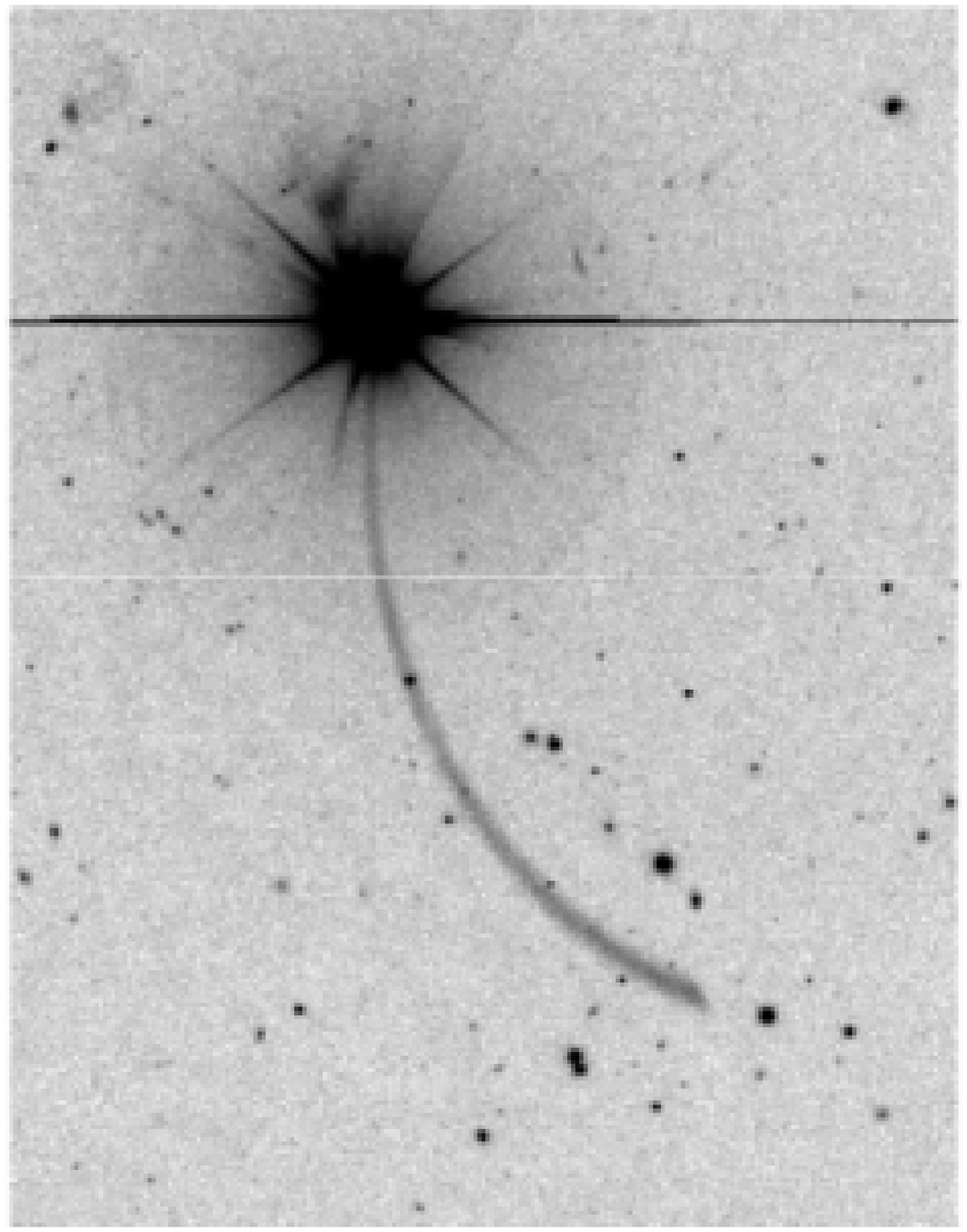}
\includegraphics[height=0.3\textwidth]{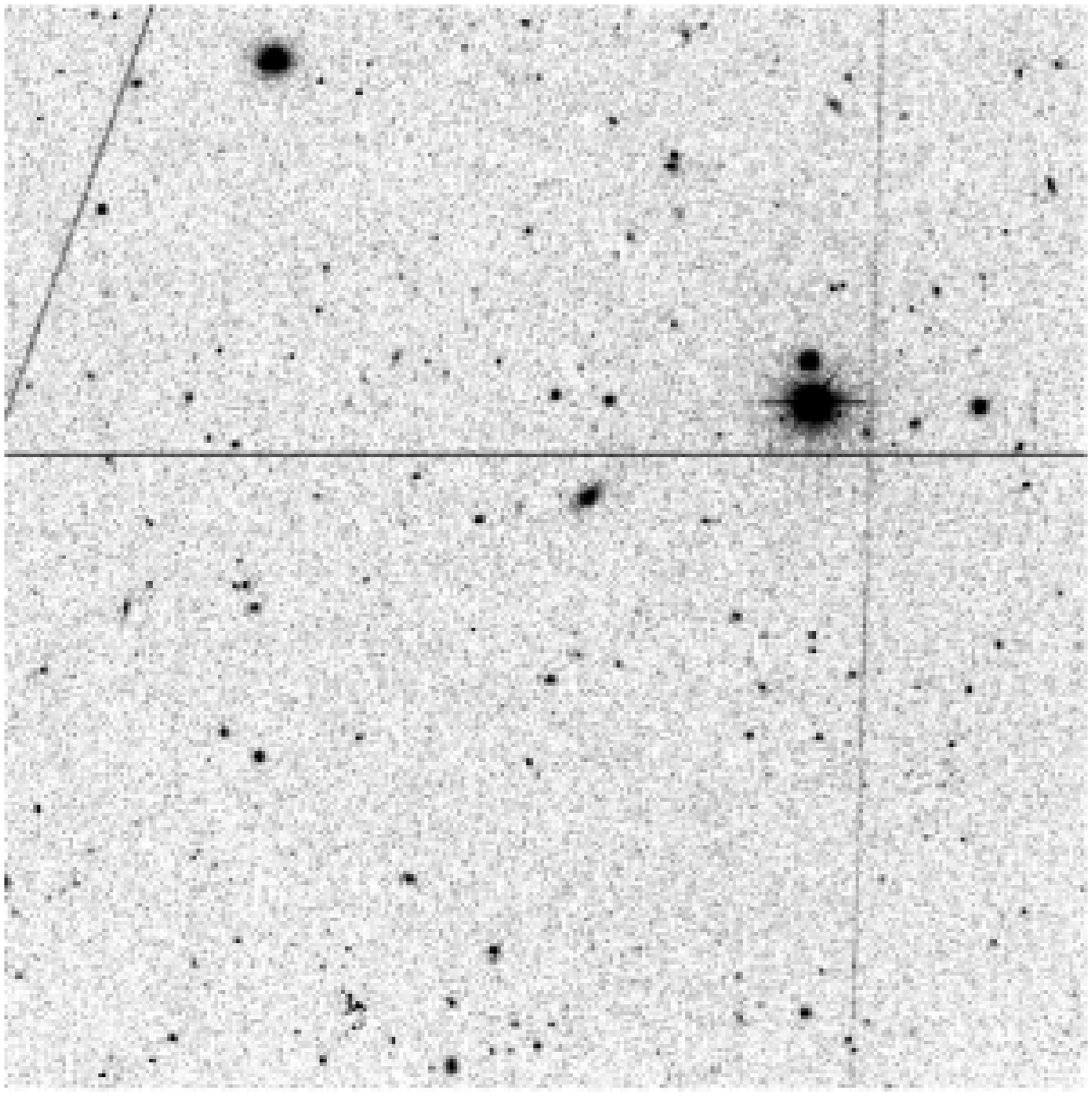}\\
\smallskip
\includegraphics[width=0.3\textwidth]{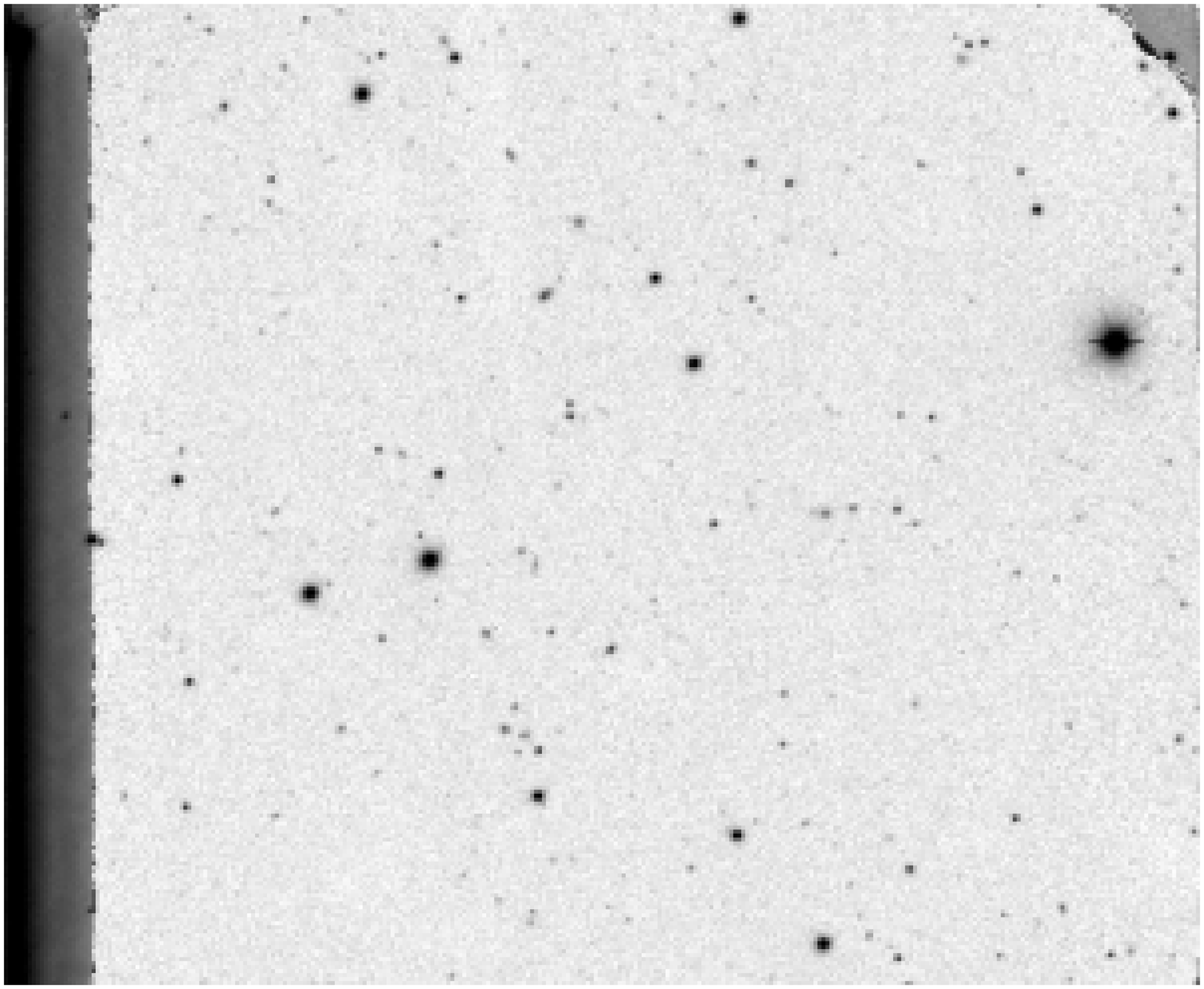}
\includegraphics[width=0.3\textwidth]{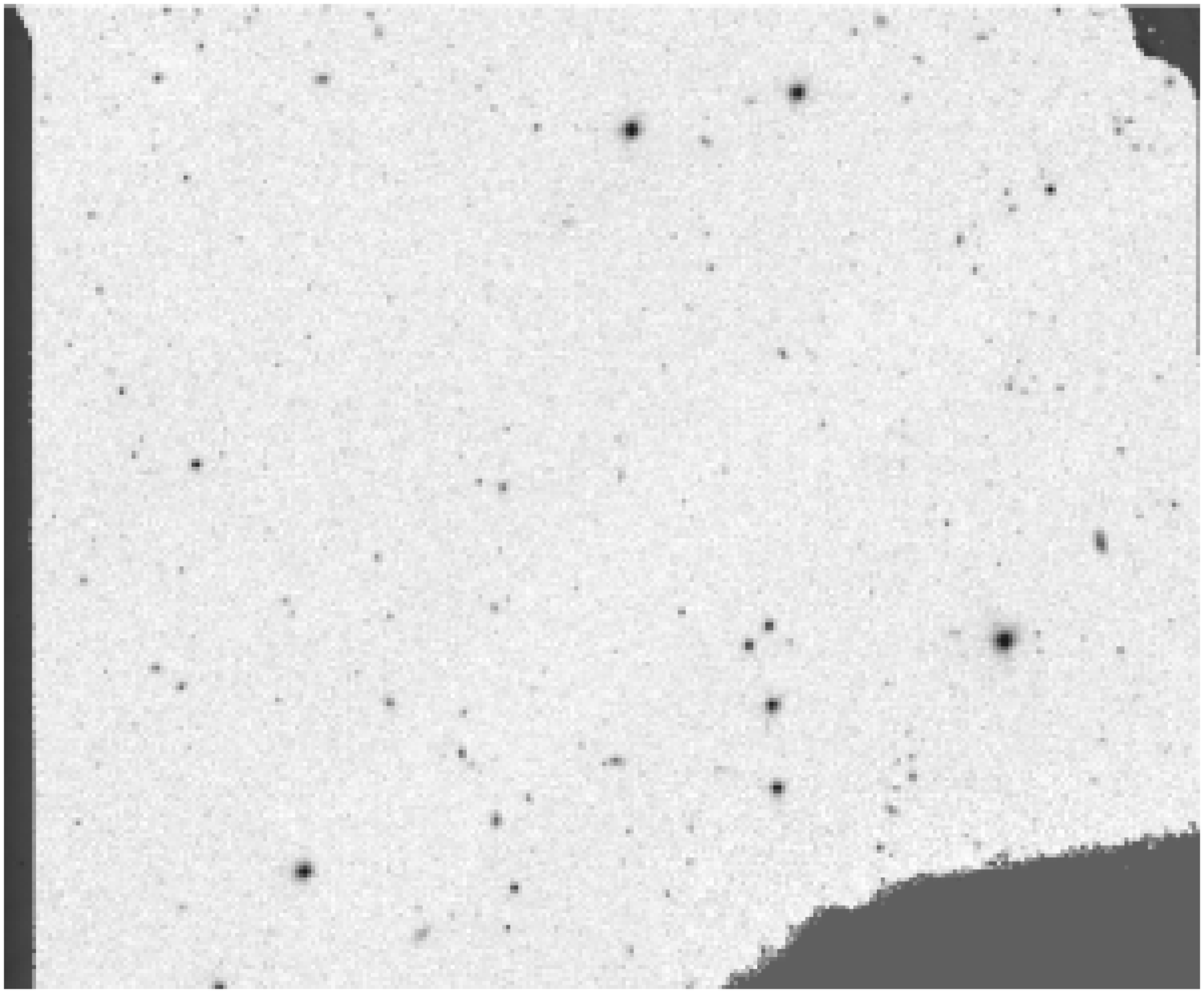}
\caption{Examples of defects to be masked before mosaicking. {\em Top panels}
show emission features: optical reflections and satellite trails. {\em Bottom
panels} exemplify ``vignetting'' features: the unidentified shade causing
residual gaps in the final mosaics (left) and the shadow of the guide star
probe (right). 
For clarity sake we have inverted colors of the shades. North is up, east is
left.}
\label{fig:defects}
\end{figure*}

During co-addition, 
all frames belonging to a given mosaic were registered 
to the same astrometric map. Different mosaics have 
different astrometric maps, using the mosaic center as 
tangent point for re-projection.
Bad pixels and ``defects'' (e.g. optical reflections 
or crossing of satellites, see the top panels in Fig. \ref{fig:defects}) were
masked and excluded from the final mosaics. 

Because of the complex observing strategy, the effective exposure time
varies across the field: the majority of the area has a 75\%
depth, and smaller regions with 25, 50 and 100\% depths exist.
We have therefore built weight maps to be associated with each scientific 
image during catalog extraction. We also include a {\em coverage flag}
in the extracted catalogs, describing the actual depth at the 
position of the detected objects.

The entire $I$ band field is shown in Fig. \ref{fig:whole_I}, 
where we have scaled down and combined the 9 sub-areas, for 
display purposes only. The $z$ band mosaic is in Fig. \ref{fig:whole_z}.
Figure \ref{fig:whole_weights} includes exposure maps. 
Unfortunately gaps in RA between different pointings are left. 
This problem is mainly caused by shadowing of the E-W edges of VIMOS CCDs
in ESIS images (see the bottom left panel in Fig. \ref{fig:defects}). The width
of the shade varies from frame to frame, and consequently so does the width of
the residual gap in the final mosaics.
Note that this is different from shadowing by the guiding star probe (bottom
right panel). The actual reason for this loss of pixels is not clear.

\section{Quality of data products}\label{sect:quality}

Observations were carried out under very different atmospheric conditions, over
a  $\sim2$ years period of time. The gaussian FWHM of point-like sources is therefore
not constant across the ESIS field, but there are regions with seeing 
$<0.8$ arcsec and others with $\textrm{FWHM}\gtrsim1.0$ arcsec.
The final catalogs were extracted on the mosaics without smoothing
to the worst seeing in the field, in order to take advantage of good imaging
quality, where available. 

We have tested 
how many sources would be lost by degrading the images with good seeing to a 
1 arcsec FWHM PSF, by convolving all individual frames with the appropriate 
kernel. As a result, roughly 4-5\% of the sources is lost when smoothing the 
images, mainly at the faintest magnitudes, after the turnover in the number
counts (see Sect. \ref{sect:catalogs}). 

\begin{figure*}[!ht]
\centering
\includegraphics[width=0.35\textwidth]{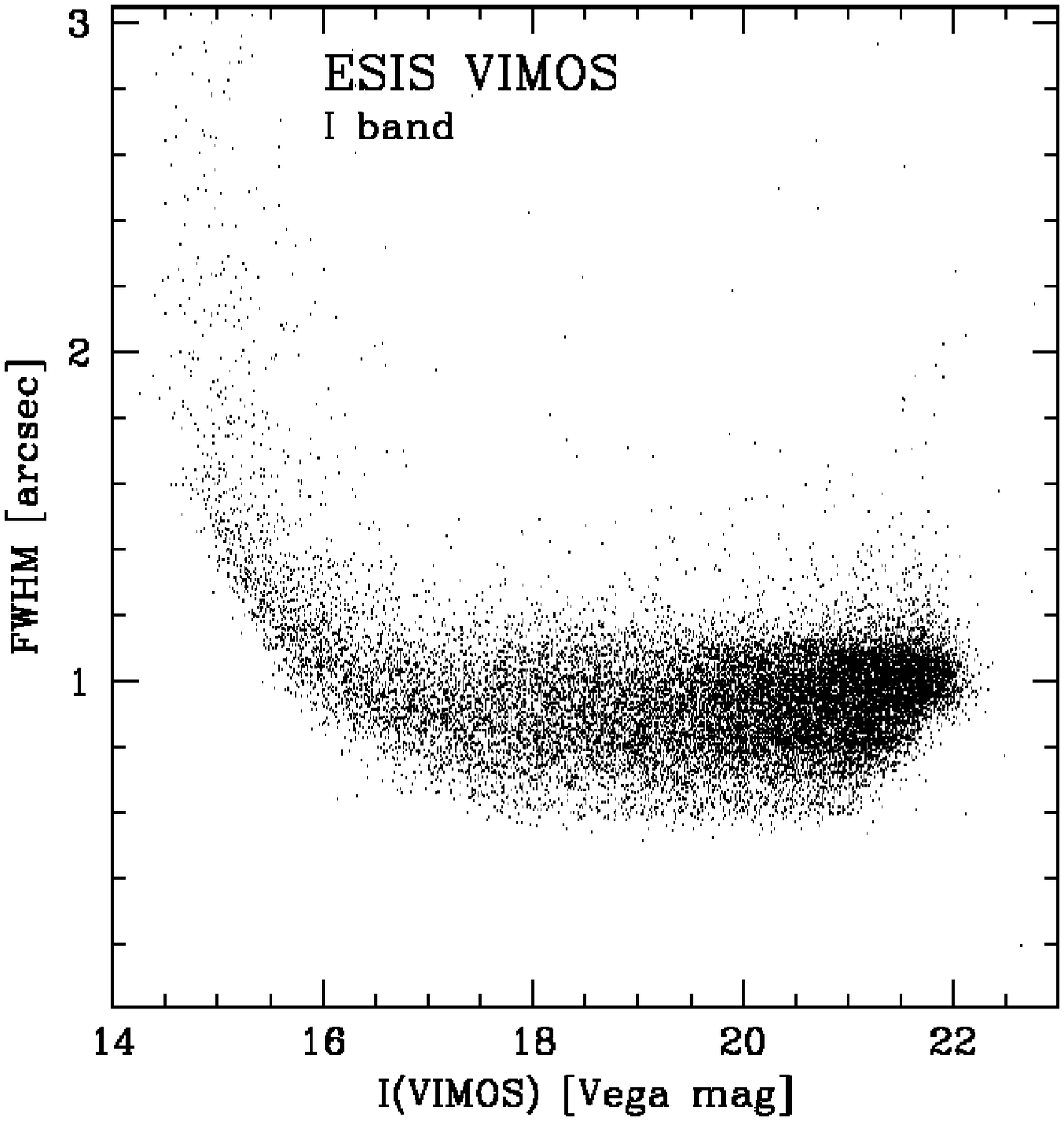}
\includegraphics[width=0.35\textwidth]{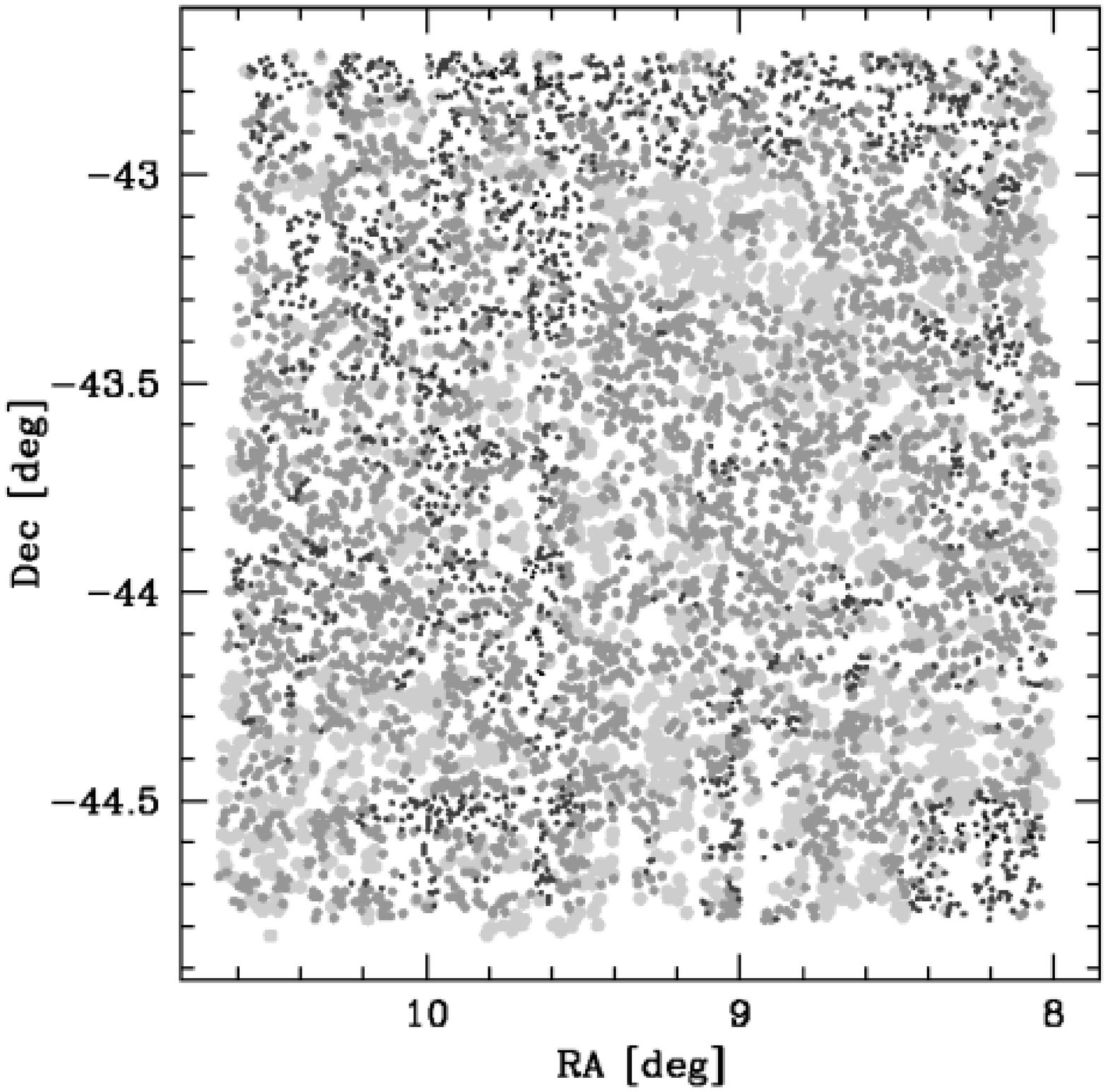}\\
\smallskip
\includegraphics[width=0.35\textwidth]{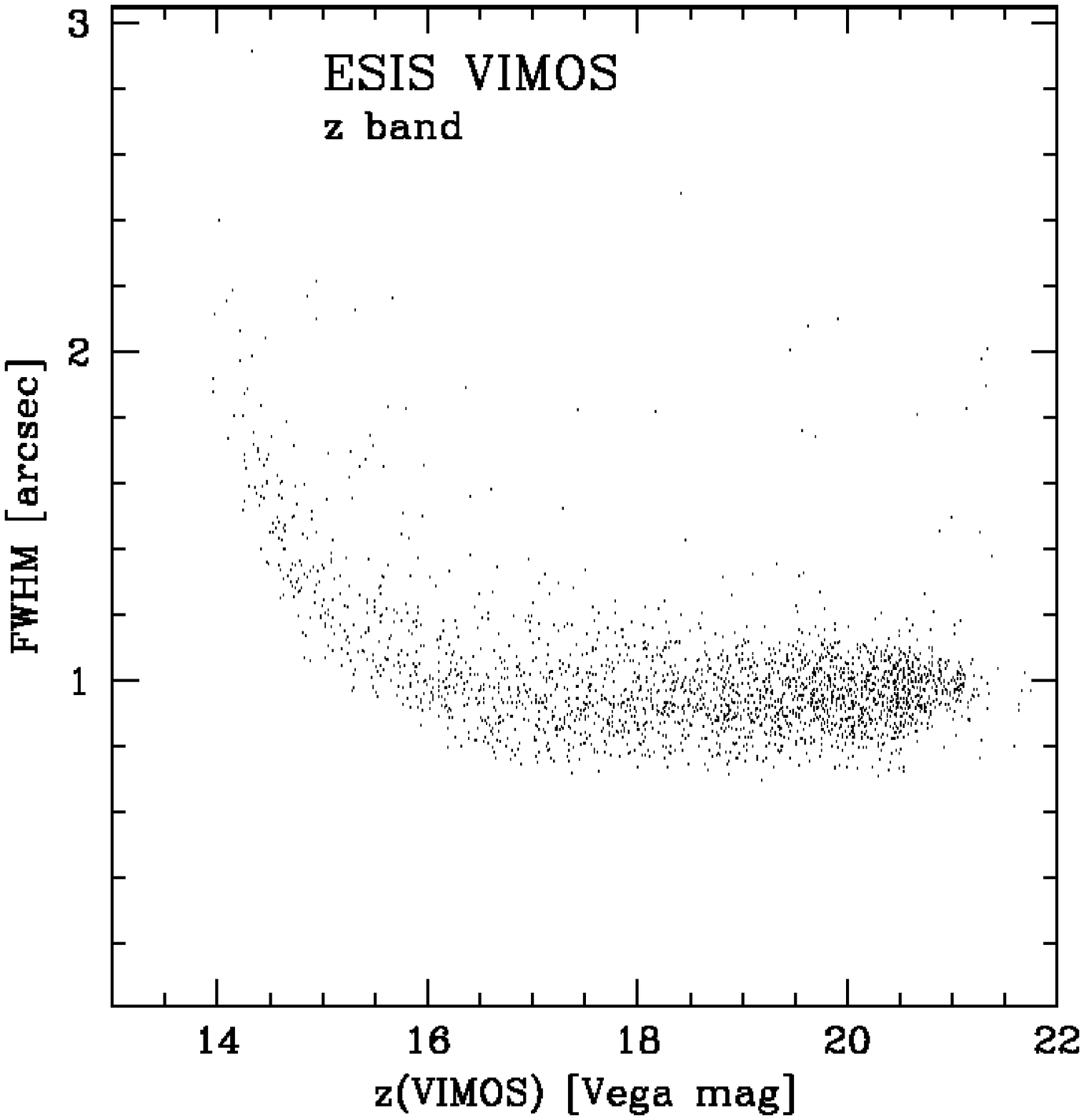}
\includegraphics[width=0.35\textwidth]{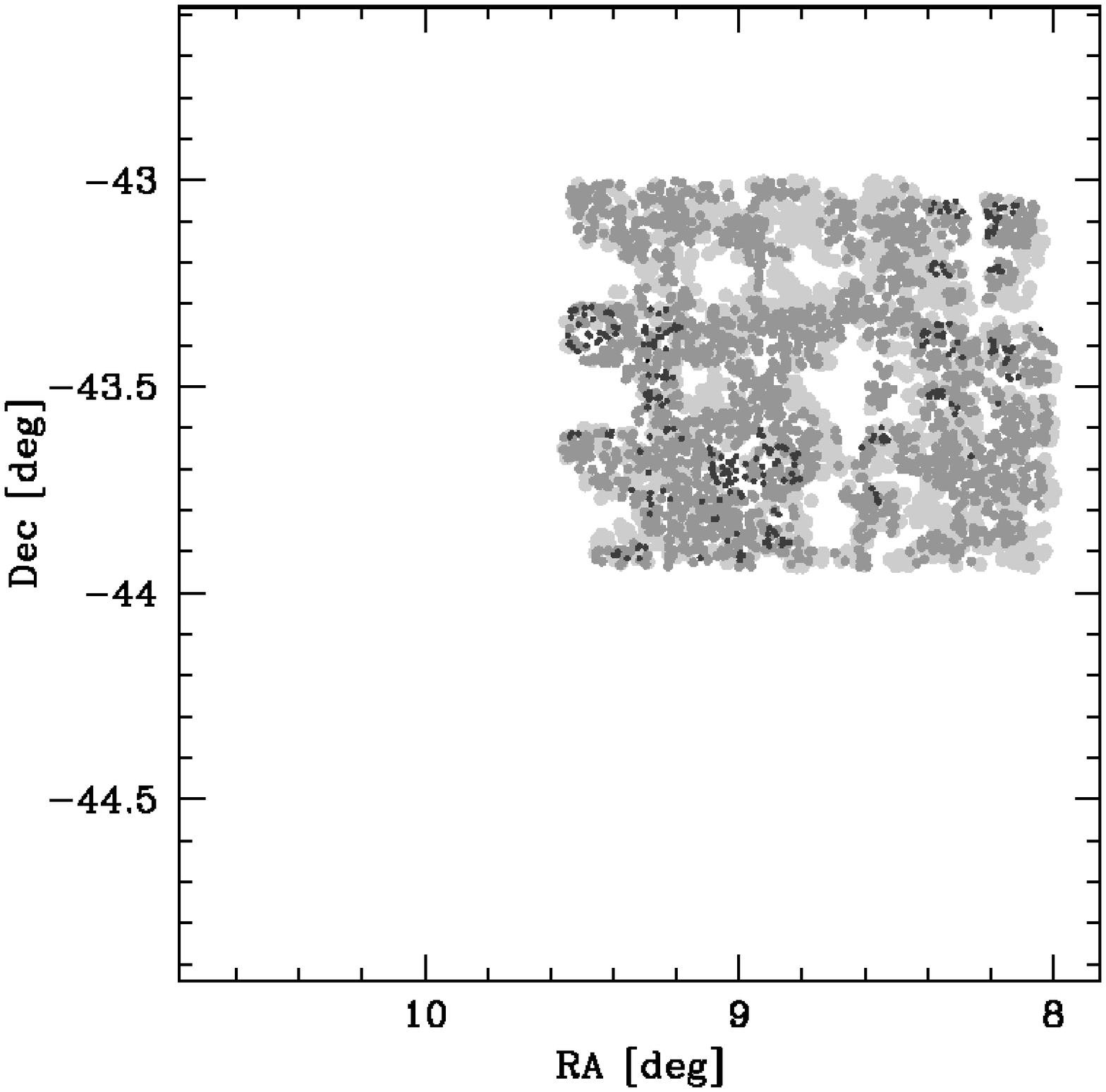}
\caption{Actual FWHM of pointlike sources in the ESIS VIMOS $I$ band (top
panels) and $z$ band (bottom panels) survey. The dependence on magnitude and
the variation of FWHM across the field are shown.
Grey levels and dot dimensions illustrate seeing properties, 
darker and smaller dots referring to better FWHM. Three bins are
considered: $\textrm{FWHM}\le0.8$, $0.8<\textrm{FWHM}\le1.0$,
$1.0<\textrm{FWHM}\le1.2$.}
\label{fig:fwhm}
\end{figure*}

Figure \ref{fig:fwhm} shows the trend of seeing as a function of magnitude
(left panels). The large scatter reflects what mentioned above.
The deviation at bright magnitudes is due to incoming saturation 
effects. The saturation thresholds for the ESIS VIMOS survey turn out to be
$I\sim16.5$ and $z\sim16$ mag $[$Vega$]$, consistent with the expectations 
of ESO's exposure time calculator\footnote{http://www.eso.org/observing/etc/} 
($I\sim16.2$ and $z\sim15.8$ for a point-like source with 
$1{\arcsec}$ seeing and $\textrm{A.M.}=1.3$).

The right hand panels of Fig. \ref{fig:fwhm} illustrate the distribution of seeing across the field.
Pointlike sources are plotted in different grey levels (darker indicating
better seeing) and dot dimensions (smaller for better FWHM). Three bins are
considered: $\textrm{FWHM}\le0.8$, $0.8<\textrm{FWHM}\le1.0$,
$1.0<\textrm{FWHM}\le1.2$.

\subsection{Astrometric accuracy}\label{sect:astrom_accuracy}

\begin{figure*}[!ht]
\centering
\includegraphics[width=0.4\textwidth]{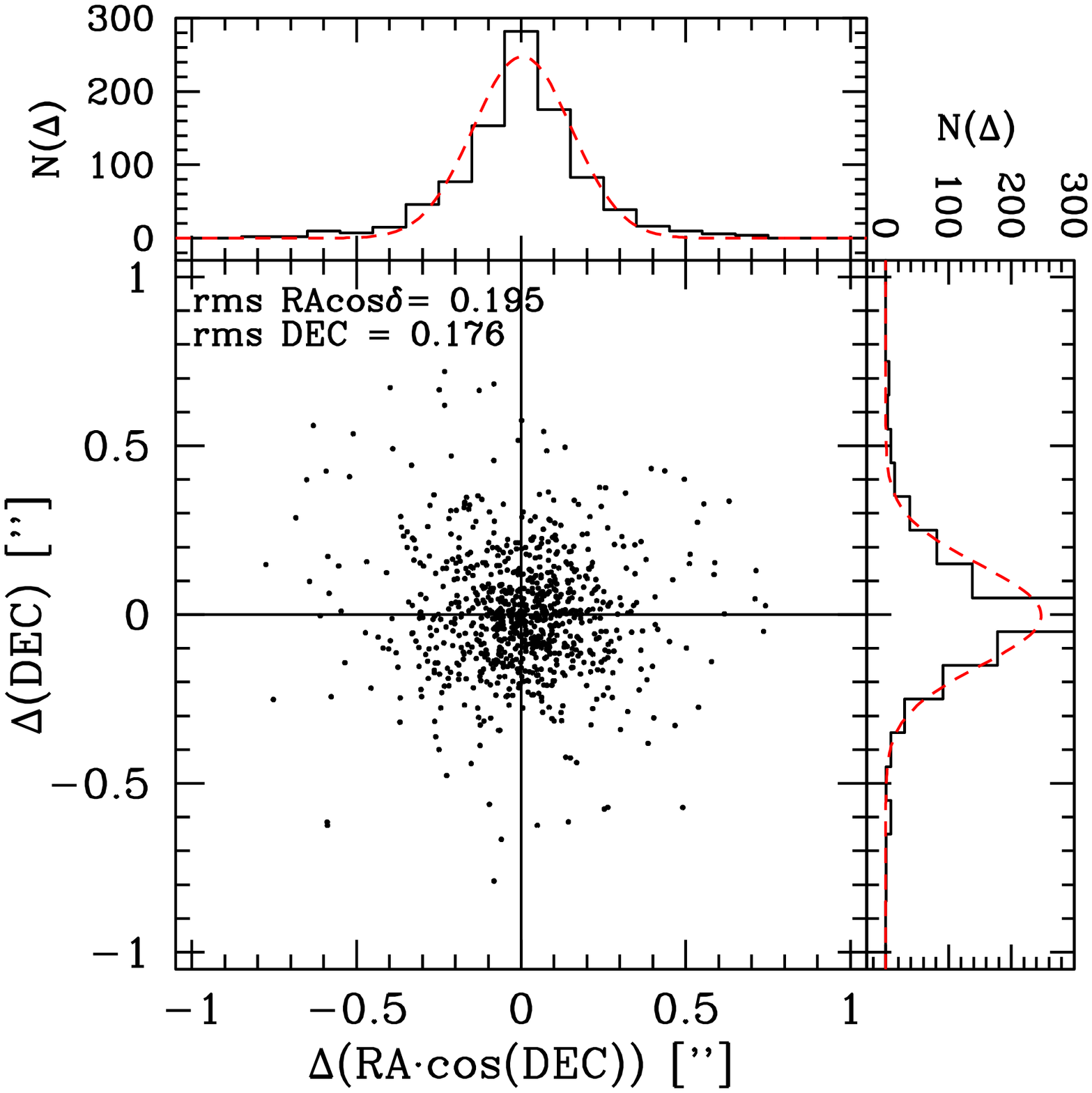}
\includegraphics[width=0.4\textwidth]{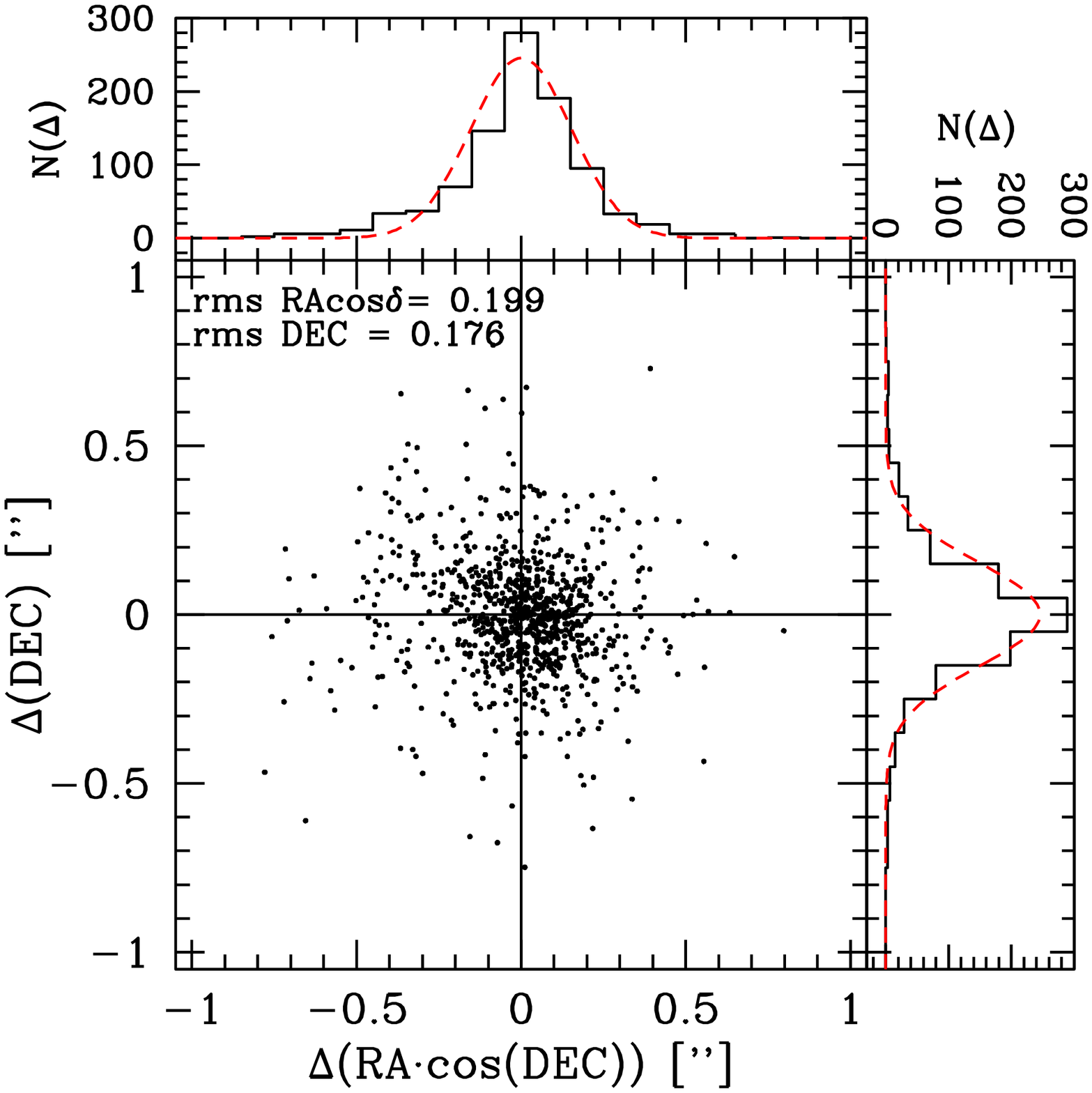} 
\includegraphics[width=0.4\textwidth]{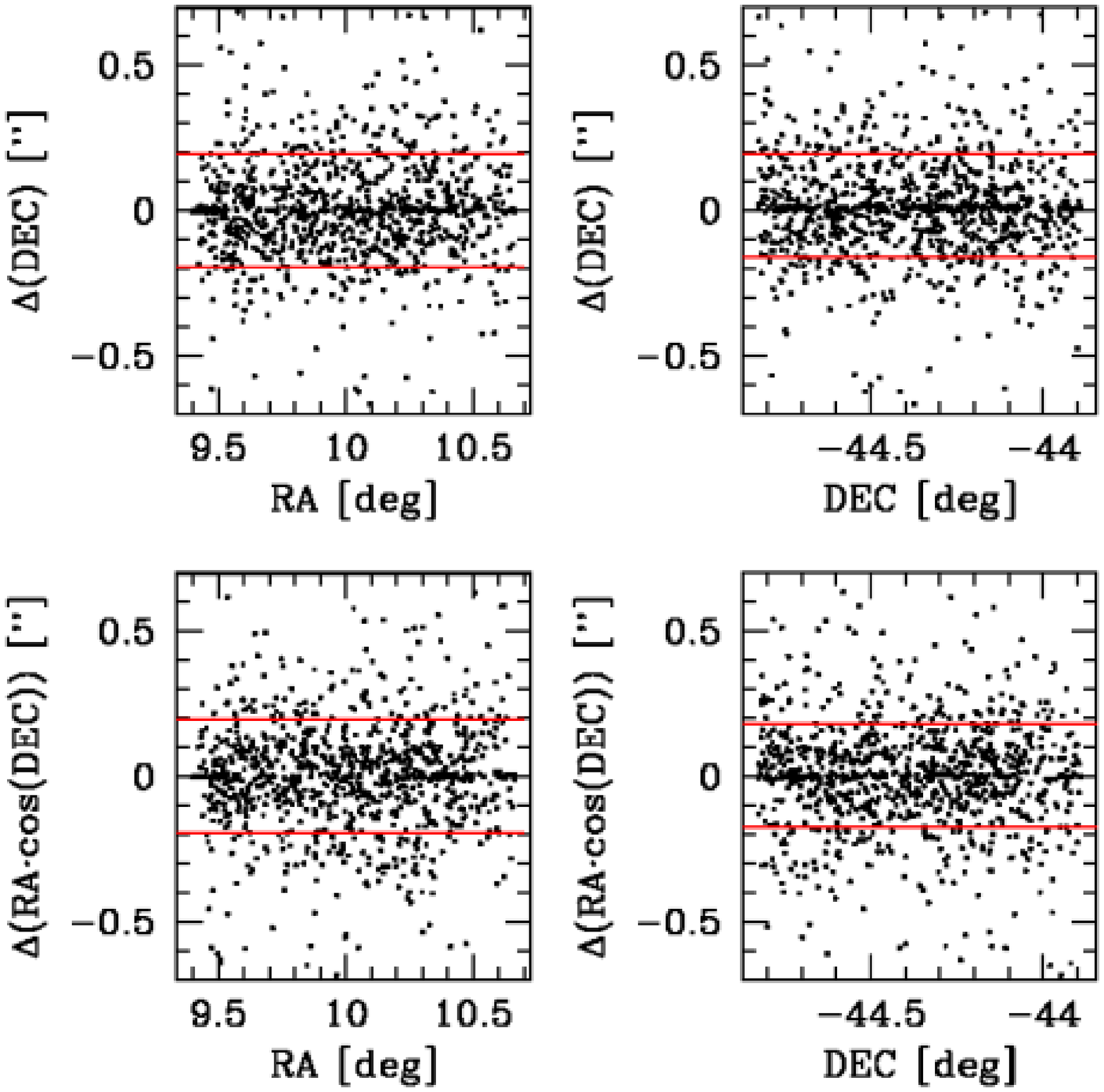}
\includegraphics[width=0.4\textwidth]{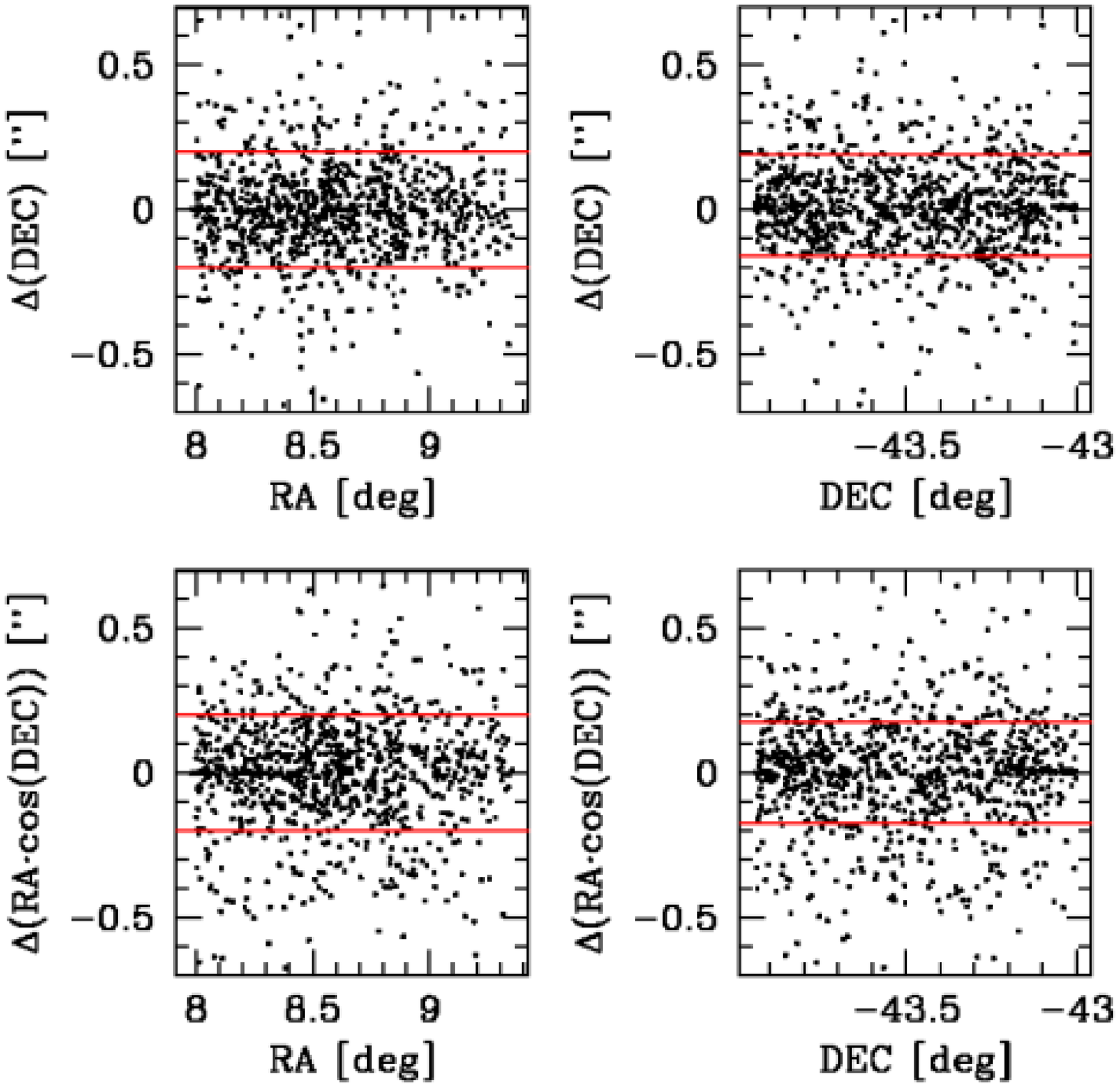} 
\includegraphics[width=0.4\textwidth]{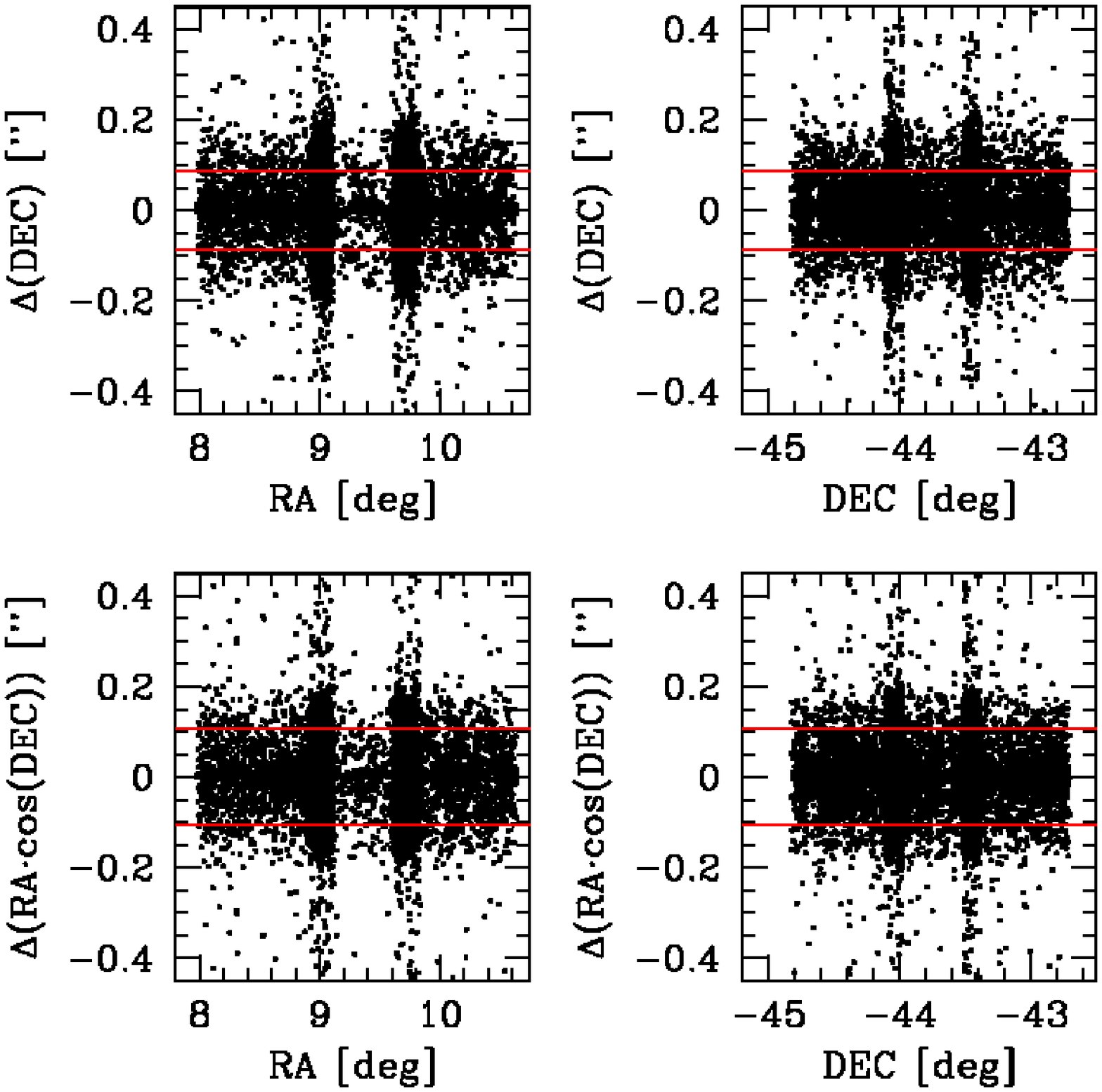}
\includegraphics[width=0.4\textwidth]{white.eps}
\caption{Astrometric accuracy of the ESIS VIMOS survey. {\em Left panels}: $I$
band. The top and central plots show the displacement of ESIS coordinates with
respect to the GSC 2.2 (STScI \& OaTO, \citeyear{gsc2}) catalog. The bottom figure shows the
relative astrometric accuracy between the 9 different $I$ band sub-areas.
{\em Right panels}: same plots for the $z$ band. Only absolute astrometry is
shown, because the $z$ band covers 1 deg$^2$ and consists of one single final
mosaic. Horizontal solid lines set the $\pm 1\sigma$ levels of 
$\Delta(coords)$.}
\label{fig:astrometry}
\end{figure*}

The coordinates of the objects detected on the final mosaics
have been compared to GSC 2.2 (STScI \& OaTO, \citeyear{gsc2}) sources, after re-centering 
the GSC catalog on the ESIS images.
Figure \ref{fig:astrometry} shows the results for one of the 
$I$ band mosaics (left panels) and the $z$ band final image (right panels).
The r.m.s. of the distribution of coordinate differences for the $I$ and $z$
bands is 0.195 and 0.199 arcsec in $\Delta(RA\cdot\cos[Dec])$ and 
0.176 arcsec in $\Delta(Dec)$. 

In the central panels in Fig. \ref{fig:astrometry} the coordinate differences 
between the ESIS and GSC catalogs are plotted against RA and Dec: no systematic
trends are detected.

As far as the $I$ band is concerned, we have checked also the coordinate match
between the 9 different mosaics, analyzing the differences in (RA,Dec) for
GSC sources in the overlap regions (left bottom panel in Fig.
\ref{fig:astrometry}). The r.m.s. of the distribution of coordinate differences
are 0.105 and 0.087 arcsec in $\Delta(RA\cdot\cos[Dec])$ and $\Delta(Dec)$
respectively.

\subsection{Photometric accuracy}\label{sect:phot_accuracy}

The photometric accuracy of the data has been tested with simulations, 
adding synthetic sources to the ESIS final mosaics (with IRAF), 
adopting a $1{\arcsec}$ seeing,
and performing source extraction as for science data.

In each band, we produced several simulated images,
by adding point-like, De Vaucouleurs or exponential
disk objects at random positions on the science frames. In each case, a different image
was produced per 0.25 mag bin, in the range 18-27 mag. A
population of $\sim8000$ sources/deg$^2$ was added each time. 

\begin{figure*}[!ht]
\centering
\includegraphics[width=0.35\textwidth]{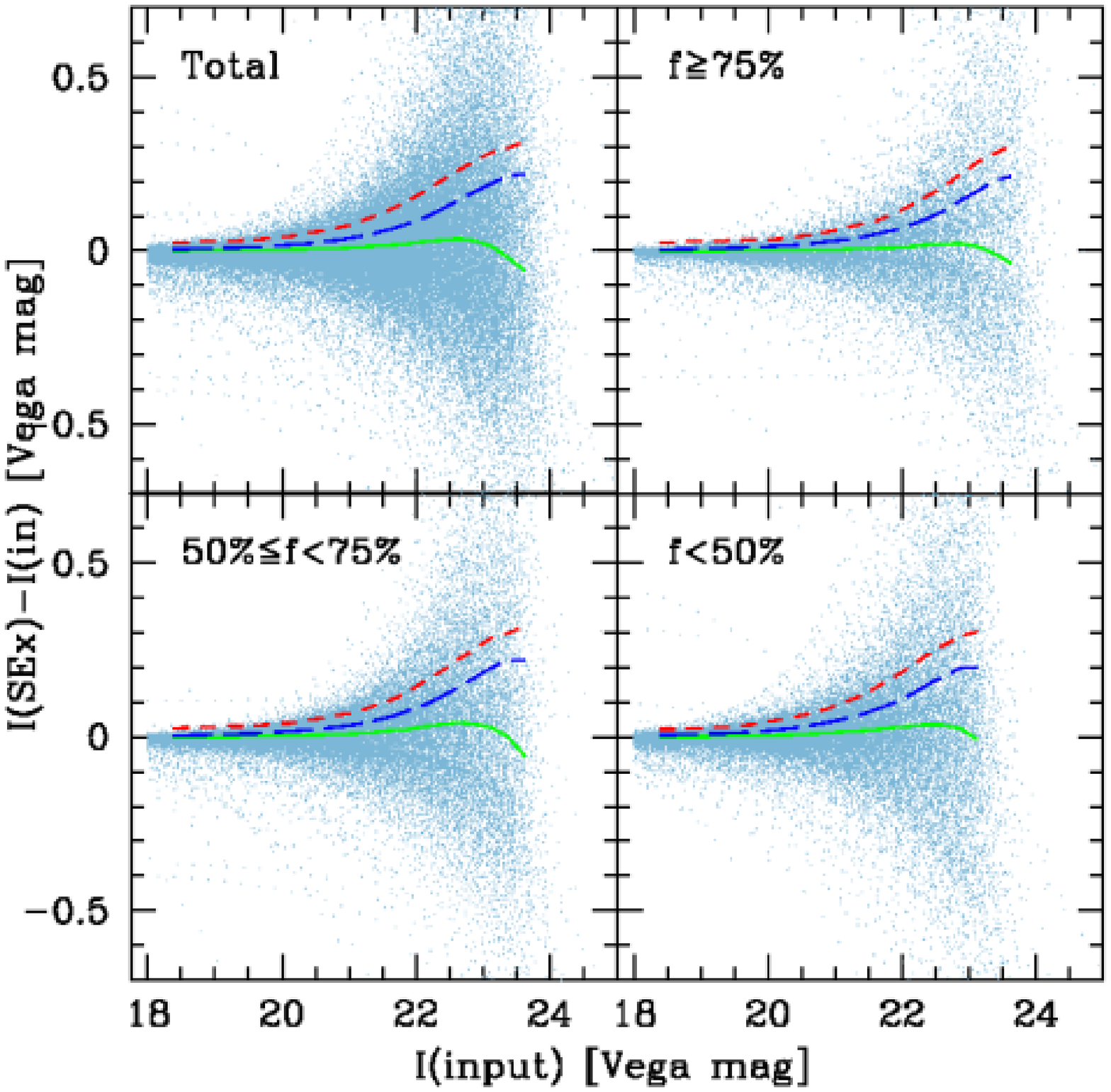}
\includegraphics[width=0.35\textwidth]{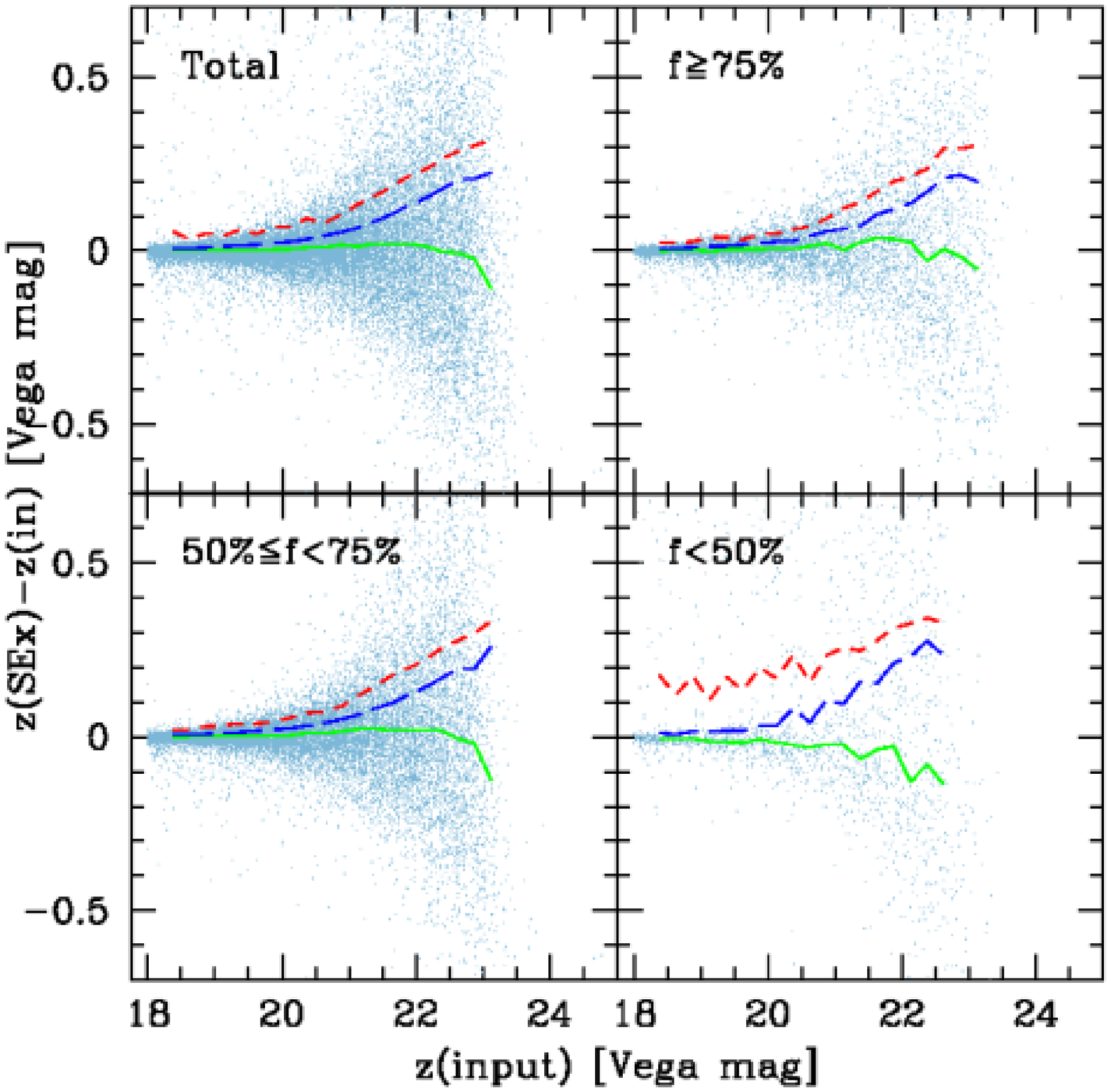}
\caption{Photometric accuracy, derived by extracting synthetic point-like
sources from ESIS VIMOS $I$ band (left) and $z$ band (right) mosaics. Different
panels refer to different effective depths. Solid, short-dashed and long-dashed
lines trace the median, the 1-$\sigma$ and the s.i.q.r. levels respectively.}
\label{fig:photom_accuracy}
\end{figure*}

Figure \ref{fig:photom_accuracy} compares
the input magnitudes of the simulated point-like sources and the values
measured by SExtractor. For each of the two bands, the four panels refer to
regions in the mosaics having effective depths $<50$\%, $50-75$\% and $\ge75$\%
(see Sect. \ref{sect:catalogs}).
The solid
lines represent the median differences, the short-dashed lines are the 1-$\sigma$
standard deviation and the long-dashed lines trace the semi-inter-quartile
ranges (s.i.q.r.). Typical s.i.q.r. uncertainties at $\textrm{mag}=21.5,\ 23.5$
(Vega) are $\sim 0.05,\ 0.20$ in the $I$ band and $\sim0.08,\ 0.22$ in the $z$
band, on average on the whole area.

Slightly larger uncertainties are found for De Vaucouleurs profiles 
and exponential disks: the corresponding s.i.q.r. are $\sim 0.07,\ 0.22$ 
at $\textrm{mag}=21.5,\ 23.5$ (Vega) in the $I$ band and $\sim 0.10,\ 0.27$ 
in $z$. The usual systematic offset for the recovered magnitudes of 
De Vaucouleurs objects is present. Its value is $\sim0.2-0.3$ magnitudes. 
This offset is well described in the literature \citep[e.g.][]{fasano1998}: 
it is due to missing flux in the outskirts of De Vaucouleurs profiles and does not 
depend on the adopted source extraction code.

\subsection{Completeness}\label{sect:completeness}

By comparing the numbers of input and detected sources in simulations, 
we derive the detection rate in the ESIS VIMOS survey.

\begin{figure*}[!ht]
\centering
\includegraphics[width=0.35\textwidth]{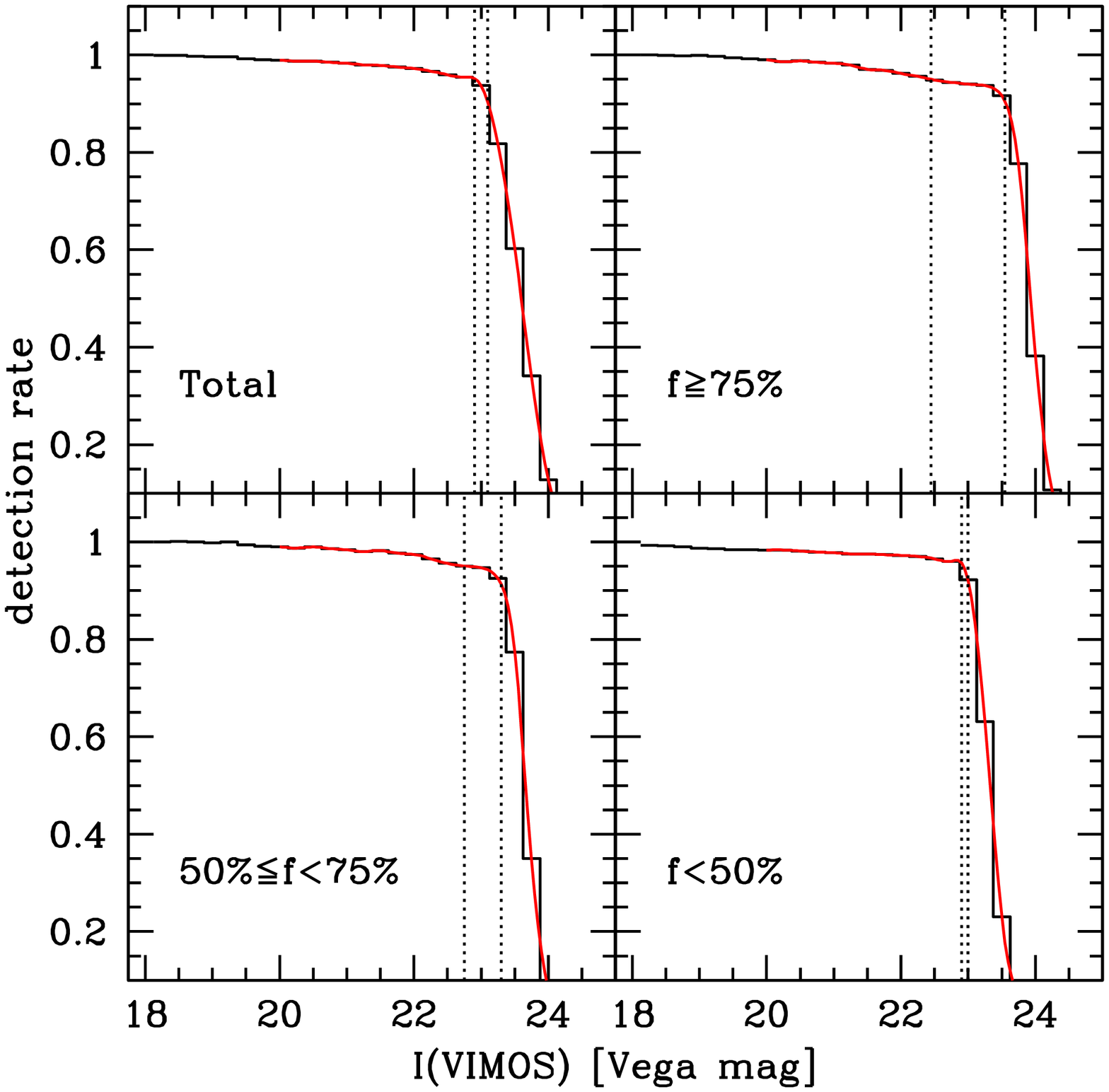}
\includegraphics[width=0.35\textwidth]{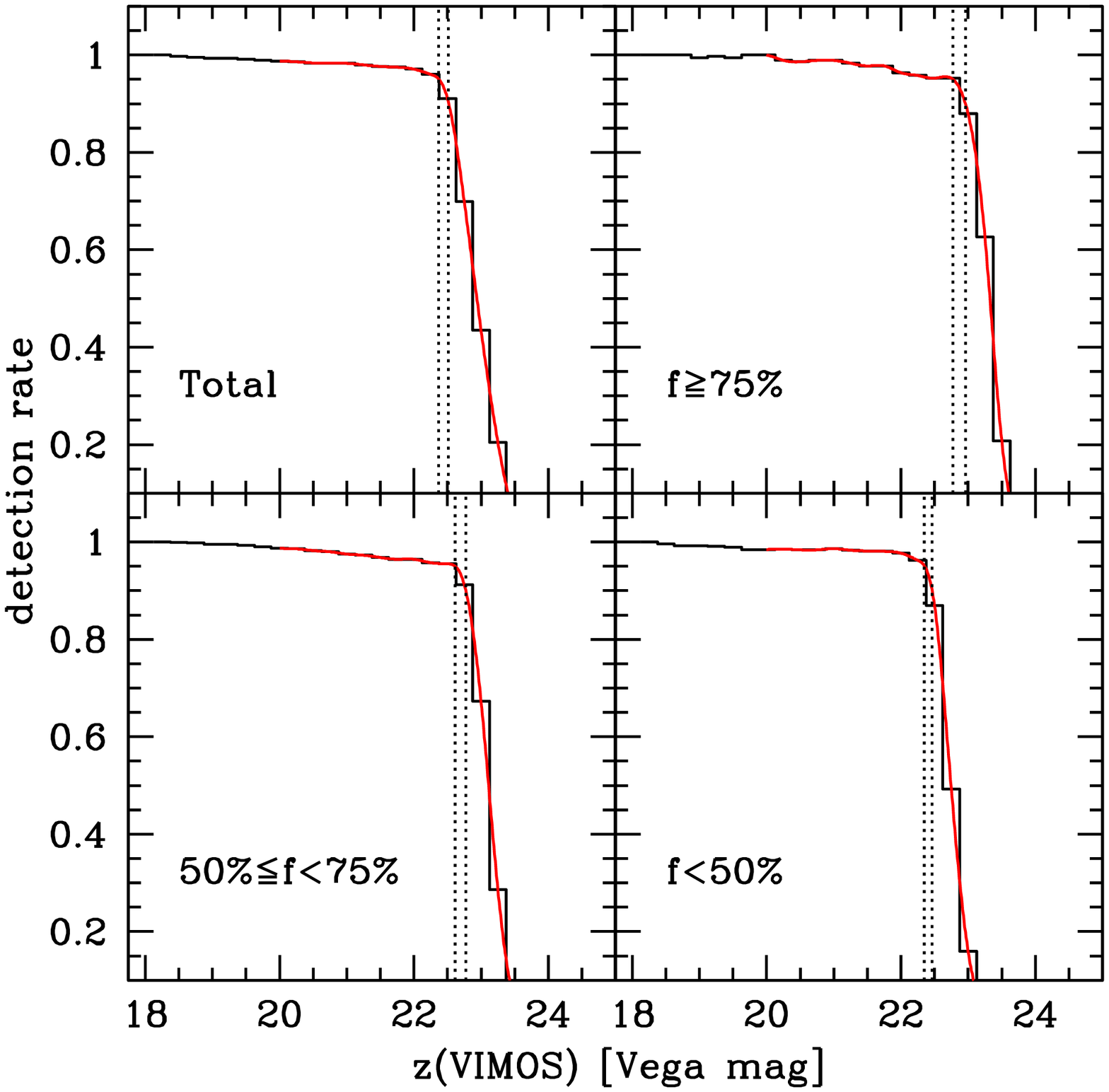}
\caption{Detection rate as a function of magnitude, from simultaed point-like
sources. The vertical dotted lines set the 90\% and 95\% completeness levels.}
\label{fig:completeness}
\end{figure*}

Figure \ref{fig:completeness} shows the results for point-like sources in 
the $I$ (left) and $z$ (right) bands, split by coverage again.
The vertical dotted lines set the 90\% and 95\% completeness levels.

On average, the 90\% level is reached at magnitude 23.1 and 22.5 (Vega) in the $I$ and
$z$ band respectively. These values increase to 23.6 and 23.0, if only the
regions with coverage $\ge 75$\% are taken into account.

As far as exponential disks and De Vaucouleurs profiles are concerned, 
the 90\% detection rate is shifted to $\sim0.3$ and $\sim0.5$ brighter 
magnitudes levels, respectively.

\begin{table*}[!ht]
\centering
\begin{tabular}{c r@{}c l l l}
\hline
\hline
\multicolumn{3}{c}{Column} & Name & Description & Units \\
\hline
&  1 & &  ID		 &  ESIS ID					  &	    \\
&  2 & &  RA		 &  Right Ascension (J2000)			  &  [deg]  \\
&  3 & &  DEC		 &  Declination (J2000) 			  &  [deg]  \\
&  4 & &  MAG\_BEST	 &  Best of MAG\_AUTO and MAG\_ISOCOR		  &  [mag]  \\
&  5 & &  MAGERR\_BEST   &  RMS error for MAG\_BEST			  &  [mag]  \\
&  6 & &  MAG\_AUTO	 &  Kron-like elliptical aperture magnitude	  &  [mag]  \\
&  7 & &  MAGERR\_AUTO   &  RMS error for Kron-like magnitude		  &  [mag]  \\
&  8 & &  MAG\_ISOCOR	 &  Corrected isophotal magnitude		  &  [mag]  \\
&  9 & &  MAGERR\_ISOCOR &  RMS error for corrected isophotal magnitude   &  [mag]  \\
& 10 & &  MAG\_APER1	 &  Aperture magnitude (1.230 arcsec diam.) 	  &  [mag]  \\
& 11 & &  MAG\_APER2	 &  Aperture magnitude (2.460 arcsec diam.) 	  &  [mag]  \\
& 12 & &  MAG\_APER3	 &  Aperture magnitude (3.280 arcsec diam.) 	  &  [mag]  \\
& 13 & &  MAG\_APER4	 &  Aperture magnitude (4.715 arcsec diam.) 	  &  [mag]  \\
& 14 & &  MAG\_APER5	 &  Aperture magnitude (6.675 arcsec diam.) 	  &  [mag]  \\
& 15 & &  MAGERR\_APER1  &  RMS error for APER1 magnitude      		  &  [mag]  \\
& 16 & &  MAGERR\_APER2  &  RMS error for APER2 magnitude      		  &  [mag]  \\
& 17 & &  MAGERR\_APER3  &  RMS error for APER3 magnitude      		  &  [mag]  \\
& 18 & &  MAGERR\_APER4  &  RMS error for APER4 magnitude      		  &  [mag]  \\
& 19 & &  MAGERR\_APER5  &  RMS error for APER5 magnitude      		  &  [mag]  \\
& 20 & &  SN		 &  Local S/N ratio				  &	    \\
& 21 & &  KRON\_RADIUS   &  Kron aperture in units of A or B		  &	    \\
& 22 & &  FWHM\_IMAGE	 &  FWHM assuming a gaussian core		  &  [pixel]\\
& 23 & &  CLASS\_STAR	 &  Star/Galaxy classifier output		  &	    \\
& 24 & &  FLUX\_RADIUS   &  Half-flux radius				  &  [pixel]\\
& 25 & &  FLAGS 	 &  SExtractor extraction flags			  &	    \\
& 26 & &  IMAFLAGS\_ISO  &  Coverage flag 				  &	    \\
\hline
\end{tabular}
\caption{Contents of ESIS VIMOS $I$ and $z$ band catalogs.}
\label{tab:catalog_description}
\end{table*}

\begin{table}[!ht]
\centering
\begin{tabular}{c c c c c}
\hline
\hline
 & \multicolumn{2}{c}{$I$ band} & \multicolumn{2}{c}{$z$ band} \\
mag. 		& \multicolumn{2}{c}{$\log N$} & \multicolumn{2}{c}{$\log N$} \\
$[$Vega$]$	& Tot. & Ext. & Tot. & Ext. \\
\hline
   14.0 &   1.915 &   0.711  &  1.991  &  0.610  \\
   14.5 &   2.164 &   1.267  &  2.343  &  1.087  \\
   15.0 &   2.390 &   1.410  &  2.388  &  1.724  \\
   15.5 &   2.497 &   1.736  &  2.565  &  2.254  \\
   16.0 &   2.526 &   1.736  &  2.579  &  2.223  \\
   16.5 &   2.655 &   2.069  &  2.656  &  2.366  \\
   17.0 &   2.738 &   2.216  &  2.792  &  2.597  \\
   17.5 &   2.870 &   2.430  &  2.964  &  2.750  \\
   18.0 &   3.024 &   2.675  &  3.113  &  2.937  \\
   18.5 &   3.181 &   2.945  &  3.261  &  3.143  \\
   19.0 &   3.361 &   3.188  &  3.437  &  3.357  \\
   19.5 &   3.513 &   3.409  &  3.580  &  3.536  \\
   20.0 &   3.659 &   3.623  &  3.709  &  3.687  \\
   20.5 &   3.795 &   3.788  &  3.891  &  3.888  \\
   21.0 &   3.938 &   3.938  &  4.017  &  4.017  \\
   21.5 &   4.087 &   4.087  &  4.135  &  4.135  \\
   22.0 &   4.216 &   4.216  &  4.232  &  4.232  \\
   22.5 &   4.338 &   4.338  &  4.264  &  4.264  \\
   23.0 &   4.410 &   4.410  &  4.180  &  4.180  \\
   23.5 &   4.409 &   4.409  &  3.951  &  3.951  \\
   24.0 &   4.270 &   4.270  &  3.433  &  3.433  \\
   24.5 &   3.901 &   3.901  &    --   &    --   \\
   25.0 &   3.008 &   3.008  &    --   &    --   \\
\hline
\end{tabular}
\caption{ESIS VIMOS $I$ and $z$ band number counts in units of 
$[$deg$^{-2}$ mag$^{-1}]$. No completeness correction was applied.}
\label{tab:number_counts}
\end{table}

\section{Catalogs}\label{sect:catalogs}

The total exact area covered by the $I$ band mosaics is 4.11 deg$^2$. Excluding 
the gaps and areas around very bright stars, the catalog covers 3.89 deg$^2$.
We define the coverage flag as the fraction $f$ of nominal exposure time (900s
for  the $I$ band, 1800s in $z$) spent on source; 25\%, 64\% and 96\% of the
sources in the $I$ band catalog lie in regions with $f>0.75,\ 0.50,\ 0.25$
respectively.  

The $z$ band survey includes a 1.08 deg$^2$ area.
After removal of regions not covered and big stars, the catalog includes 0.98 deg$^2$, 
20\%, 55\% and 96\% of which are above $f>0.75,\ 0.50,\ 0.25$.

Catalogs were extracted using SExtractor \citep{bertin1996} from the final
mosaics, weighting the data with coverage maps. A total of 312929 and 57926
sources were measured in the $I$ and $z$ band respectively.

The list and description of columns included in the released catalogs are
provided in Tab. \ref{tab:catalog_description}. All magnitudes are given in 
Vega units. Data are provided as electronic tables associated to this work.

\begin{figure*}[!ht]
\centering
\includegraphics[width=0.35\textwidth]{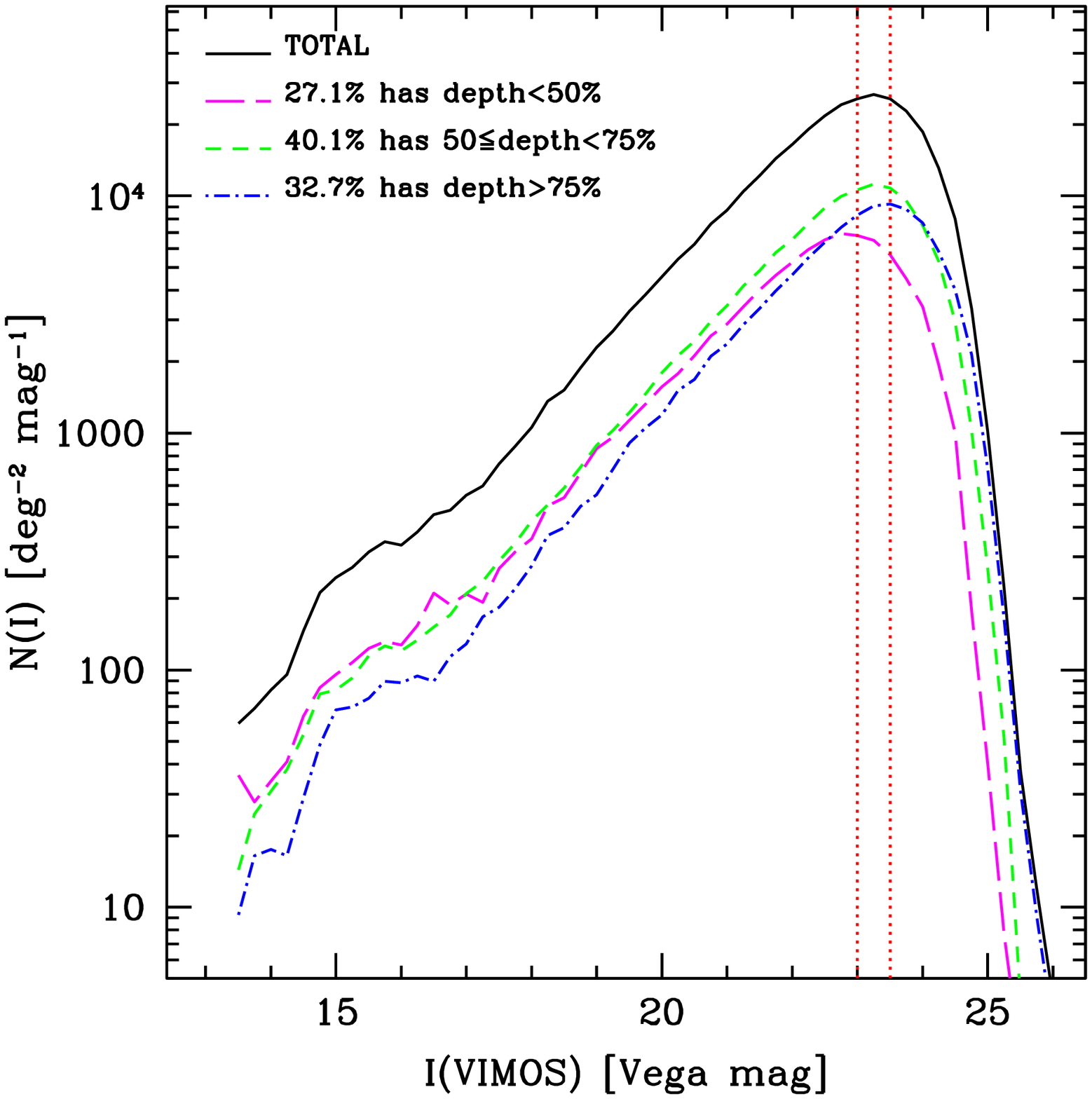}
\includegraphics[width=0.35\textwidth]{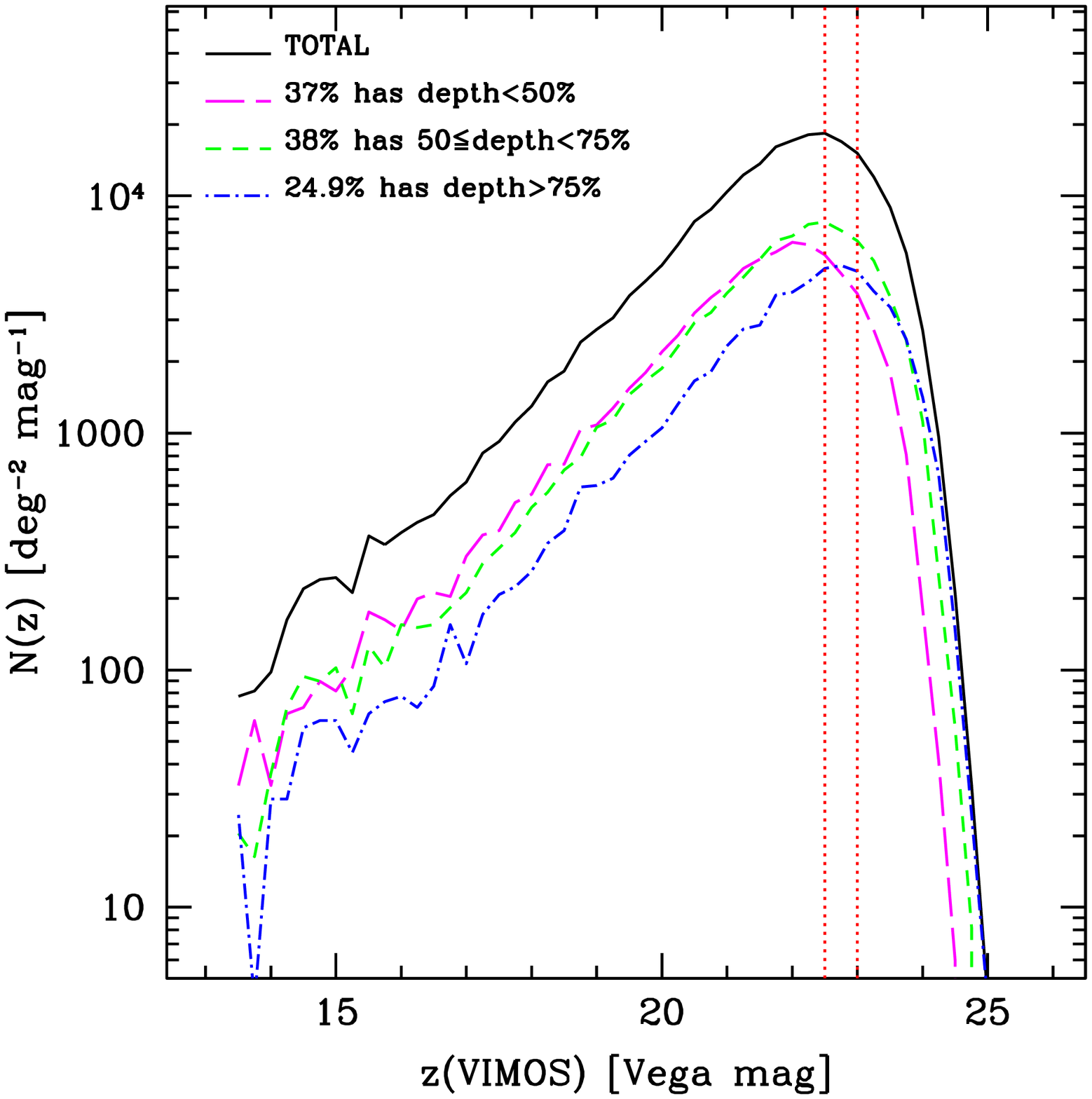}\\
\includegraphics[width=0.35\textwidth]{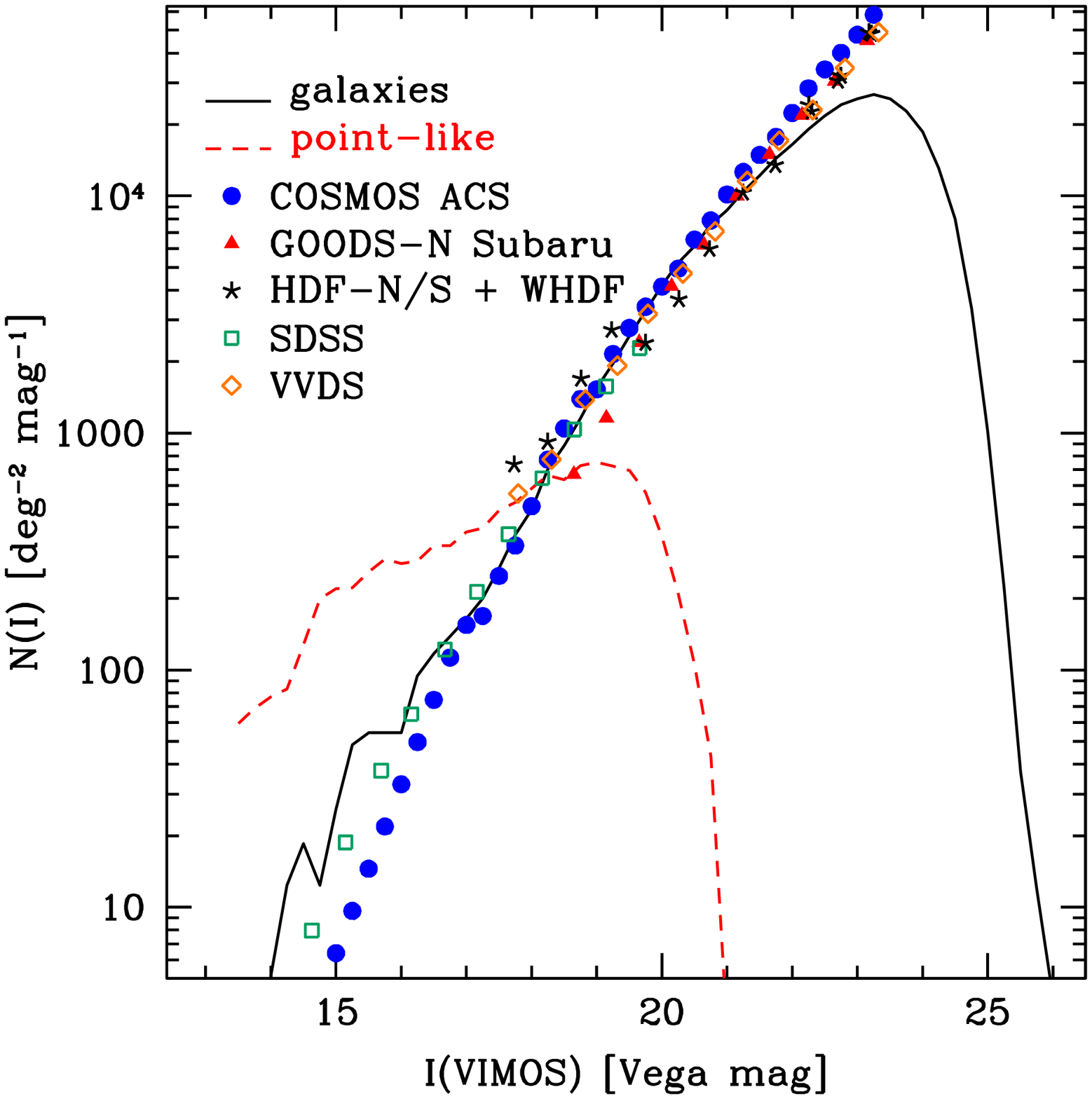}
\includegraphics[width=0.35\textwidth]{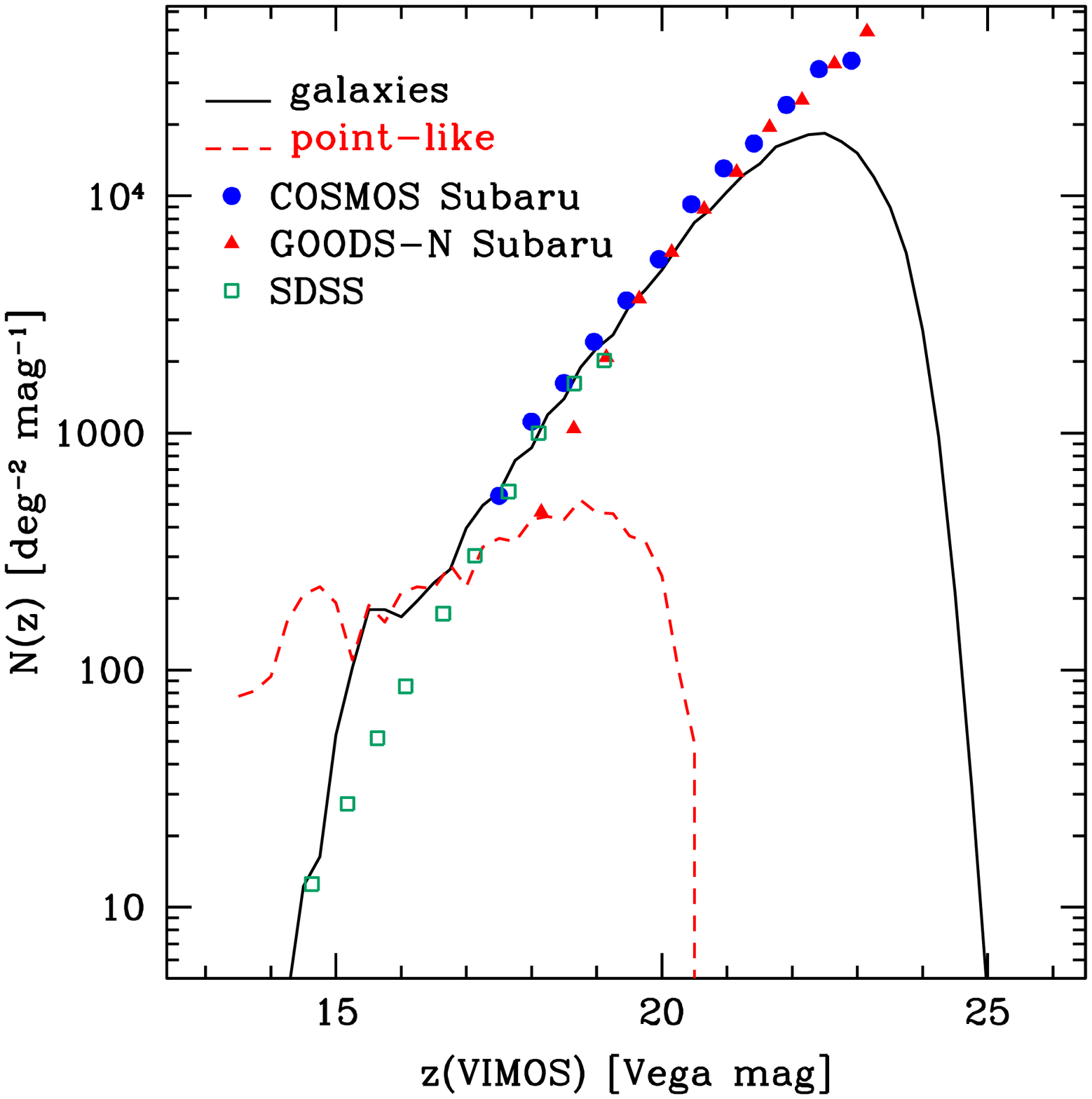}
\caption{Observed source number counts for the ESIS VIMOS survey, split into
effective depth bins (top panels) and into point-like vs. extended sources
(bottom panels). Literature data belong to: COSMOS
\citep{scoville2007,taniguchi2007}, GOODS-N \citep{capak2004}, VVDS
\citep{mccraken2003}, HDFN, HDFS, William Herschel \citep[WHDF][]{metcalfe2001},
SDSS \citep{yasuda2001}.}  
\label{fig:number_counts}
\end{figure*}

Figures \ref{fig:number_counts} shows the ESIS VIMOS number counts, as
split by effective depth (top panels) and into point-like and extended sources
(bottom panels). The latter are compared to data from the COSMOS
\citep{scoville2007,taniguchi2007}, GOODS-N \citep{capak2004}, VVDS \citep{mccraken2003},
HDFN, HDFS, William Herschel \citep[WHDF][]{metcalfe2001} and
SDSS\citep{yasuda2001} surveys. The ESIS data are fairly consistent with the
literature; discrepancies can be due to cosmic variance at the bright
magnitudes, to a different criterion in galaxy/point-like sources separation at
intermediate fluxes, and to incoming ESIS incompleteness at the faint end.
We warn also that saturation effects might play a role at the brightest fluxes
(see Sect. \ref{sect:quality}).

\begin{figure*}[!ht]
\centering
\includegraphics[width=0.35\textwidth]{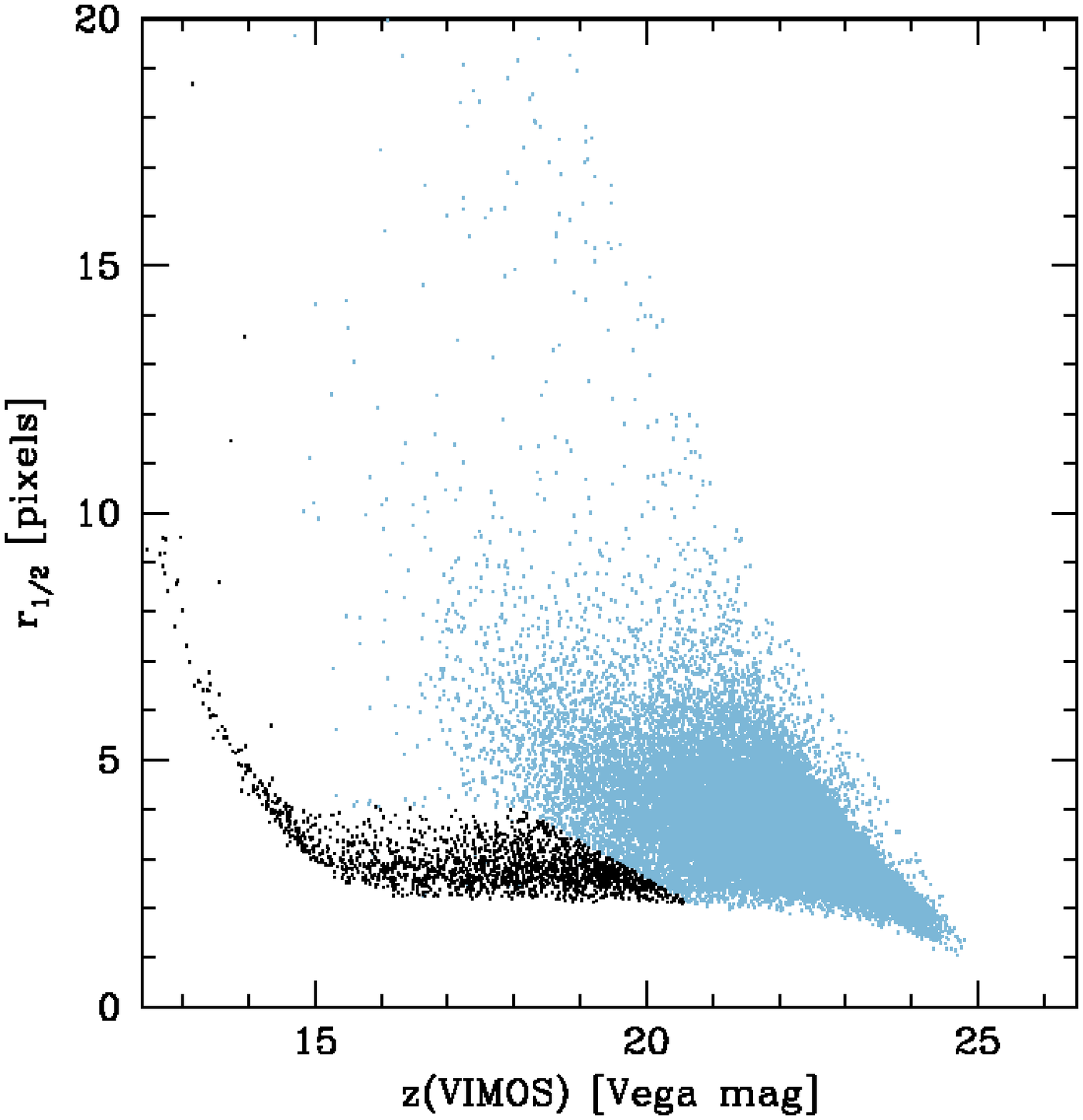}
\includegraphics[width=0.35\textwidth]{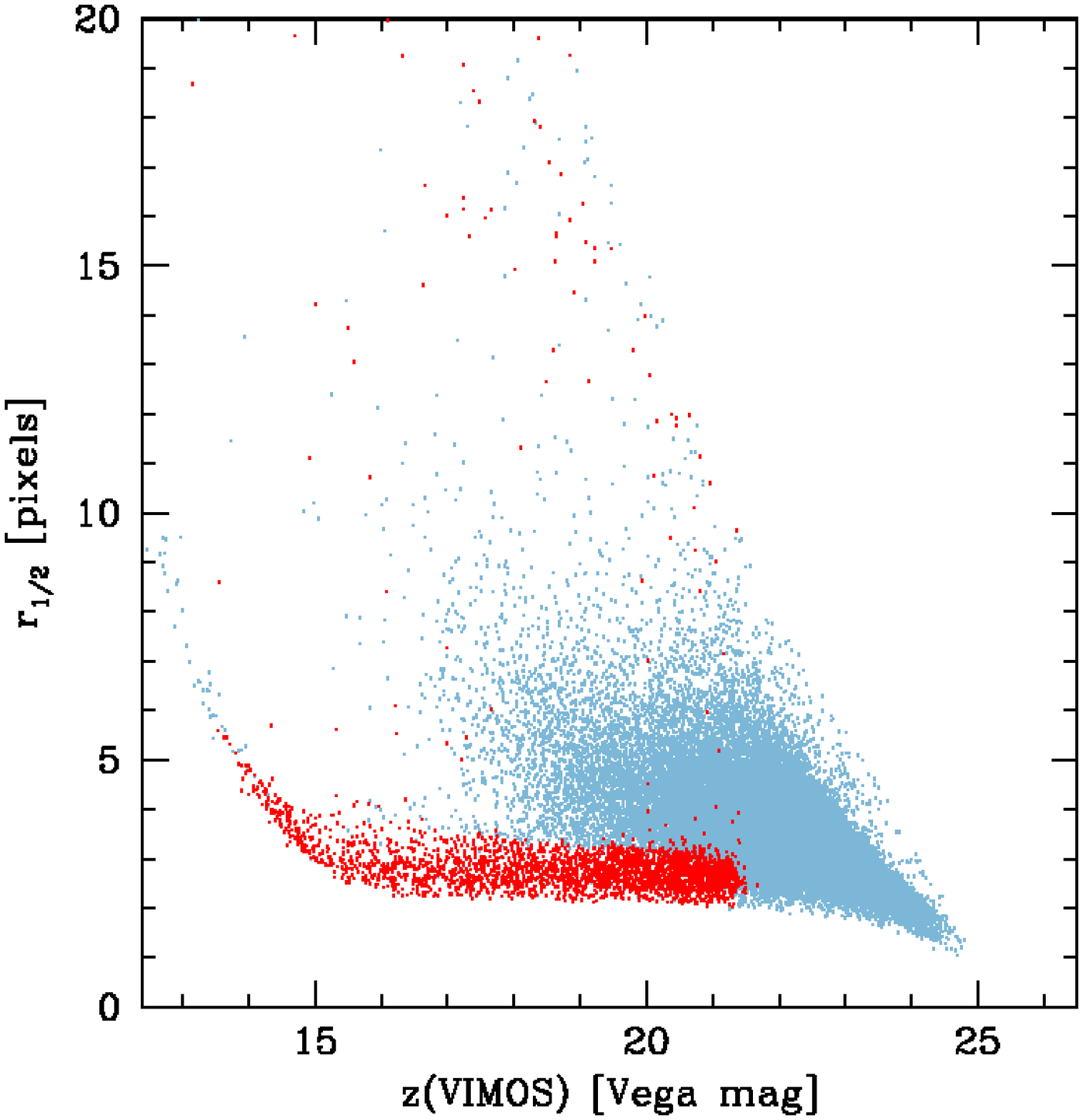}
\caption{Selection of point-like sources based on the half-flux radius or
SExtractor's stellarity flag (right-hand panel), for the $z$ band. Darker dots
represent point-like objects.}
\label{fig:stellarity}
\end{figure*}

The data collected from the literature used the SExtractor's stellarity flag to
distinguish these two classes, while here the half-flux radius is exploited
\citep[see for example][]{berta2006}.
Figure \ref{fig:stellarity} shows the different selection of point-like
objects for the $z$ band, based on a simple cut in the stellarity flag ($>0.95$
in this case, right panel) or based on the dependence of the half-flux radius
on magnitude (left panel). 
Note again the effects of pixel saturation: the bright point-like objects 
deviate from the locus defined by half-flux radius. The SExtractor stellarity
flag misses the brightest saturated point-like sources and tends to be contaminated by 
galaxies at intermediate fluxes \citep[see also][]{berta2006}.

Table \ref{tab:number_counts} reports the observed number counts of the ESIS
VIMOS survey.

\section{High-redshift galaxies}\label{sect:selection}

Large area, multiwavelength extragalactic surveys are designed 
to study the evolution of galaxies across cosmic time. 
SWIRE-ESIS covers the whole electromagnetic spectrum from the X-rays 
to radio wavelengths and is well suited to identify large samples of any class of objects
reachable at its depth.

Of particular interest is the recently discovered class of 
massive galaxies ($M>10^{11}$ M$_\odot$) above redshift $z>1.5$, 
whose formation and evolution appears to be faster than what predicted by
pure hierarchical scenarios and apparently showing {\em downsizing} effects 
\citep[e.g.][]{cimatti2002c,daddi2004,fontana2006,bundy2005,franceschini2006}.

Thanks to IRAC observations, SWIRE-ESIS directly probes the assembled stellar
mass of galaxies up to redshift $z\sim3$. Due to its moderate depth,
SWIRE can detect only the most massive tail of the
high-redshift galaxy stellar mass function \citep[e.g. $M>2\times10^{11}$
M$_\odot$ at $z\sim3$,][]{berta2007b}.

Here we describe how the availability of $I$ and $z$ band data, when combined with
the other multiwavelength observations, provides good constraints in the selection of
high-$z$ massive galaxies.

For the current purpose, we have matched the SWIRE Spitzer catalog 
\citep{lonsdale2003,lonsdale2004} to ESIS $B,\ V,\ R,\ I$ and $z$
\citep[][and this work]{berta2006}, near-IR $J$, $K_s$ \citep{dias2007}, 
and GALEX GR4\footnote{http://galex.stsci.edu/GR4/} data, using 
a closest neighbor criterion with a 1 arcsec radius \citep[see][]{berta2006}. 
We use total magnitudes in all bands.
The SWIRE 3$\sigma$ depths are 2.2, 3.2, 25.8, 22.6, 200 $\mu$Jy at 3.6, 4.5, 5.8, 8.0 and 24 $\mu$m;
the ESIS {\em BVR} 95\% completeness levels are reached at magnitudes $B\, V\sim25$ and $R\sim24.5$ (Vega);
the $J,\ K_s$ catalogs are 95\% complete at 19.82, 18.73 mag (Vega) respectively.

The ESIS VIMOS $I$ band area includes 219877 SWIRE sources, detected in at least
one IRAC band. Among these, 142776 (65\%) have at least one optical detection 
({\em BVRIz}), 137206 are detected in the $I$ band, 
52620 in {\em BVR} (one band at least, over 1.5 deg$^2$), 
35670 in $z$ (over $\sim 1$ deg$^2$), and 30355 are detected in the GALEX DR4
ELAIS-S1 survey. 
The $z$ band area, covered by all bands ({\em BVRIz}, {\em JK$_s$}, 
Spitzer, GALEX, X-rays and radio), includes 63770 SWIRE sources, 
45873 ($\sim72$\%) of which detected in at least one optical band.
Table \ref{tab:multilambda_stats} summarizes these statistics.

\begin{table}
\centering
\begin{tabular}{l c r}
\hline
\hline
Class & area & number \\
 & $[$deg$^2]$ & \\
\hline
SWIRE ($I$-band area)& 4.11 & 219877 \\
SWIRE+{\em BVRIz} (one band) & 3.89 & 142776\\
SWIRE+{\em BVR} (one band) & 1.5 & 52620 \\
SWIRE+$I$ & 3.89 & 137206\\
SWIRE+$z$ & 0.98 & 35670 \\
SWIRE+GALEX & 4.11 & 30355 \\
24$\mu$m & 4.11 & 12770\\
\hline
SWIRE ($z$-band area)& 0.98 & 63770 \\
SWIRE+{\em BVRIz} (one band) & 0.98 & 45873\\
SWIRE+{\em BVR} (one band) & 0.98 & 39883 \\
SWIRE+$I$ & 0.98 & 41337\\
SWIRE+$z$ & 0.98 & 35670 \\
SWIRE+GALEX & 0.98 & 8884 \\
24$\mu$m & 0.98 & 3733\\
\hline
4.5$\mu$m-peak (total) & 4.11 & 938\\  
4.5$\mu$m-peak (interlopers) & 4.11 & 353\\ 
5.8$\mu$m-peak & 4.11 & 973 \\  
\hline
\end{tabular}
\caption{Multi-wavelength detection statistics in the ELAIS-S1 
area.}
\label{tab:multilambda_stats}
\end{table}

\subsection{Selection of IR-peakers}\label{sect:ir_pekaers}

\begin{figure*}[!ht]
\centering
\includegraphics[width=0.35\textwidth]{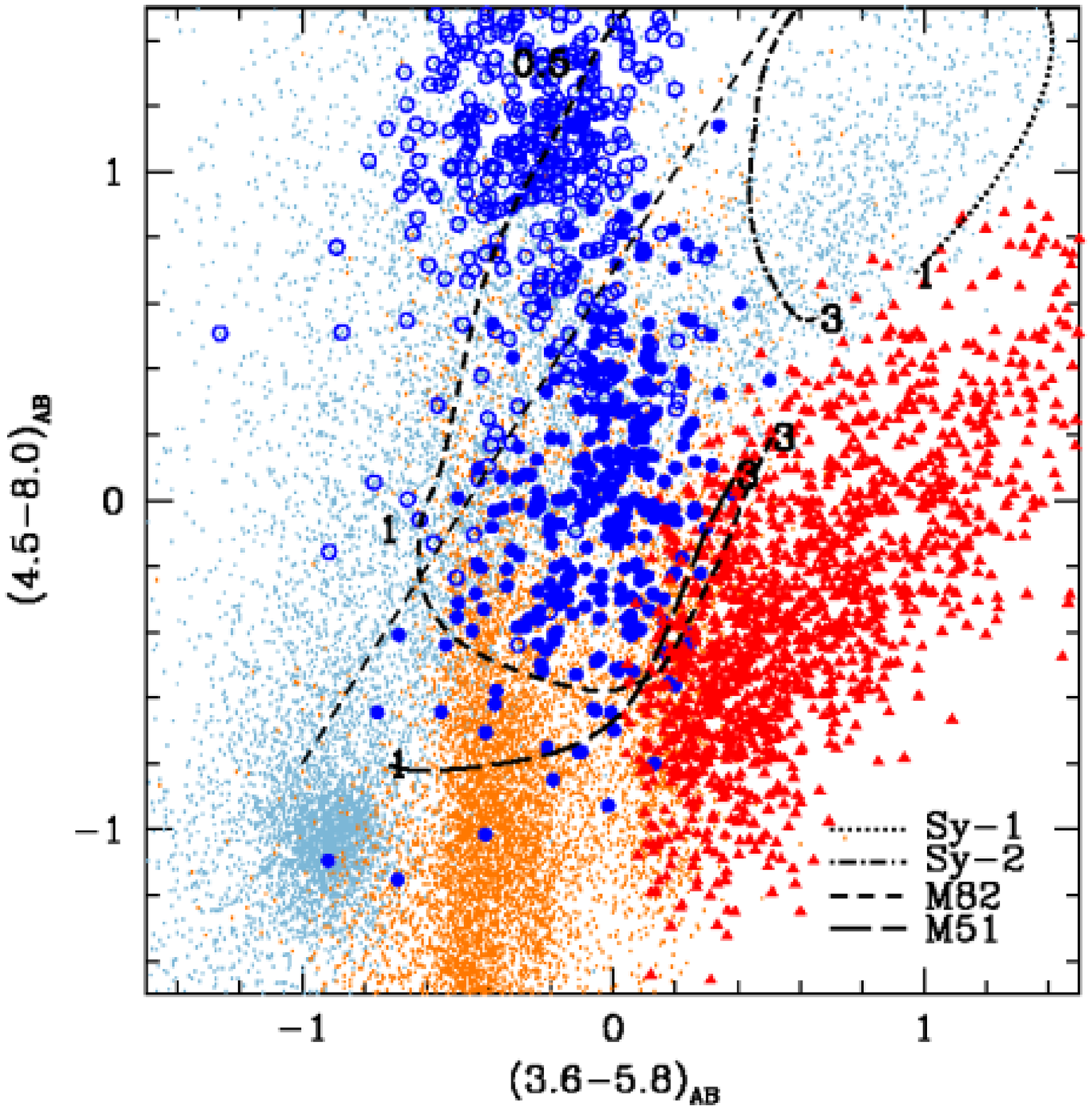}
\includegraphics[width=0.35\textwidth]{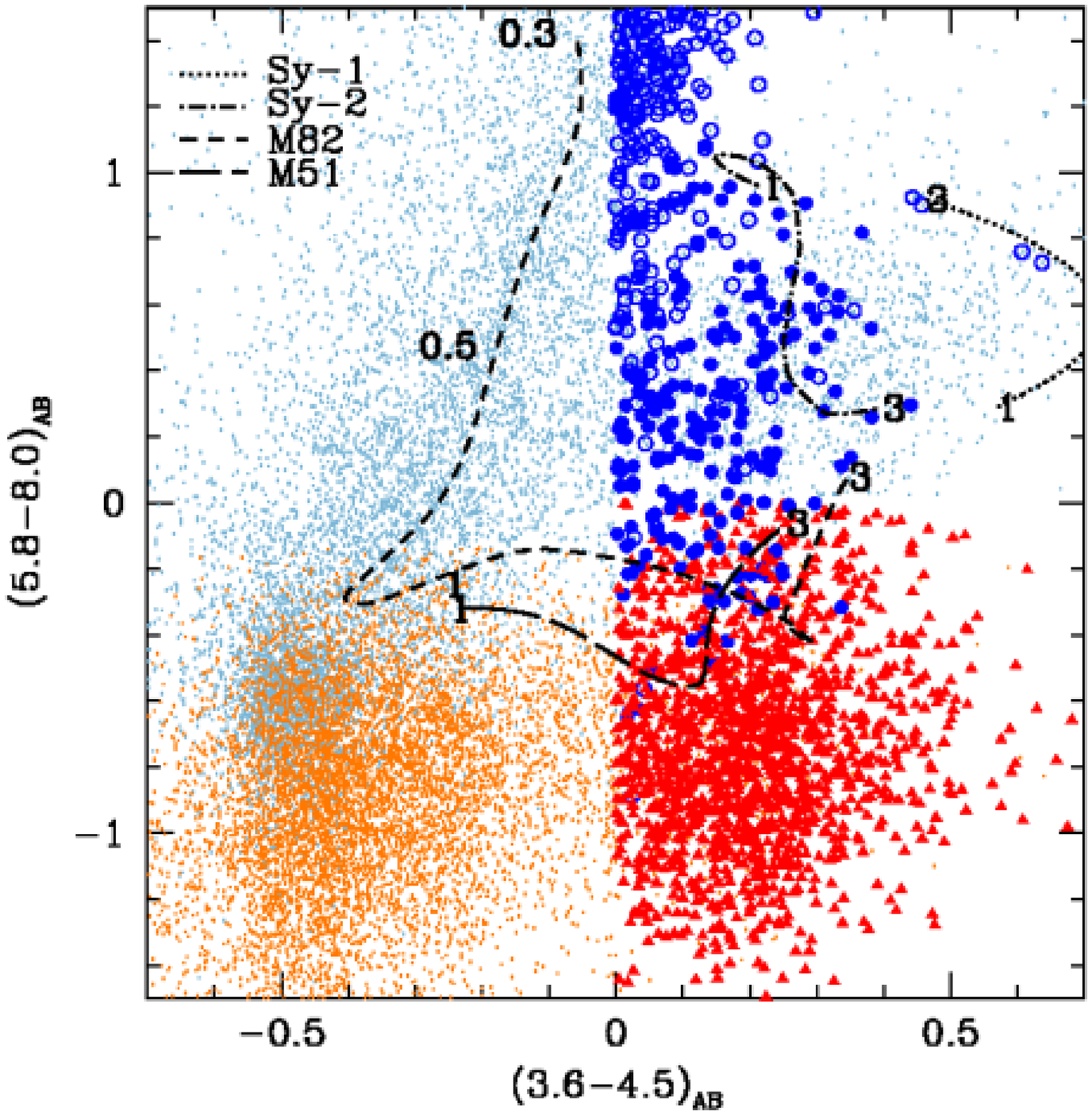}
\includegraphics[width=0.35\textwidth]{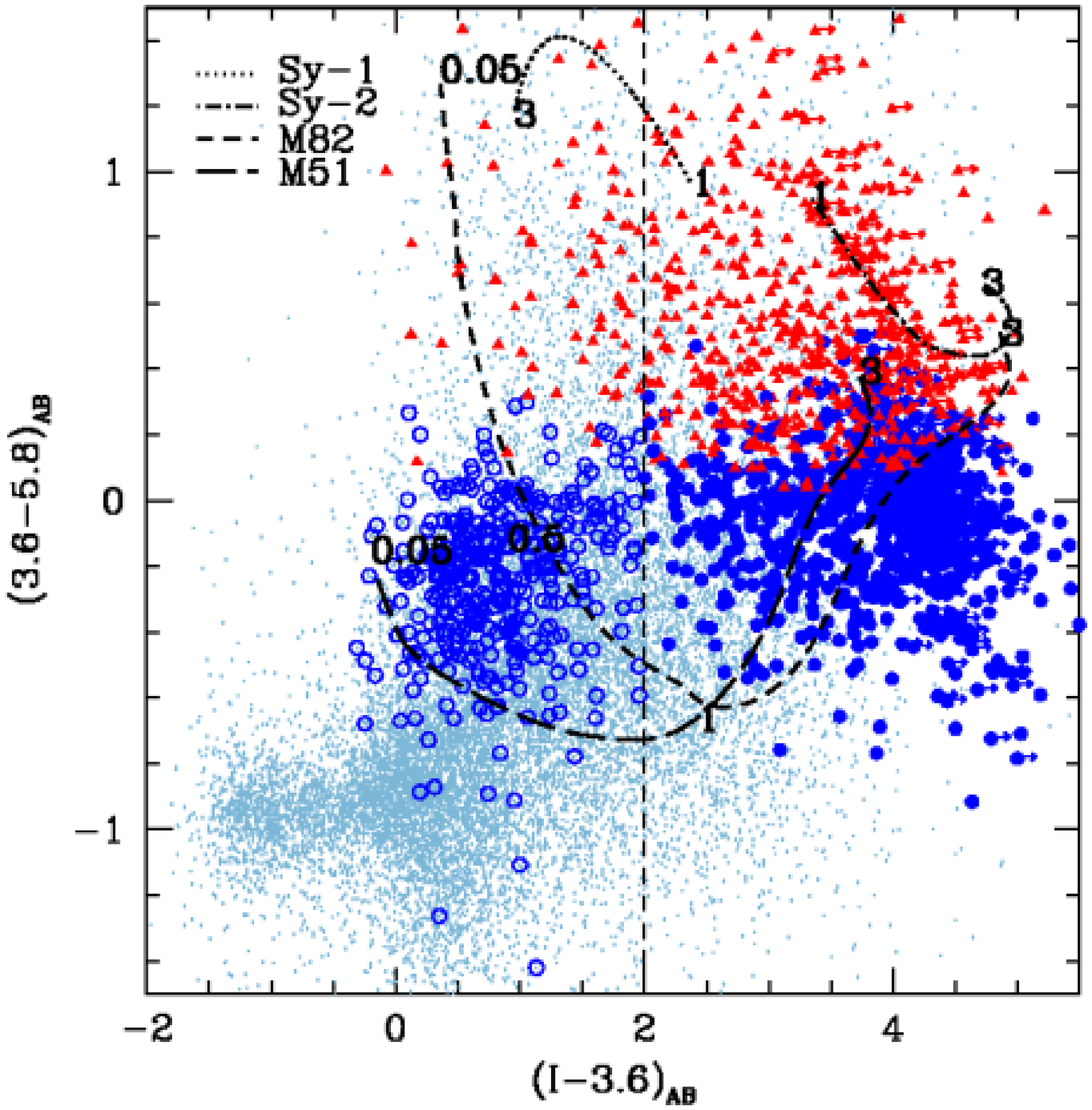}
\includegraphics[width=0.35\textwidth]{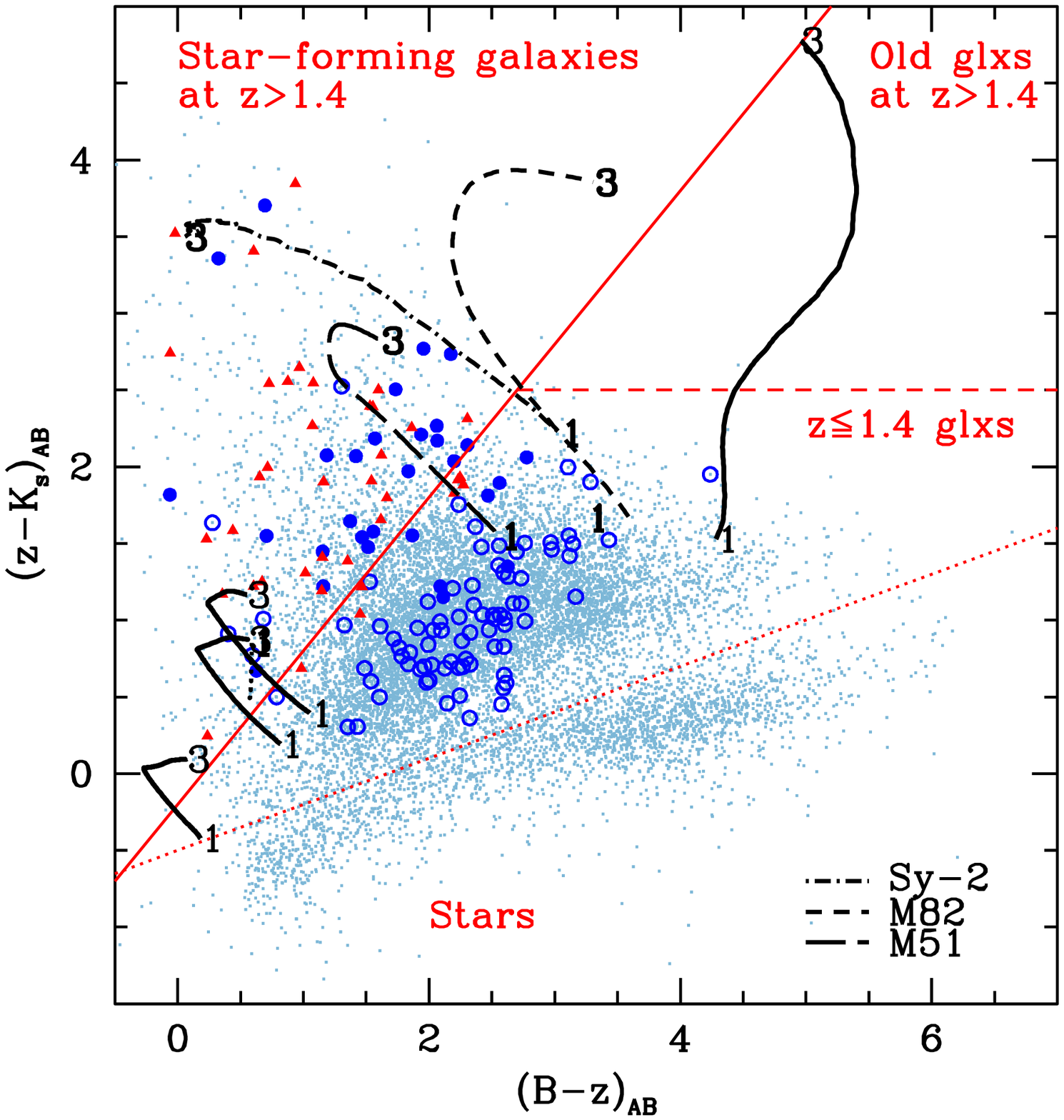}
\caption{Color-color plots adopted in the selection of IR-peakers. Filled
triangles and circles represent 5.8$\mu$m-peakers and 4.5$\mu$m-peakers
respectively. Open circles are 4.5$\mu$m-peakers identified as low-redshift
interlopers on the basis of $(I-3.6)_{AB}$. Small dots
trace the general SWIRE-ESIS population; orange dots are general sources with no
8.0$\mu$m detection (upper limits are used). Redshift tracks for different templates
are plotted: M82 (dusty starburst) and M51 (spiral) belong to the GraSil library \citep{silva1998};
the Sy1 and Sy2 templates represent Mrk231 and IRAS19254-7245
\citep{fritz2006,berta2003}. The black thick solid lines in the $BzK$ diagram (bottom-right panel)
represent three GraSil spiral models with no extinction (ages 0.2, 1.0, 2.0 Gyr) and
one GraSil elliptical (age 3 Gyr). Redshift values are reported along tracks.
The diagonal and vertical dashed lines in the left-hand diagrams represent the 
color conditions to identify low-redshift interlopers (see text).}
\label{fig:cc_plots_IB}
\end{figure*}

Several selection criteria have been defined to identify high redshift galaxies,
exploiting optical and near-IR colors (e.g. Ly-break galaxies, LBGs, Steidel et 
al. \citeyear{steidel1996}, \citeyear{steidel1999}; BX-BM, Adelberger et
al. \citeyear{adelberger2004}; distant red galaxies, DRGs, Franx et al.
\citeyear{franx2003}; extremely red objects, EROs, Cimatti et al.
\citeyear{cimatti2002c};  
distant obscured galaxies, DOGs, Dey et al. \citeyear{dey2008}; $BzK$, Daddi et al.
\citeyear{daddi2004}).
LBGs, BX-BM's and DOGs classes consist of very faint objects that require 
$U$-band and/or deeper optical observations than those available in the ELAIS-S1 field.
EROs and DRGs in ELAIS-S1 are being discussed in a forthcoming paper by 
Dias et al. (2008, in prep.).

Recently, \citet{berta2007b} and \citet{lonsdale2006} have shown that IRAC data
allow the selection of high-$z$ (massive) galaxies by identifying 
the restframe 1.6$\mu$m stellar peak \citep{sawicki2002,simpson1999} in the SEDs of
galaxies, redshifted to $z=1.5-3.0$. We called these objects {\em IR-peakers}
(informally known also as ``IR-bump galaxies'').

\citet{berta2007b} have presented the selection of IR-peakers and 
studied their stellar content. Here we refine and reinforce the selection using $I$ band data.
This allows us not only to avoid low-redshift interlopers, but also to extend the
search for these sources over the whole ESIS-VIMOS area ($\sim4$ deg$^2$),
instead of restricting it to the central square degree where optical {\em BVR} 
imaging is available to date.

The main IR-peak selection is based in the detection of the 1.6$\mu$m peak in
the IRAC domain. Two different classes of galaxies are identified, 
those for which the SED peaks in the 4.5$\mu$m  ($z=1-2$) and those 
in the  5.8$\mu$m band ($z=2-3$):

\begin{equation}\label{eq:4.5_peakers}
S_\nu(3.6)<S_\nu(4.5)>S_\nu(5.8)
\end{equation}
\begin{equation}\label{eq:5.8_peakers}
S_\nu(3.6)<S_\nu(4.5)<S_\nu(5.8)>S_\nu(8.0)
\end{equation}

5.8$\mu$m-peakers include also sources not detected at 8.0$\mu$m, but
whose 8.0$\mu$m upper limit is still consistent with Eq. \ref{eq:5.8_peakers}.
This selection is shown in the top panels of Fig. \ref{fig:cc_plots_IB}, in
the classic IRAC color-color diagrams \citep{lacy2004,stern2005}. Triangles and circles depict 5.8$\mu$m- and 4.5$\mu$m-peakers. Open symbols
represent low-redshift interlopers, identified on the basis of optical-IR colors, as described below. For clarity's
sake, we do not plot upper limit arrows for IR-peakers not detected at 8.0$\mu$m.
Redshift tracks for a variety of templates are shown, up to $z=3$ (see labels on plot).
Such a selection has also the advantage to avoid any contamination from
AGNs, which are characterized by a restframe near-IR power-law-like (monotonic)
SED. At the SWIRE 5.8$\mu$m 3$\sigma$ depth (25.8 $\mu$Jy),
973 5.8$\mu$m-peak galaxies and 938 4.5$\mu$m-peakers 
were identified (see Tab. \ref{tab:multilambda_stats}).

As shown by \citet{berta2007b}, unfortunately 4.5$\mu$m-peakers suffer a
significant
contamination from low redshift interlopers, when  the 4.5$\mu$m flux is boosted
by a strong 3.3$\mu$m PAH emission. On the other hand, in the case of
5.8$\mu$m-peakers, two photometric bands sample the SED on the blue side of the
1.6$\mu$m peak, therefore the conditions $S_\nu(3.6)<S_\nu(4.5)<S_\nu(5.8)$ provide a robust criterion to
avoid interlopers. 

Optical data provide a significant help in breaking the degeneracy and 
identifying interlopers in the 4.5$\mu$m-peak sample.
The bottom-left panel in Fig. \ref{fig:cc_plots_IB} shows the position of 
IR-peakers in the $(I-3.6)_{AB}$ vs. $(3.6-5.8)_{AB}$ space, over the whole 
ESIS-VIMOS area. According to template redshift tracks, those sources having
\begin{equation}\label{eq:interlopers}
(I-3.6)<2 \ \ [AB]
\end{equation}
are likely to be $z<1$ interlopers. This result is consistent with what found
by \citet{berta2007b}, based on $(K_s-3.6)$ and $(R_C-3.6)$ colors.

Thanks to $I$ band data, interlopers are then identified on IRAC color-color 
diagrams (open circles). 
Here, the diagonal dashed line depicts the
condition $(4.5-8.0)_{AB}=1.5\times(3.6-5.8)_{AB} +0.7$, already found in \citet{berta2007b}, 
and defined thanks to $K_s$ band data.
A total of 353 4.5$\mu$m-peakers at redshift $z>1$ are left.

The availability of $z$ band data over $\sim1$ deg$^2$ allows us to 
build the $BzK$ diagram \citep{daddi2004} and verify where
IR-peakers lie in this color-space (bottom-right diagram of Fig. 
\ref{fig:cc_plots_IB}). Only those IR-peakers detected in the $BzK$+IRAC bands
are taken into account, including 38 5.8$\mu$m-peakers and 30 4.5$\mu$m-peakers (excluding 
interlopers). 
The $BzK$ diagram confirms that the IR-peak selection effectively identifies
objects above $z\simeq1.4$, with our sources lying in the star-forming locus. 
The majority of $(I-3.6)<2$ interlopers fall in the $z<1.4$ region, with
a few sources contaminating the high-$z$ area. Roughly 30\% and 40\% of 
$BzK$ 4.5$\mu$m- and 5.8$\mu$m-peak galaxies are detected by MIPS at 24$\mu$m 
at the SWIRE depth. 

None of the $BzK$-detected IR-peakers falls in the passive
$z>1.4$ galaxy class. The median $K_s$ magnitude of the general $BzK$ population 
detected in the ESIS area is 20.14 $[$AB$]$. $BzK$ galaxies with $(z-K_s)>2.5$ 
have a median $K_s=20.65$ $[$AB$]$. A passive $BzK$ galaxy 
(with median $K_s$ magnitude), characterized by $(z-K_s)_{AB}>2.5$ and 
$(B-z)_{AB}\gtrsim2.7$, would be detected at $z>23.1$ and $B>25.8$ $[$AB$]$.
Unfortunately, this goes beyond the limit of the ESIS survey.

Finally, in addition to the confirmation 
coming from the test based of the $BzK$ criterion, 
it is worth to recall that
independent studies of IR-peakers based on optical Keck spectroscopy
\citep{berta2007a} and IRS mid-IR spectroscopy \citep{weedman2006} showed
that the IRAC-based broad-band selection successfully identifies galaxies 
above $z\sim1.5$.

\begin{figure*}[!ht]
\centering
\includegraphics[width=0.30\textwidth]{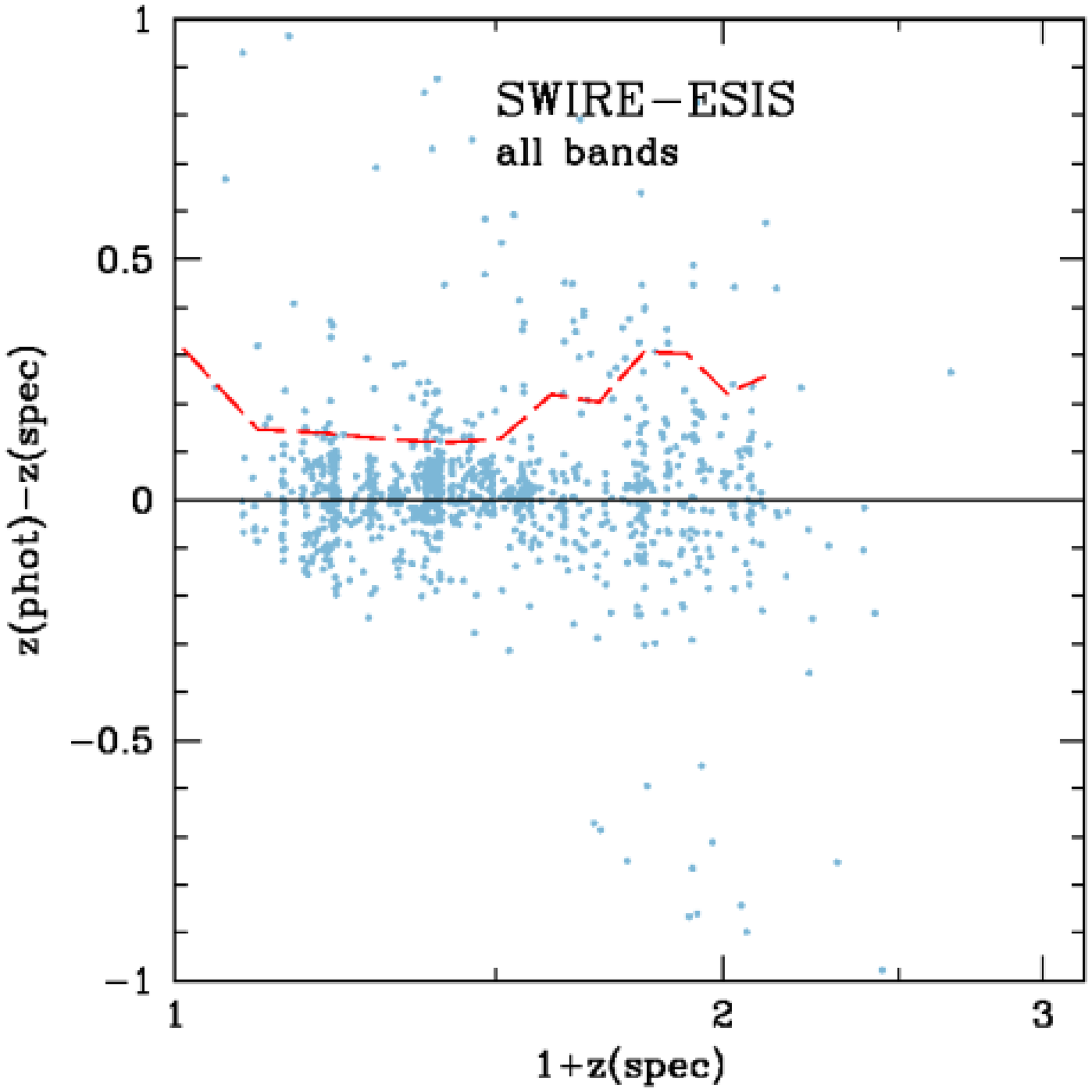}
\includegraphics[width=0.30\textwidth]{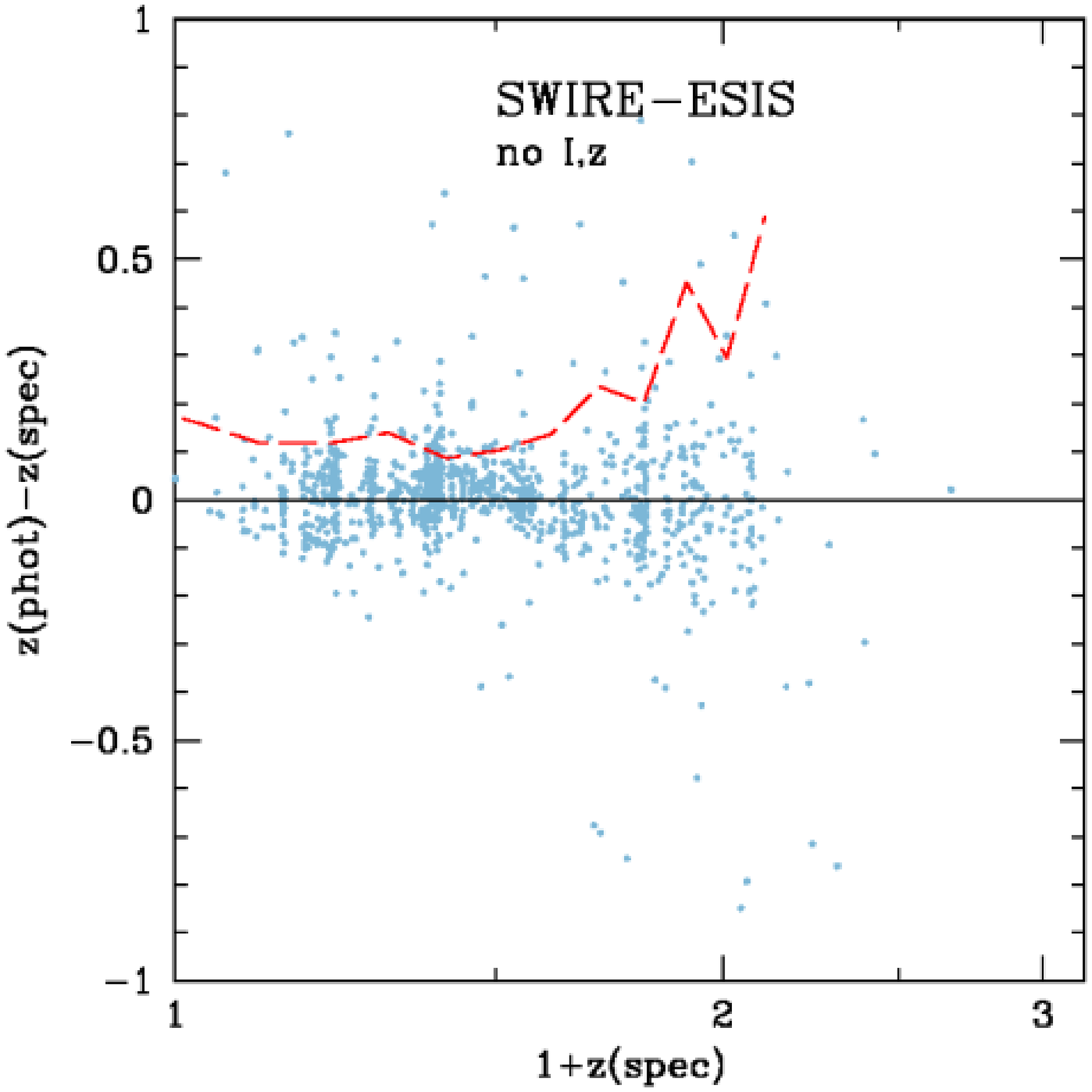}
\includegraphics[width=0.30\textwidth]{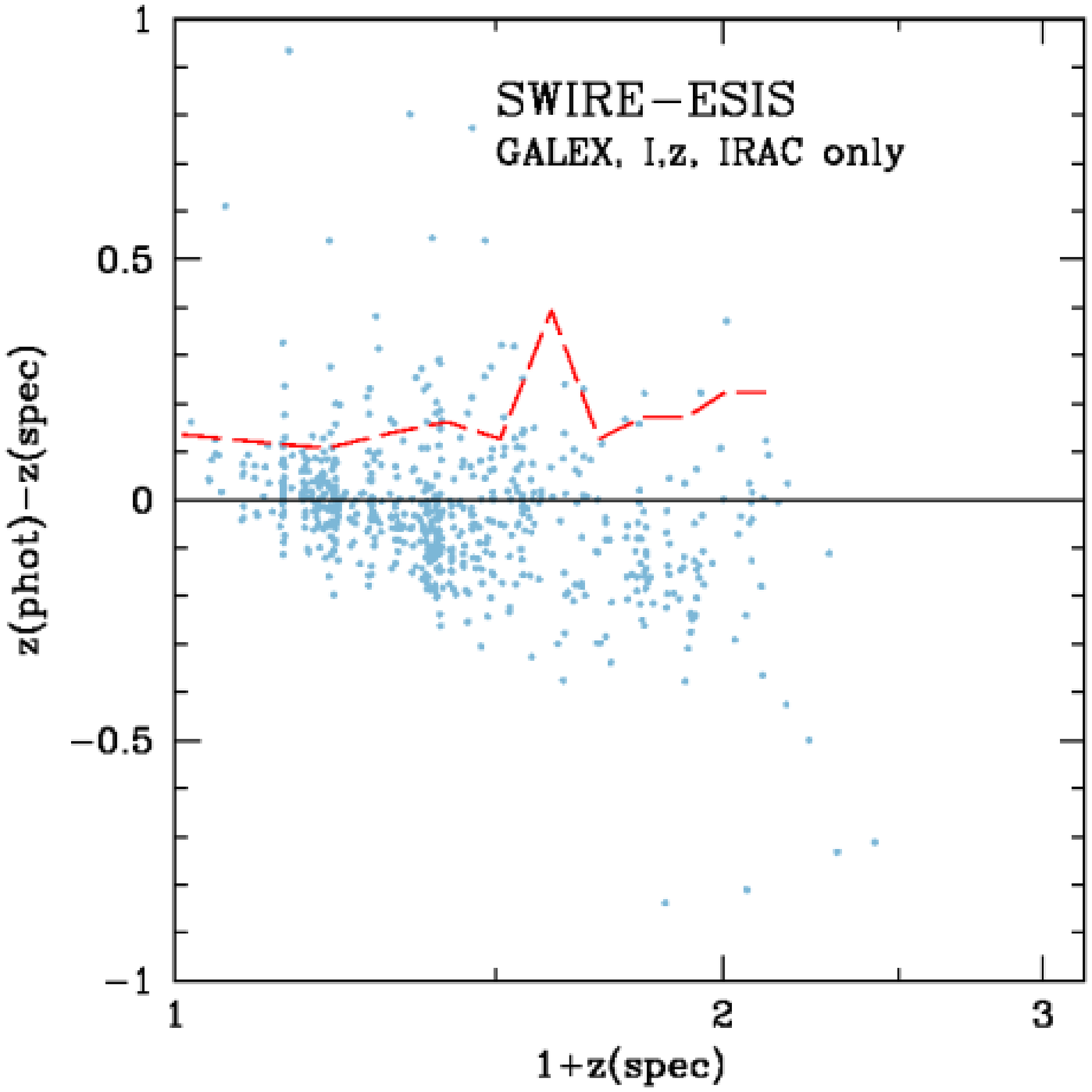}
\caption{Photometric redshift tests on galaxies in the SWIRE-ESIS field, run with EAZY code
(G. Brammer, P.~G. van Dokkum \& P. Coppi, {\em ApJ} submitted). {\em Left}: all photometric bands; {\em center}: all
bands excepted $I$ and $z$; {\em right}: GALEX, $I$ and IRAC only. The dashed
lines mark the $1\sigma$ scatter in $z_{phot}-z_{spec}$. Spectroscopic data
belong to Feruglio et al. (sub.), Sacchi et al. (in prep.) and \citet{lafranca2004}.}
\label{fig:eazy_test}
\end{figure*}

\subsection{Photometric redshift}\label{sect:photo_z}

When dealing with a very large number of sources, as in the case of our wide area
survey, the estimate of galaxy redshifts is mainly based on photometric
techniques. 

Let us check the advantage of introducing VIMOS $I$ and $z$ bands in the
estimate of photometric redshifts for the 
{\em general} galaxy population in the SWIRE-ESIS survey, and explore whether
the data in hand are sufficient to produce reliable photometric redshifts in
those areas not covered by other optical (e.g. {\em BVR}) observations.

To this aim, we have exploited the capabilities of the newly public photometric
redshift code EAZY (G. Brammer, P.~G. van Dokkum \& P. Coppi, {\em ApJ} submitted), which combines a maximum
likelihood (ML) algorithm and Bayesian constraints on the expected redshift
distribution of galaxies.
Bayesian techniques in photometric redshift estimates have been thoroughly
studied and applied to large datasets in the past with good results 
\citep[e.g.][]{benitez2000}. An interesting comparison between Bayesian
and pure ML techniques was performed by \citet{hildebrandt2008}, who show the advantage of the
former, especially in avoiding aliases, minimizing the incidence of interlopers,
and providing a reliable estimate of the redshift uncertainties.
The EAZY code has the advantage of being very user-friendly and specifically
designed for multi-wavelength surveys.

Figure \ref{fig:eazy_test} shows the results of these tests, comparing
photometric and spectroscopic (Feruglio et al., sub.; Sacchi et al., in prep.; La Franca et al.
\citeyear{lafranca2004}) redshifts. 
In the left panel, photometric redshift were estimated by using all the
available UV-optical-IR data (GALEX, ESIS, $J,\ K_s$, IRAC), while the central
panel does not account for $I$ and $z$ data. Finally the right-hand plot shows
photometric redshifts obtained including only GALEX, $I$ band and IRAC fluxes
(i.e. those data available over the whole ELAIS-S1 area).

When introducing $I$ and $z$ --- in combination with all the other available
data --- one would expect a significant improvement in the photometric redshift
estimate. Actually, these data are effective in constraining photometric
redshifts, only above redshift $z\sim0.8$, when the Balmer break lies in or red
ward the $R$ band. Unfortunately, only few spectroscopic redshifts are available
in this redshift domain, in ELAIS-S1. Nevertheless, one can appreciate some
improvement in the scatter of $z_{phot}-z_{spec}$ for $z>0.8$, where $\sigma_{\Delta z}$
reduces from 0.36 to 0.25. No significant improvement is seen at lower 
redshifts ($\sigma_{\Delta z}\sim0.15$).

On the other hand, when considering only GALEX, $I$ band and IRAC data, the 
result reverses. At low redshift the match between photometric and
spectroscopic $z$ is reasonably good, and the scatter is not dramatically larger
than in the previous cases ($\sigma_{\Delta z}\simeq0.17$). The code tends to systematically underestimate the
redshift, but the drift is smaller than $\Delta (z)\simeq 0.05$. 
At redshifts larger than $z\sim0.6$,
the results become rather poor: the scatter increases, and the systematic shift
is no longer negligible. At $z>0.8$ we have a systematic $\Delta z\simeq -0.15$.
This effect can probably be explained with the difficulty
to detect galaxies at higher redshift with GALEX. An obvious exception is the case of 
IR-peakers, for which the IRAC detection of 1.6$\mu$m stellar peak constrains
the redshift \citep[e.g.][]{berta2007b,berta2007a,weedman2006}.

The fraction of catastrophic outliers\footnote{Catastrophic outliers 
are defined here as sources having $\Delta z>3 \sigma_{\Delta z}$.} increases 
from 5\% to 10\%, when using only GALEX, $I$ and IRAC bands.

Generally speaking, \citet{rowanrobinson2008}
showed that at least 5 different photometric bands are needed, in order to obtain
a reliable photometric redshift estimate, including 2 IRAC fluxes, and (for
example) 3 optical. ELAIS-S1 will benefit of three additional optical bands over
the whole field, once the ESIS-WFI data processing will be completed and presented in a
forthcoming paper.

\section{Summary}\label{sect:summary}

In the era of wide-area cosmological surveys,
we have presented $I$ and $z$ band wide field imaging of ELAIS-S1,
carried out with the VIMOS camera on VLT. 
These observations are part of the {\em ESO-Spitzer Imaging 
extragalactic Survey} (ESIS) and cover $\sim4$ deg$^2$ in the $I$ band and 
$\sim1$ deg$^2$ in $z$. The nominal exposure time is 1800s and 3600s in $I$ and $z$
respectively, but the coverage is not uniform, due to the very complex 
observing pattern.

As a result of accurate data processing, nine independent mosaics for 
the $I$ band and one single, 1 deg$^2$ wide frame in the $z$ band were produced.
The final catalogs include more than 300000
$I$ band sources and 50000 $z$ band objects.
Equatorial coordinates of VIMOS objects are constrained within $\sim0.2$ r.m.s. arcsec, when 
compared to the GSC 2.2 catalog. Relative astrometric accuracy between the nine 
$I$ band images is as good as $\sim0.1$ arcsec r.m.s.

Photometric uncertainties were estimated through simulations, and turn out to be 
$\sim 0.05,\ 0.20$ mag at $I=21.5,\ 23.5$ and slightly larger in the $z$ band.
The survey reaches a 90\% average completeness at 
23.1 and 22.5 (Vega magnitudes) in the $I$ and
$z$ band respectively. 

The ESIS VIMOS survey provides optical $I$ counterparts for $>60$\% 
(137206/219877) of SWIRE/Spitzer sources in the ELAIS-S1 area.

We have exploited the new $I$ and $z$ band data --- in combination to the other 
multi-wavelength observations available in ELAIS-S1 --- 
to search for high-redshift ($z=1-3$) galaxies.
Our selection is based on the detection of 
the 1.6$\mu$m stellar peak in the IRAC wavelength domain \citep[IR-peakers][]{berta2007b}.
The $I$ band data give a valuable constraint on IR-peakers: requiring
$(I-3.6)_{AB}>2$ avoids low-redshift ($z\lesssim0.6$) interlopers. 
The availability of VIMOS data over 
$\sim4$ deg$^2$ allows the selection to be extended over the whole SWIRE-ESIS 
field. A total of 973 5.8$\mu$m-peak galaxies at $z=2-3$ 
were found over $\sim4$ deg$^2$ at the SWIRE 5.8$\mu$m 3$\sigma$ depth (25.8 $\mu$Jy).

The $z$ band data were used to perform the $BzK$ \citep{daddi2004} selection over 
$\sim1$ deg$^2$ and verify the consistency of the IR-peakers class.
Our high-$z$ objects lie in the locus of $z>1.4$ star forming galaxies. 
Apparently, no passive IR-peakers are detected at SWIRE-ESIS depth, 
over the central ESIS 1 deg$^2$ field.

At the SWIRE depth, these galaxies have stellar masses $M>10^{11}$ M$_\odot$ at $z>2$.
The identification of a significant number of these rare high-$z$ sources is possible only 
thanks to the wide area probed by SWIRE-ESIS.
This survey is giving a significant contribution --- among others --- in 
studying the very massive tail of the stellar mass function at $z>2$,
and constraining the physical properties of its constituents \citep[e.g.][]{berta2007b,lonsdale2006}.

Updates on the ESIS project are available on the web page {\tt http://www.astro.unipd.it/esis}.


\begin{acknowledgements}
We wish to thank the anonymous referee for their useful comments 
which improved the presentation of our results.

S.B. was supported by University of Padova and INAF-OaPD grants.
S.R. was supported by a University of Padova grant.
We are grateful to G. Brammer for providing the EAZY code before its 
public release.

The Spitzer Space Telescope is operated by the Jet 
Propulsion Laboratory, California Institute of Technology, 
under contract with NASA. 
SWIRE was supported by NASA through the Spitzer Legacy Program under 
contract 1407 with the Jet Propulsion Laboratory.

We acknowledge NASA's support for construction,
operation, and science analysis for the GALEX mission, developed in cooperation
with the Centre National d'\'Etudes Spatiales of France and the Korean Ministry of
Science and Technology.

\end{acknowledgements}


\bibliographystyle{aa}
\bibliography{biblio_esis_vimos}   


\begin{table*}
\centering
\scriptsize
\begin{tabular}{l c c c c c || l c c c c c}
\hline
\hline
Field & OB & R.A. & Dec. & Date & A.M. & Field & OB & R.A. & Dec. & Date & A.M.\\
number &  & $[$hh:mm:ss$]$ & $[^{\circ}:^{\prime}:{\arcsec}]$ & $[$yyyy-mm-dd$]$ & & number &  & $[$hh:mm:ss$]$ & $[^{\circ}:^{\prime}:{\arcsec}]$ & $[$yyyy-mm-dd$]$ & \\
\hline
Field 1  &   A  	     &   00:32:48.00	& -44:24:00.0	&  2003-07-08	&	     1.509  &	Field 26 &   A  	     &   00:38:26.10	& -43:30:45.0	&  2003-11-19	&	     1.258  \\
     	 &   B  	     &   00:32:59.30	& -44:21:36.0	&  2003-11-18	&	     1.063  &	     	 &   B  	     &   00:38:37.24	& -43:28:21.0	&  2003-11-18	&	     1.066  \\
     	 &   C  	     &   00:33:10.59	& -44:19:12.0	&  2003-07-27	&	     1.454  &	     	 &   C  	     &   00:38:48.35	& -43:25:57.0	&  2003-11-16	&	     1.090  \\
Field 2  &   A  	     &   00:34:13.80	& -44:24:00.0	&  2003-07-08	&	     1.434  &	Field 27 &   A  	     &   00:39:50.63	& -43:30:45.0	&  2003-11-21	&	     1.383  \\
     	 &   B  	     &   00:34:25.10	& -44:21:36.0	&  2003-11-21	&	     1.239  &	     	 &   B  	     &   00:40:01.76	& -43:28:21.0	&  2003-07-10	&	     1.092  \\
     	 &   C  	     &   00:34:36.38	& -44:19:12.0	&  2003-07-27	&	     1.386  &       	 &   C  	     &   00:40:12.88	& -43:25:57.0	&  2003-11-16	&	     1.165  \\
Field 3  &   A  	     &   00:35:39.59	& -44:24:00.0	&  2003-07-08	&	     1.366  &	Field 28 &   A  	     &   00:41:15.15	& -43:30:45.0	&  2004-08-11	&	     1.064  \\
     	 &   B  	     &   00:35:50.90	& -44:21:36.0	&  2003-11-17	&	     1.270  &	     	 &   B  	     &   00:41:26.29	& -43:28:21.0	&  2004-08-21	&	     1.177  \\
     	 &   C  	     &   00:36:02.18	& -44:19:12.0	&  2003-07-27	&	     1.330  &	     	 &   C  	     &   00:41:37.41	& -43:25:57.0	&  2004-08-21	&	     1.068  \\
Field 4  &   A  	     &   00:37:05.39	& -44:24:00.0	&  2003-07-08	&	     1.314  &	Field 29 &   A  	     &   00:32:48.00	& -43:13:00.0	&  2003-08-03	&	     1.059  \\
     	 &   B  	     &   00:37:16.69	& -44:21:36.0	&  2003-11-18	&	     1.085  &	     	 &   B  	     &   00:32:59.08	& -43:10:36.0	&  2003-12-19	&	     1.286  \\
     	 &   C  	     &   00:37:27.98	& -44:19:12.0	&  2003-07-27	&	     1.232  &	     	 &   C  	     &   00:33:10.15	& -43:08:12.0	&  2004-07-17	&	     1.201  \\
Field 5  &   A  	     &   00:38:31.19	& -44:24:00.0	&  2003-07-08	&	     1.268  &	Field 30 &   A  	     &   00:34:12.11	& -43:13:00.0	&  2003-12-16	&	     1.086  \\
     	 &   B  	     &   00:38:42.49	& -44:21:36.0	&  2003-07-09	&	     1.268  &	     	 &   B  	     &   00:34:23.19	& -43:10:36.0	&  2003-12-21	&	     1.219  \\
     	 &   C  	     &   00:38:53.78	& -44:19:12.0	&  2003-11-16	&	     1.384  &	     	 &   C  	     &   00:34:34.26	& -43:08:12.0	&  2004-07-19	&	     1.104  \\
Field 6  &   A  	     &   00:39:56.99	& -44:24:00.0	&  2003-07-08	&	     1.225  &	Field 31 &   A  	     &   00:35:36.23	& -43:13:00.0	&  2003-12-16	&	     1.389  \\
     	 &   B  	     &   00:40:08.29	& -44:21:36.0	&  2003-11-18	&	     1.065  &	     	 &   B  	     &   00:35:47.31	& -43:10:36.0	&  2004-08-21	&	     1.364  \\
     	 &   C  	     &   00:40:19.57	& -44:19:12.0	&  2003-07-31	&	     1.069  &	     	 &   C  	     &   00:35:58.37	& -43:08:12.0	&  2004-08-13	&	     1.389  \\
Field 7  &   A  	     &   00:41:22.78	& -44:24:00.0	&  2004-07-12	&	     1.207  &	Field 32 &   A  	     &   00:37:00.34	& -43:13:00.0	&  2003-12-18	&	     1.175  \\
     	 &   B  	     &   00:41:34.09	& -44:21:36.0	&  2004-07-14	&	     1.074  &	     	 &   B  	     &   00:37:11.42	& -43:10:36.0	&  2004-07-22	&	     1.109  \\
     	 &   C  	     &   00:41:45.37	& -44:19:12.0	&  2004-08-21	&	     1.063  &	     	 &   C  	     &   00:37:22.49	& -43:08:12.0	&  2004-07-22	&	     1.067  \\
Field 8  &   A  	     &   00:32:48.00	& -44:06:15.0	&  2003-07-08	&	     1.170  &	Field 33 &   A  	     &   00:38:24.46	& -43:13:00.0	&  2003-07-08	&	     1.060  \\
     	 &   B  	     &   00:32:59.24	& -44:03:51.0	&  2003-11-18	&	     1.064  &	     	 &   B  	     &   00:38:35.54	& -43:10:36.0	&  2003-07-10	&	     1.075  \\
     	 &   C  	     &   00:33:10.47	& -44:01:27.0	&  2003-07-27	&	     1.542  &	     	 &   C  	     &   00:38:46.60	& -43:08:12.0	&  2003-11-16	&	     1.242  \\
Field 9  &   A  	     &   00:34:13.37	& -44:06:15.0	&  2003-07-08	&	     1.143  &	Field 34 &   A  	     &   00:39:48.57	& -43:13:00.0	&  2003-11-21	&	     1.461  \\
     	 &   B  	     &   00:34:24.61	& -44:03:51.0	&  2003-07-09	&	     1.144  &	     	 &   B  	     &   00:39:59.65	& -43:10:36.0	&  2003-07-10	&	     1.065  \\
     	 &   C  	     &   00:34:35.84	& -44:01:27.0	&  2003-07-27	&	     1.271  &	     	 &   C  	     &   00:40:10.72	& -43:08:12.0	&  2003-11-16	&	     1.133  \\
Field 10 &   A  	     &   00:35:38.73	& -44:06:15.0	&  2003-07-08	&	     1.121  &	Field 35 &   A  	     &   00:41:12.69	& -43:13:00.0	&  2004-08-11	&	     1.073  \\
     	 &   B  	     &   00:35:49.98	& -44:03:51.0	&  2003-07-09	&	     1.116  &	     	 &   B  	     &   00:41:23.77	& -43:10:36.0	&  2004-08-20	&	     1.196  \\
     	 &   C  	     &   00:36:01.21	& -44:01:27.0	&  2003-07-27	&	     1.188  &	     	 &   C  	     &   00:41:34.83	& -43:08:12.0	&  2004-08-21	&	     1.144  \\
Field 11 &   A  	     &   00:37:04.10	& -44:06:15.0	&  2003-07-08	&	     1.101  &	Field 36 &   A  	     &   00:32:48.00	& -42:55:15.0	&  2003-08-01	&	     1.112  \\
     	 &   B  	     &   00:37:15.35	& -44:03:51.0	&  2003-11-18	&	     1.099  &	     	 &   B  	     &   00:32:59.03	& -42:52:51.0	&  2003-08-03	&	     1.086  \\
     	 &   C  	     &   00:37:26.57	& -44:01:27.0	&  2003-07-27	&	     1.157  &	     	 &   C  	     &   00:33:10.04	& -42:50:27.0	&  2003-11-18	&	     1.117  \\
Field 12 &   A  	     &   00:38:29.47	& -44:06:15.0	&  2003-07-08	&	     1.086  &	Field 37 &   A  	     &   00:34:11.71	& -42:55:15.0	&  2003-11-19	&	     1.054  \\
     	 &   B  	     &   00:38:40.71	& -44:03:51.0	&  2003-07-09	&	     1.085  &	     	 &   B  	     &   00:34:22.74	& -42:52:51.0	&  2003-11-18	&	     1.064  \\
     	 &   C  	     &   00:38:51.94	& -44:01:27.0	&  2003-11-16	&	     1.293  &	     	 &   C  	     &   00:34:33.75	& -42:50:27.0	&  2003-11-18	&	     1.139  \\
Field 13 &   A  	     &   00:39:54.83	& -44:06:15.0	&  2003-07-08	&	     1.074  &	Field 38 &   A  	     &   00:35:35.42	& -42:55:15.0	&  2003-11-19	&	     1.054  \\
     	 &   B  	     &   00:40:06.08	& -44:03:51.0	&  2003-11-18	&	     1.060  &	     	 &   B  	     &   00:35:46.45	& -42:52:51.0	&  2003-11-19	&	     1.067  \\
     	 &   C  	     &   00:40:17.31	& -44:01:27.0	&  2003-11-16	&	     1.069  &	     	 &   C  	     &   00:35:57.46	& -42:50:27.0	&  2003-11-19	&	     1.058  \\
Field 14 &   A  	     &   00:41:20.20	& -44:06:15.0	&  2004-08-18	&	     1.133  &	Field 39 &   A  	     &   00:36:59.13	& -42:55:15.0	&  2003-11-19	&	     1.076  \\
     	 &   B  	     &   00:41:31.45	& -44:03:51.0	&  2004-07-14	&	     1.065  &	     	 &   B  	     &   00:37:09.70	& -42:52:51.0	&  2003-11-16	&	     1.053  \\
     	 &   C  	     &   00:41:42.68	& -44:01:27.0	&  2004-08-21	&	     1.061  &	     	 &   C  	     &   00:37:21.17	& -42:50:27.0	&  2003-11-19	&	     1.081  \\
Field 15 &   A  	     &   00:32:48.00	& -43:48:30.0	&  2003-12-17	&	     1.201  &	Field 40 &   A  	     &   00:38:22.84	& -42:55:15.0	&  2003-11-19	&	     1.205  \\
     	 &   B  	     &   00:32:59.19	& -43:46:06.0	&  2003-12-21	&	     1.138  &	     	 &   B  	     &   00:38:33.86	& -42:52:51.0	&  2003-11-16	&	     1.452  \\
     	 &   C  	     &   00:33:10.36	& -43:43:42.0	&  2004-07-18	&	     1.110  &	     	 &   C  	     &   00:38:44.88	& -42:50:27.0	&  2003-11-16	&	     1.534  \\
Field 16 &   A  	     &   00:34:12.94	& -43:48:30.0	&  2003-12-16	&	     1.251  &	Field 41 &   A  	     &   00:39:46.55	& -42:55:15.0	&  2003-11-19	&	     1.369  \\
     	 &   B  	     &   00:34:24.13	& -43:46:06.0	&  2004-07-25	&	     1.058  &	     	 &   B  	     &   00:39:57.57	& -42:52:51.0	&  2003-11-18	&	     1.080  \\
     	 &   C  	     &   00:34:35.31	& -43:43:42.0	&  2004-07-19	&	     1.059  &	     	 &   C  	     &   00:40:08.59	& -42:50:27.0	&  2003-11-16	&	     1.104  \\
Field 17 &   A  	     &   00:35:37.89	& -43:48:30.0	&  2003-12-17	&	     1.313  &	Field 42 &   A  	     &   00:41:10.26	& -42:55:15.0	&  2004-07-12	&	     1.163  \\
     	 &   B  	     &   00:35:49.07	& -43:46:06.0	&  2004-09-06	&	     1.064  &	     	 &   B  	     &   00:41:21.28	& -42:52:51.0	&  2004-08-20	&	     1.161  \\
     	 &   C  	     &   00:36:00.25	& -43:43:42.0	&  2004-09-06	&	     1.060  &	     	 &   C  	     &   00:41:32.29	& -42:50:27.0	&  2004-08-12	&	     1.148  \\
Field 18 &   A  	     &   00:37:02.83	& -43:48:30.0	&  2004-09-06	&	     1.109  &	Field 43 &   A  	     &   00:32:48.00	& -44:41:45.0	&  2004-06-21	&	     1.152  \\
     	 &   B  	     &   00:37:14.02	& -43:46:06.0	&  2004-08-15	&	     1.164  &	     	 &   B  	     &   00:32:59.36	& -44:39:21.0	&  2004-06-23	&	     1.127  \\
     	 &   C  	     &   00:37:25.19	& -43:43:42.0	&  2004-08-21	&	     1.073  &	     	 &   C  	     &   00:33:10.70	& -44:36:57.0	&  2004-06-23	&	     1.106  \\
Field 19 &   A  	     &   00:38:27.77	& -43:48:30.0	&  2003-11-19	&	     1.312  &	Field 44 &   A  	     &   00:34:14.23	& -44:41:45.0	&  2004-06-21	&	     1.127  \\
     	 &   B  	     &   00:38:38.96	& -43:46:06.0	&  2003-07-09	&	     1.063  &	     	 &   B  	     &   00:34:25.59	& -44:39:21.0	&  2004-06-26	&	     1.351  \\
     	 &   C  	     &   00:38:50.14	& -43:43:42.0	&  2003-11-16	&	     1.077  &	     	 &   C  	     &   00:34:36.94	& -44:36:57.0	&  2004-06-26	&	     1.290  \\
Field 20 &   A  	     &   00:39:52.72	& -43:48:30.0	&  2003-11-21	&	     1.324  &	Field 45 &   A  	     &   00:35:40.47	& -44:41:45.0	&  2004-06-21	&	     1.095  \\
     	 &   B  	     &   00:40:03.90	& -43:46:06.0	&  2003-07-10	&	     1.140  &	     	 &   B  	     &   00:35:51.83	& -44:39:21.0	&  2004-06-26	&	     1.134  \\
     	 &   C  	     &   00:40:15.08	& -43:43:42.0	&  2003-11-16	&	     1.200  &	     	 &   C  	     &   00:36:03.17	& -44:36:57.0	&  2004-07-25	&	     1.082  \\
Field 21 &   A  	     &   00:41:17.66	& -43:48:30.0	&  2004-08-20	&	     1.137  &	Field 46 &   A  	     &   00:37:06.70	& -44:41:45.0	&  2004-06-27	&	     1.123  \\
     	 &   B  	     &   00:41:28.85	& -43:46:06.0	&  2004-07-24	&	     1.069  &	     	 &   B  	     &   00:37:18.06	& -44:39:21.0	&  2004-08-15	&	     1.207  \\
     	 &   C  	     &   00:41:40.02	& -43:43:42.0	&  2004-08-21	&	     1.063  &	     	 &   C  	     &   00:37:29.41	& -44:36:57.0	&  2004-08-12	&	     1.098  \\
Field 22 &   A  	     &   00:32:48.00	& -43:30:40.0	&  2003-12-17	&	     1.121  &	Field 47 &   A  	     &   00:38:32.94	& -44:41:45.0	&  2004-06-27	&	     1.104  \\
     	 &   B  	     &   00:32:59.13	& -43:28:21.0	&  2003-12-19	&	     1.457  &	     	 &   B  	     &   00:38:44.30	& -44:39:21.0	&  2004-08-12	&	     1.114  \\
     	 &   C  	     &   00:33:10.25	& -43:25:57.0	&  2004-07-17	&	     1.117  &	     	 &   C  	     &   00:38:55.64	& -44:36:57.0	&  2004-08-20	&	     1.097  \\
Field 23 &   A  	     &   00:34:12.53	& -43:30:45.0	&  2003-12-16	&	     1.147  &	Field 48 &   A  	     &   00:39:59.17	& -44:41:45.0	&  2004-06-27	&	     1.089  \\
     	 &   B  	     &   00:34:23.66	& -43:28:21.0	&  2004-07-17	&	     1.061  &	     	 &   B  	     &   00:40:10.53	& -44:39:21.0	&  2004-08-20	&	     1.115  \\
     	 &   C  	     &   00:34:34.78	& -43:25:57.0	&  2004-07-19	&	     1.066  &	     	 &   C  	     &   00:40:21.88	& -44:36:57.0	&  2004-08-12	&	     1.133  \\
Field 24 &   A  	     &   00:35:37.05	& -43:30:45.0	&  2003-12-18	&	     1.110  &	Field 49 &   A  	     &   00:41:25.41	& -44:41:45.0	&  2004-08-21	&	     1.226  \\
     	 &   B  	     &   00:35:48.19	& -43:28:21.0	&  2004-08-21	&	     1.142  &	         &   B  	     &   00:41:36.77	& -44:39:21.0	&  2004-08-21	&	     1.128  \\
     	 &   C  	     &   00:35:59.30	& -43:25:57.0	&  2004-08-13	&	     1.240  &	  	 &   C  	     &   00:41:48.11	& -44:36:57.0	&  2004-08-21	&	     1.069  \\
Field 25 &   A  	     &   00:37:01.58	& -43:30:45.0	&  2003-12-18	&	     1.287  &		 &		     &			&		&		&		    \\
     	 &   B  	     &   00:37:12.71	& -43:28:21.0	&  2004-07-22	&	     1.057  &		 &		     &			&		&		&		    \\
     	 &   C  	     &   00:37:23.83	& -43:25:57.0	&  2004-08-13	&	     1.066  &		 &		     &			&		&		&		    \\
\hline
\end{tabular}
\normalsize
\caption{Log file of $I$ band ESIS VIMOS observations.}
\label{tab:log_I}
\end{table*}

\begin{table*}
\centering
\scriptsize
\begin{tabular}{l c c c c c}
\hline
\hline
Field & OB & R.A. & Dec. & Date & A.M.\\
number &  & $[$hh:mm:ss$]$ & $[^{\circ}:^{\prime}:{\arcsec}]$ & $[$yyyy-mm-dd$]$ & \\
\hline
Field 15  &    A	&      00:32:48.00  &	-43:48:30.0 &	 2003-12-17 &  1.250 \\
  	  &    B	&      00:32:59.19  &	-43:46:06.0 &	 2003-12-21 &  1.174 \\
  	  &    C	&      00:33:10.36  &	-43:43:42.0 &	 2004-07-18 &  1.088 \\
Field 16  &    A	&      00:34:12.94  &	-43:48:30.0 &	 2003-12-16 &  1.311 \\
  	  &    B	&      00:34:24.13  &	-43:46:06.0 &	 2004-07-25 &  1.063 \\
  	  &    C	&      00:34:35.31  &	-43:43:42.0 &	 2004-07-19 &  1.064 \\
Field 17  &    A	&      00:35:37.89  &	-43:48:30.0 &	 2003-12-17 &  1.386 \\
  	  &    B	&      00:35:49.07  &	-43:46:06.0 &	 2004-09-06 &  1.076 \\
  	  &    C	&      00:36:00.25  &	-43:43:42.0 &	 2004-09-06 &  1.058 \\
Field 18  &    A	&      00:37:02.83  &	-43:48:30.0 &	 2004-09-06 &  1.088 \\
  	  &    B	&      00:37:14.02  &	-43:46:06.0 &	 2004-08-15 &  1.130 \\
  	  &    C	&      00:37:25.29  &	-43:43:46.8 &	 2004-08-21 &  1.098 \\
Field 22  &    A	&      00:32:48.00  &	-43:30:45.0 &	 2003-12-17 &  1.154 \\
  	  &    B	&      00:32:59.13  &	-43:28:21.0 &	 2003-12-19 &  1.561 \\
  	  &    C	&      00:33:10.25  &	-43:25:57.0 &	 2004-07-17 &  1.093 \\
Field 23  &    A	&      00:34:12.53  &	-43:30:45.0 &	 2003-12-16 &  1.186 \\
  	  &    B	&      00:34:23.66  &	-43:28:21.0 &	 2004-07-17 &  1.057 \\
  	  &    C	&      00:34:34.78  &	-43:25:57.0 &	 2004-07-19 &  1.058 \\
Field 24  &    A	&      00:35:37.05  &	-43:30:45.0 &	 2003-12-18 &  1.138 \\
  	  &    B	&      00:35:48.19  &	-43:28:21.0 &	 2004-08-21 &  1.181 \\
  	  &    C	&      00:35:59.30  &	-43:25:57.0 &	 2004-08-13 &  1.191 \\
Field 25  &    A	&      00:37:01.58  &	-43:30:45.0 &	 2003-12-18 &  1.372 \\
  	  &    B	&      00:37:12.71  &	-43:28:21.0 &	 2004-07-22 &  1.061 \\
  	  &    C	&      00:37:23.83  &	-43:25:57.0 &	 2004-08-13 &  1.059 \\
Field 29  &    A	&      00:32:48.00  &	-43:13:00.0 &	 2003-08-03 &  1.055 \\
  	  &    B	&      00:32:59.08  &	-43:10:36.0 &	 2003-12-19 &  1.355 \\
  	  &    C	&      00:33:10.15  &	-43:08:12.0 &	 2004-07-17 &  1.159 \\
Field 30  &    A	&      00:34:12.11  &	-43:13:00.0 &	 2003-12-16 &  1.107 \\
  	  &    B	&      00:34:23.19  &	-43:10:36.0 &	 2003-12-21 &  1.273 \\
  	  &    C	&      00:34:34.26  &	-43:08:12.0 &	 2004-07-19 &  1.082 \\
Field 31  &    A 	&     00:35:36.23   &  -43:13:00.0  &	2003-12-16  & 1.478  \\
  	  &    B 	&     00:35:47.31   &  -43:10:36.0  &	2004-08-21  & 1.293  \\
  	  &    C 	&     00:35:58.37   &  -43:08:12.0  &	2004-08-13  & 1.314  \\
Field 32  &    A 	&     00:37:00.34   &  -43:13:00.0  &	2003-12-18  & 1.219  \\
 	  &    B 	&     00:37:11.42   &  -43:10:36.0  &	2004-07-22  & 1.086  \\
  	  &    C 	&     00:37:22.49   &  -43:08:12.0  &	2004-07-22  & 1.058  \\
\hline
\end{tabular}
\normalsize
\caption{\label{tab:log_z} Log file of $z$ band ESIS VIMOS observations.}
\end{table*}


\begin{appendix}

\section{Treatment of fringing in VIMOS $I$ and $z$ band
frames}\label{sect:app_fringing}

As mentioned in Section \ref{sect:datared}, the VIMOS $I$ and $z$ band frames
suffer a strong fringing that needs to be corrected before photometric
calibration and mosaicking. Here we describe the main causes of fringing 
and the procedure adopted to clean ESIS VIMOS images.

\subsection{Fringing effects}\label{sect:fringing}

At wavelengths shorter than $\sim$7000\AA, the silicon layer of VIMOS CCDs
absorbs photons efficiently, close to the surface of the chip. 
On the other hand, in the $I$ and $z$ bands the coefficient of absorption
of silicon falls rapidly with increasing wavelength.
Consequently a significant number of low energy photons 
passes through the silicon, which acts as a thin film against
the higher index of refraction of the substrate: light is 
reflected back and forth between the two silicon boundaries
\citep[e.g.][]{malumuth2003}. 

The crossing and re-crossing waves mutually interfere, 
reinforce or cancel in amplitude, depending on their wavelength 
relative to the film thickness.
If the thickness of the Si layer varies across the chip, 
then ``fringes'' are formed, following paths of constant thickness. 

If the CCD is illuminated by mono-chromatic light, with a proper wavelength, 
sharp fringes are produced. If a uniform broad-band source is used, the range of
wavelengths washes out the appearance of fringes, and their pattern 
will have a reduced amplitude.

The night sky 
is dominated by emission lines, particularly at the red end of the optical
domain. The fringe pattern formed on VIMOS images in the $I$ and $z$ bands is
hence dominated by the combination of the thin-film interference for the different 
monochromatic sky lines. 

The observed fringe patterns therefore depend not only on
the silicon thickness across the chip, but also on which lines 
are strong in the night sky. 

Since the composition and conditions of the 
sky should not vary by large amounts across one night, at a first look the
observed fringe patterns do not vary strongly from one image 
to another.  
Nevertheless, the amplitude of the
fringes, and their strength relative to the uniform sky background may vary
significantly. Any continuum variation
independent on the line emission does change the uniform background relative to
the fringe pattern. Some examples are excess scattered light from the moon or
from the sun near twilight.

In addition to changes in the
continuum background, also the line strengths may vary, as a consequence of 
changes in the high-altitude atmospheric conditions (e.g. temperature,
solar-wind particles). In this case, if the line ratios change significantly,
not only the amplitude of fringes changes, but the whole fringe pattern
may shift. Even if the fringe pattern is mostly stable,
second-order changes make it very difficult to apply a
single pattern to all images taken during a full observing night \citep[see also
the description of fringing by Elixir,][]{magnier2004}.

\subsection{De-fringing}\label{sect:de_fringe}

The additive fringing component described in Section \ref{sect:fringing}
must be removed from all science images, by subtracting a template fringe frame,
adequately scaled to the amplitude of fringes in each image.

As the light from stars and galaxies is mainly dominated by 
continuum emission, with a negligible fraction from emission lines, 
their contribution to fringing is minimal (see Sect. \ref{sect:fringing}), 
and fringes are basically produced by the night sky background.

The ideal {\em fringe pattern} frame would be obtained from blank sky images 
taken in the same conditions as science frames. 
The optimal estimation of the fringe pattern is thus obtained from the science
images themselves: by combining all frames obtained at different pointings
across each observing night, it is possible to obtain an improved sky flat
field \citep[see][]{berta2006} containing the fringe information. 
The procedure adopted to extract the fringe pattern is as follows:

\begin{itemize}
\item first of all, mask all astronomical objects from the science images,
by using the {\tt objmasks} task in the MSCRED IRAF package.
\item very bright stars and extended galaxies 
produce extended halos on the images, which need to be masked manually.
\item all science frames are combined together, using
object-masks. The result is an {\em improved sky flat field}, without any
astronomical source, showing strong fringes.
\item the {\em fringe pattern} is then isolated by dividing the improved sky flat
field by a boxcar-median frame.
\end{itemize}

The fringe pattern frame is finally subtracted from each science image, 
using an appropriate scaling factor in order to account for the possible 
changes in fringes intensity, relative to the background, and minimize the
residuals.

If sky conditions were perfectly stable across the whole observing night,
one could use the same fringe pattern to correct all images taken during the
night. In practice --- with the exception of few fortunate nights --- 
sky conditions are far from being perfectly consistent for the whole duration of
observations and it is necessary to produce several (three or more) fringe
frames per night, splitting it in contiguous shorter bits. As a consequence, the 
number of frames used to produce each pattern decreases. 

In the worst cases, it is possible that the frames used to produce a given
fringe pattern are not pointing in significantly different regions of the sky. This case
is particularly critical if very bright stars lie in the field of view, because
there exist regions of the chip where fringes are not sampled, as a consequence of heavy
masking of big haloes. 

An example of the de-fringing procedure is depicted in Fig. \ref{fig:defringing},
where we show one VIMOS frame before (left panel) and after (right panel)
subtracting the fringe pattern (central panel).

After fringe removal, it is finally possible to apply a super-sky-flat ({\em
ss-flat}) \citep[see][]{berta2006} in order to correct for residual second order 
inhomogeneities in the background. Super-sky-flat frames were produced by
combining all de-fringed science frames and masking astronomical sources.
Again, because of sky variability, several {\em ss-flat} frames were needed
during each night.

\end{appendix}

\end{document}